\documentclass[12pt]{article}

\ifx\pdfoutput\undefined
\usepackage[dvips,bookmarks]{hyperref}
\else
\usepackage{hyperref}
\fi
\hypersetup{colorlinks=false,bookmarksopen,bookmarksnumbered,citecolor=blue,
   pdfstartview=FitH}

\usepackage[pdftex]{graphicx}	
\usepackage{latexsym}

\usepackage{amssymb,amsfonts,amsmath}
\usepackage{graphicx} 
\usepackage{indentfirst}

 \usepackage{bbm}

\parskip=\medskipamount

\newcommand {\cC}{{\cal C}}
\newcommand {\cD}{{\cal D}}
\newcommand {\cE}{{\cal E}}

\newcommand {\cJ}{{\cal J}}
\newcommand {\cK}{{\cal K}}
\newcommand {\cL}{{\cal L}}
\newcommand {\cM}{{\cal M}}
\newcommand {\cN}{{\cal N}}

\newcommand {\cR}{{\cal R}}
\newcommand {\cS}{{\cal S}}

\newcommand {\cV}{{\cal V}}

\newcommand {\cX}{{\cal X}}

\def\a{\alpha}

\def\b{\beta}
\def\c{\chi}
\def\d{\delta}
\def\e{\epsilon}
\def\f{\phi}
\def\g{\gamma}
\def\G{\Gamma}

\def\j{\psi}
\def\k{\kappa}
\def\l{\lambda}
\def\m{\mu}

\def\q{\theta}
\def\r{\rho}
\def\s{\sigma}
\def\t{\tau}

\def\x{\xi}
\def\z{\zeta}
\def\D{\Delta}
\def\F{\Phi}
\def\J{\Psi}
\def\L{\Lambda}
\def\O{\Omega}
\def\P{\Pi}

\def\S{\Sigma}
\def\U{\Upsilon}
\def\X{\Xi}

\def\rd{{\rm d}}
\def\ri{{\rm i}}
\def\re{{\rm e}}

\newcommand{\ad}{{\dot{\alpha}}}                           
\newcommand{\ve}{\varepsilon}                            
\newcommand{\cDB}{{\bar\cD}}                            

\newcommand{\pa}{\partial}                           
\newcommand{\hf}{\frac12}

\newcommand{\vf}{\varphi}

\newcommand{\be}{\begin{equation}}
\newcommand{\ee}{\end{equation}}
\newcommand{\bea}{\begin{eqnarray}}
\newcommand{\eea}{\end{eqnarray}}
\newcommand{\non}{\nonumber}
\newcommand{\bm}[1]{\mbox{\boldmath$#1$}}

\def\double #1{#1{\hbox{\kern-2pt $#1$}}}

\newcommand{\teb}{{\bar{\theta}}}
\newcommand{\qb}{{\bar{\theta}}}

\newif\ifdtup

\newcommand{\bsubeq}{\begin{subequations}}
\newcommand{\esubeq}{\end{subequations}}

\numberwithin{equation}{section}

\begin{document}
\begin{titlepage}
\begin{flushright}
October, 2017\\
\end{flushright}
\vspace{5mm}

\begin{center}
{\Large \bf Two-form supergravity,  superstring couplings, 
and Goldstino superfields  in three dimensions 
}\\ 
\end{center}

\begin{center}

{\bf 
Evgeny I. Buchbinder${}^{a}$, Jessica Hutomo${}^{a}$, 
Sergei M. Kuzenko${}^{a}$ 
and
\\
Gabriele Tartaglino-Mazzucchelli${}^{b}$
} \\
\vspace{5mm}

\footnotesize{
${}^{a}${\it School of Physics and Astrophysics M013, The University of Western Australia\\
35 Stirling Highway, Crawley W.A. 6009, Australia}}  
~\\
\vspace{4mm}
\footnotesize{
${}^{b}${\it Instituut voor Theoretische Fysica, KU Leuven,\\
Celestijnenlaan 200D, B-3001 Leuven, Belgium}
}
\vspace{2mm}
~\\
Email: \texttt{evgeny.buchbinder@uwa.edu.au, 20877155@student.uwa.edu.au,
sergei.kuzenko@uwa.edu.au, 
gabriele.tartaglino-mazzucchelli@kuleuven.be}\\
\vspace{2mm}

\end{center}

\begin{abstract}
\baselineskip=14pt
We develop off-shell formulations for $\cN=1$ and $\cN=2$ anti-de Sitter 
supergravity theories in three spacetime dimensions that contain gauge two-forms in the auxiliary field sector. 
These formulations are shown to  allow consistent couplings of supergravity  
to the Green-Schwarz superstring with $\cN=1$ or $\cN=2$ 
spacetime supersymmetry. In addition to being $\k$-symmetric, 
the Green-Schwarz superstring actions constructed
are also invariant under super-Weyl transformations of the target space. 
We also present a detailed study of models for spontaneously broken local supersymmetry in three dimensions obtained by coupling the known 
off-shell $\cN=1$ and $\cN=2$ supergravity theories to nilpotent 
Goldstino superfields. 
\end{abstract}

\vfill

\vfill
\end{titlepage}

\newpage
\renewcommand{\thefootnote}{\arabic{footnote}}
\setcounter{footnote}{0}

\tableofcontents{}
\vspace{1cm}
\bigskip\hrule

\allowdisplaybreaks


\section{Introduction}
\setcounter{equation}{0}

The Green-Schwarz superstring action with $\cN=1$ or $\cN=2$ 
supersymmetry \cite{GreenSchwarz} exists for spacetime dimensions $D=3,4,6$ and  10. 
However its light-cone quantisation breaks Lorentz invariance unless either $D=10$
(see, e.g., \cite{GSW}), which corresponds to critical superstring theory, 
or $D=3$ \cite{MT1,MT2}. Due to the exceptional status of the $D=3$ case, 
it is of interest to study in more detail three-dimensional (3D) superstring actions 
in supergravity backgrounds. 
In order for such a  coupling to supergravity to be consistent, 
the superstring action must possess a local fermionic invariance 
(known as the $\kappa$-symmetry) which was first discovered 
in the cases of  massive \cite{deAL1,deAL2} and 
massless  \cite{Siegel:1983hh} superparticles.\footnote{For 
reviews of various aspects of 
the $\kappa$-symmetry, see, e.g.,  \cite{Sezgin:1993xg,Sorokin}.}
The $\k$-symmetry, in its turn, requires the superstring action to include 
a Wess-Zumino term associated with a closed super 3-form in curved superspace
such that (i) it is the field strength of a gauge super 2-form;  
and (ii) it reduces to  a non-vanishing invariant super 3-form in the flat superspace limit. The latter requirement means that only certain supergravity formulations
are suitable to describe string propagation in curved superspace.
The constraints on the geometry of curved $D=3,4,6,10$ superspace,  
which are required for the coupling of supergravity  
to the Green-Schwarz superstring, were studied about
thirty years ago 
\cite{Henneaux:1984mh,Witten:1985nt,Grisaru:1985fv,Bergshoeff:1985su}. 
Nevertheless, there still remain some open questions and unexplored cases, 
as can be seen from the recent work by Tseytlin and Wulff  \cite{Wulff:2016tju} 
that determined the precise constraints imposed on the 10D target superspace geometry by the requirement of classical $\kappa$-symmetry of the Green-Schwarz superstring. 
In regard to the 3D case, it should be kept in mind that at the time when Refs. 
\cite{Grisaru:1985fv,Bergshoeff:1985su} were written, those
off-shell formulations for $\cN=1$ and $\cN=2$ supergravity theories, 
which are suitable to describe consistent superstring propagation, 
had not been described in the literature. 
One such theory, the so-called $\cN=2$
two-form supergravity, was formulated six years ago \cite{KT-M11}.
A new $\cN=2$ supergravity theory will be given in the present paper. 

The present work aims at developing:
(i)  $\cN=1$ and $\cN=2$ anti-de Sitter (AdS) 
supergravity theories  that contain gauge two-forms in the auxiliary field sector; 
(ii) consistent couplings of these supergravity theories to the Green-Schwarz superstring 
with $\cN=1$ or $\cN=2$ supersymmetry; and (iii) models for spontaneously broken 3D supergravity obtained by coupling the off-shell
$\cN=1$ or $\cN=2$ supergravity theories to Goldstino superfields. 
The first two goals are related to the above discussion. 
As to point (iii), it requires additional comments.

In the last three years, there has been considerable  interest in 
models for spontaneously broken $\cN=1$ local supersymmetry 
in four dimensions
 \cite{ADFS,DFKS,BFKVP,HY,K15,KW,SvdWW,BMST,FKRR,BHKMS,KMcAT-M,BK17}, 
 including the models for off-shell supergravity coupled to nilpotent Goldstino superfields. 
 One of the reasons for this interest is that 
 a positive contribution to the cosmological constant is generated once the local supersymmetry becomes spontaneously broken. 
 For instance, if the supergravity multiplet is coupled to an irreducible Goldstino superfield
 \cite{LR,SW,KTyler,K15,BHKMS} (with the Volkov-Akulov Goldstino \cite{VA,AV} 
being the only independent component field of the superfield), 
a {\it universal} positive contribution to the cosmological constant is 
generated,\footnote{The gravitino becomes massive in accordance 
with the super-Higgs effect \cite{VS,VS2,DZ}.}
which is proportional to $f^2$, with 
the parameter $f$ setting the scale of supersymmetry breaking. 
The same positive contribution is generated by the reducible Goldstino 
superfields used in the models studied in \cite{BFKVP,HY,KMcAT-M}.\footnote{The 
notion of irreducible and reducible Goldstino superfields was introduced
in \cite{BHKMS}.} There is one special reducible Goldstino superfield,
the nilpotent three-form multiplet introduced in \cite{FKRR,BK17}, which yields 
a dynamical positive contribution to the cosmological constant. 

Since our universe is characterised by a positive cosmological constant, 
and a theoretical explanation for this positivity is required, 4D supergravity theories 
with nilpotent Goldstino superfields deserve further studies. 
In this respect, it is also of some interest to construct models 
for spontaneously broken local 3D  $\cN=1$ and $\cN=2$ supersymmetry
that are obtained  by coupling off-shell 3D supergravity to nilpotent superfields.
This is one of the objectives of the present work.

This paper is organised as follows. Sections 2 and 3 provide thorough discussions 
of the $\cN=1$ and $\cN=2$ off-shell supergravity theories, respectively. 
Section 4 describes consistent couplings of the two-form supergravity theories 
to the Green-Schwarz superstring with $\cN=1$ or $\cN=2$ 
spacetime supersymmetry. The nilpotent Goldstino superfields and their couplings
to various off-shell supergravity theories are presented in section 5.
Here we introduce only those reducible Goldstino superfields that are defined 
in the presence of  conformal supergravity without making use of any conformal compensator.  Section 6 contains concluding comments and 
a brief discussion of the results obtained. 
The main body of the paper is accompanied by three technical appendices which
are devoted to the analysis of the component structure of several Goldstino superfield 
models in the flat superspace limit.


\section{Two-form multiplet in $\cN=1$ supergravity}
\label{N=1}
\setcounter{equation}{0}

In this section we describe two off-shell formulations for $\cN=1$ AdS supergravity, with $4+4$ off-shell degrees of freedom,
which differ from each other by their auxiliary fields. One of them
is known since the late 1970s, see \cite{GGRS} for a review, 
and its auxiliary field is a scalar. 
The other formulation is obtained by replacing the auxiliary scalar field 
with the field strength of a gauge two-form, which requires the use of a different 
compensating supermultiplet.  As was pointed out in 
\cite{Grisaru:1985fv,Bergshoeff:1985su}, the latter formulation is required for 
consistent coupling to the Green-Schwarz superstring. 
However, the technical details of this formulation have not been described 
in the literature, to the best of our knowledge. 

We follow the notation and make use of the results of \cite{KLT-M11}.
Every supergravity theory will be realised as a super-Weyl invariant coupling of conformal supergravity to a compensating supermultiplet.

\subsection{Conformal  supergravity}\label{section2.1}

Consider a curved $\cN=1$ 
superspace, $\cM^{3|2}$, parametrised by local real coordinates
$z^{M}=(x^m,\q^{\mu})$, with $m=0,1,2$ and $\mu=1,2$,
of which $x^m$ are bosonic and $\q^{\mu}$ fermionic. 
We introduce a preferred  basis of one-forms
$E^A=(E^a,E^\a)$ and its dual basis $E_A=(E_a,E_\a)$, 
\bea
E^A=\rd z^ME_{M}{}^A~,
\qquad E_A = E_A{}^M  \pa_M ~,
\label{beins}
\eea
which will be referred to as the supervielbein and its inverse, respectively.

The superspace structure group is ${\rm SL}(2,{\mathbb R})$, 
the double cover of the connected Lorentz group ${\rm SO}_0(2,1)$. 
The covariant derivatives have the form:
\bea
\cD_{A}&=& (\cD_a, \cD_\a )= E_{A}+\O_A~,
\label{23cd}
\eea
where
\bea
\O_A=\hf\O_{A}{}^{bc}M_{bc}=-\O_{A}{}^b M_b=\hf\O_{A}{}^{\b\g}M_{\b\g}
\label{2.444}
\eea
is the Lorentz connection.
The Lorentz generators with two vector indices ($M_{ab} =-M_{ba}$),  one vector index ($M_a$)
and two spinor indices ($M_{\a\b} =M_{\b\a}$) are related to each other by the rules:
$M_a=\hf \ve_{abc}M^{bc}$ and $M_{\a\b}=(\g^a)_{\a\b}M_a$.
These generators 
act on a vector $V_c$ 
and a spinor $\J_\g$ 
as follows:
\bea
M_{ab}V_c=2\eta_{c[a}V_{b]}~, ~~~~~~
M_{\a\b}\J_{\g}
=\ve_{\g(\a}\J_{\b)}~.
\label{generators}
\eea

The covariant derivatives are characterised by graded commutation relations 
\bea
{[}\cD_{{A}},\cD_{{B}}\}&=&
T_{ {A}{B} }{}^{{C}}\cD_{{C}}
+\hf R_{{A} {B}}{}^{{cd}}M_{{cd}}~,
\label{algebra-0}
\eea
where $T_{ {A}{B} }{}^{{C}}$ and $R_{{A} {B}}{}^{{cd}}$ are
the torsion and curvature tensors, respectively. 
To describe supergravity, the covariant derivatives 
have to obey certain torsion constraints \cite{GGRS} such that 
the algebra \eqref{algebra-0} takes the form 
\bsubeq
\bea
\{\cD_\a,\cD_\b\}&=&
2\ri\cD_{\a\b}
-4\ri \cS M_{\a\b}
~,~~~~~~~~~
\label{N=1alg-1}
\\
{[}\cD_{a},\cD_\b{]}
&=& (\g_a)_\b{}^{\g}\Big[
\cS\cD_{\g}
-
C_{\g\d\r}M^{\d\r} \Big]
-\frac{2}{3}\Big[
\cD_{\b}\cS\d_a^c
-2\ve_{ab}{}^{c}(\g^b)_{\b\g}\cD^{\g}\cS\Big]M_c
~,~~~~~~~~~
\label{N=1alg-3/2-2}
\\
{[}\cD_{a},\cD_b{]}
&=&
\ve_{abc}\Big{\{}
\ri \Big{[}\frac{1}{2}(\g^c)_{\a\b}C^{\a\b\g}
-\frac{2}{3}(\g^c)^{\b\g}\cD_{\b}\cS\Big{]}\cD_\g
\non\\
&&
~~~~~~
+\Big{[}
\frac{\ri}{2}(\g^c)^{\a\b}(\g^d)^{\g\d}\cD_{(\a}C_{\b\g\d)}
+\Big(
\frac{2\ri}{3}\cD^2\cS
+4\cS^2\Big)\eta^{cd}
\Big{]}M_d
\Big{\}}
~.
~~~~~~~~~~~~
\label{N=1alg-2}
\eea
\esubeq
Here the scalar $\cS$ is real, 
while the symmetric spinor $C_{\a\b\g}=C_{(\a\b\g)}$ is {\it imaginary}.
The dimension-2 Bianchi identities imply that 
\bea
\cD_{\a}C_{\b\g\d}&=&
\cD_{(\a}C_{\b\g\d)}
-\ri\ve_{\a(\b}\cD_{\g\d)}\cS \quad \Longrightarrow \quad 
\cD^\g C_{\a\b\g} = -\frac{4\ri }{3} \cD_{\a\b} \cS
~.
\eea
Throughout this section we make use of the definition $\cD^2 := \cD^\a \cD_\a$.

The definition of the torsion and curvature tensors, eq.  \eqref{algebra-0},
can be recast in the language of superforms, 
which will be used in section \ref{section4}. 
Starting from 
the Lorentz connection $\O_A$ given by  \eqref{2.444}, we introduce
 the connection one-form 
\bea
\O = E^C \O_C ~, \qquad 
\O V_A = \O_A{}^B V_B  = E^C \O_{CA}{}^B V_B~, 
\qquad V_A =(V_a, \J_\a)~.
\eea
Then the torsion and curvature two-forms are 
\bsubeq
\bea
T^C&:=&\hf E^B\wedge E^AT_{AB}{}^C
=-\rd E^C+E^B\wedge\O_B{}^C
~,
\\
R_C{}^D&:=&
\hf E^B\wedge E^AR_{AB}{}_C{}^D
=\rd\O_C{}^D-\O_C{}^E\wedge\O_E{}^D
~.
\eea
\esubeq

The  gauge group of conformal supergravity includes local transformations of the form
\bea
\d_\cK\cD_A=[\cK,\cD_A]~,~~~~~~
\cK=\x^C E_C+\hf K^{cd} M_{cd}
~,
\label{SUGRA-gauge-group1}
\eea
with  the gauge parameters $\x^C(z) $ and $K^{bc}(z)$ 
obeying natural reality conditions but otherwise arbitrary.
Here the supervector field $\x= \x^C E_C$ describes a general coordinate 
transformation, and  $K^{cd} $ a local Lorentz transformation.
The transformation \eqref{SUGRA-gauge-group1} 
acts on  a tensor superfield $T$ as follows:
\bea
\d_\cK T=\cK T ~.
\label{SUGRA-gauge-group2}
\eea

The algebra of covariant derivatives is invariant under 
super-Weyl transformations 
\bsubeq \label{2.10}
\bea
\d_\s\cD_\a&=&
\hf \s\cD_\a + \cD^{\b}\s M_{\a\b}
~,
\\
\d_\s\cD_a&=&
\s\cD_a
+\frac{\ri}{ 2}(\g_a)^{\g\d}\cD_{\g} \s\cD_{\d}
+\ve_{abc}\cD^b\s M^{c}
~,
\eea
\esubeq
with the parameter $\s$ being a real unconstrained superfield, provided 
the torsion superfields transform as
\bea
\d_\s\cS&=&\s\cS-\frac{\ri}{4}  \cD^2\s~,~~~~~~
\d_\s C_{\a\b\g}=\frac{3}{2}\s C_{\a\b\g}-\hf  \cD_{(\a\b}\cD_{\g)}\s
~.
\label{sW}
\eea
The super-Weyl transformation of the vielbein is
\bsubeq\label{2.14ab}
\bea
\d_\s E^a&=&
-\s E^a
~,
\\
\d_\s E^\a&=&
-\hf\s E^\a
-\frac{\ri}{2}E^b(\g_b)^{\a\b}\cD_\b\s
~.
\eea
\esubeq

The  gauge group of conformal supergravity  is generated by the local transformations
\eqref{SUGRA-gauge-group1} and \eqref{2.10}.
Due to the super-Weyl invariance, 
the above geometry describes the Weyl multiplet of 
$\cN=1$ conformal supergravity \cite{vanN85}, which consists of  
the vielbein $e_m{}^a (x) $ and the gravitino $\psi_m{}^\a(x)$
(no auxiliary fields).\footnote{The super-Weyl transformation of $\cS$ implies 
that its lowest component 
$\cS|_{\q=0}$ is a pure gauge.}

A tensor superfield $T $ 
is said to be (super-Weyl) primary of weight $w$
 if its super-Weyl transformation law is 
\bea
\d_\s T = w \s T~.
\eea
Such superfields will be of primary importance in what follows.

The action for conformal supergravity 
was constructed for the first time by van Nieuwenhuizen \cite{vanN85}
using the $\cN=1$ superconformal tensor calculus. More recently, 
it was re-formulated in superspace \cite{Kuzenko:2012ew}, 
as well as within the superform approach  \cite{Kuzenko:2012ew,BKNT-M2}.
The interested reader is referred to these publications for the technical details.


\subsection{Supersymmetric action}

To construct a locally supersymmetric and super-Weyl invariant  action \cite{KLT-M11},  
one needs a real scalar Lagrangian $\cL$ 
that is super-Weyl primary of weight $+2$,
\bea
\d_\s\cL=2\s\cL~.
\label{N=1sWL}
\eea
The action is
\bea
S&=& \ri \int\rd^3x\rd^2\q \,E \,\cL
~,~~~
\qquad E= {\rm Ber}(E_M{}^A)
~.
\label{N=1Ac}
\eea
The action is super-Weyl invariant, since the super-Weyl transformation of $E$ 
proves to be  $\d_\s E=-2\s E$.

Instead of defining the action using the superspace integration, 
 an alternative approach is to construct 
 a dimensionless super 3-form $\X_3[\cL]$ which is given in terms of  
 $\cL$ and possesses the following properties: 
 (i) $\X_3 [\cL]$ is closed, $\rd \, \X_3[\cL] =0$; and
(ii) $\X_3 [\cL] $ is  super-Weyl invariant, $\d_\s \X_3 [\cL]=0$.\footnote{See 
\cite{Castellani,Hasler, Ectoplasm, GGKS}
for the construction of locally supersymmetric invariants
in $D$ spacetime dimensions by using closed super $D$-forms.}
Modulo an overall numerical factor, 
these conditions prove to completely determine $\X_3[\cL]$   to be
\bea
\X_3[\cL] &=&
\frac{\ri}{2}E^\g \wedge E^\b \wedge E^a
(\g_a)_{\b\g}\cL
+\frac{1}{4}E^\g \wedge E^b \wedge E^a\ve_{abc}
(\g^c)_{\g}{}^\d\cD_{\d}\cL
\non\\
&&
-\frac{1}{24}E^c \wedge E^b \wedge E^a\ve_{abc}
\big(\ri\cD^2+8\cS\big)\cL~.
\label{2.13}
\eea
This super 3-form was originally constructed in \cite{Ectoplasm,Becker:2003wb}, 
however its super-Weyl invariance was first described in \cite{Kuzenko:2012ew}.
The action \eqref{N=1Ac} is recast via $ \X_3[\cL]$ as follows 
\bea 
S = \int_{\cM^3} \X_3[\cL]~,
\label{ectoS}
\eea
where the integration is carried out over a spacetime  $\cM^3$ being homotopic  to the bosonic body 
of the curved superspace $\cM^{3|2}$ obtained by switching off the Grassmann variables.


\subsection{AdS supergravity} \label{section2.3}

Both AdS and Poincar\'e supergravity theories can be realised as super-Weyl invariant systems 
describing the coupling of conformal supergravity to a compensating multiplet.
The standard choice 
for compensator  is a nowhere vanishing scalar superfield $\varphi$, 
such that $\vf^{-1}$ exists, 
with the  super-Weyl transformation
\bea
\d_\s\varphi=\hf \s\varphi
~.
\eea
The  action for $\cN=1$ AdS  supergravity is given  by
\bea
S_{\rm SG}&=&
- \frac{4 } {\k}  \ri 
\int  \rd^3x\rd^2\q
\,E \,\Big\{
{\ri} \cD^\a\varphi \,\cD_\a\varphi 
-2\cS \varphi^2 
+ \l   \varphi^4
\Big\}
~,~~~
\label{N=1-scalar}
\eea
where $\k$ is the gravitational coupling constant, and 
the parameter $\l$ determines the cosmological constant.
Setting $\l=0$ in \eqref{N=1-scalar}
gives the action for $\cN=1$ Poincar\'e supergravity. 

The equation of motion for the compensator is 
\begin{subequations}\label{N=1equations}
\bea
{\mathbb S} = \l~, \qquad 
{\mathbb S}:= 
\vf^{-3} \Big( \frac{\ri}{2} \cD^2 +\cS\Big) \vf
~.
\label{N=1equations.a}
\eea
For completeness we also give the equation of motion for the gravitational superfield
(which is the $\cN=1$ supersymmetric analogue of the gravitational field)
\bea
{\mathbb C}_{\a\b\g} =0~, \qquad 
{\mathbb C}_{\a\b\g} 
:= - \frac{1}{2} \varphi^{-1}\Big( \cD_{(\a\b} \cD_{\g)}-2C_{\a\b\g} \Big)\varphi^{-2} ~,
\label{2.16b}
\eea
\end{subequations}
see \cite{KNT-M15} for the technical details. 
The specific feature of ${\mathbb S}$ and ${\mathbb C}_{\a\b\g} $ is that they 
are super-Weyl invariant. 
Note that it is possible to choose a super-Weyl gauge  
in which $\varphi=1$ and, therefore,  
${\mathbb S}$ and ${\mathbb C}_{\a\b\g} $
coincide with ${\cS}$ and ${C}_{\a\b\g} $, respectively. 
In this gauge, the equations \eqref{N=1equations}  describe,  locally, 
the $\cN=1$ AdS superspace \cite{KLT-M12}.

The action \eqref{N=1-scalar} can readily be reduced 
to components. In the super-Weyl gauge $\varphi=1$
we obtain
\bea
S_{\rm SG}
=
\frac{1}{\k}\int\rd^3x\, e\,
\Big{\{}
\frac{1}{2}\cR
-4S^2
+8S\l
\Big\}
+{\rm fermions}
~,\qquad e=\det(e_m{}^a)~,
\eea
where $e_m{}^a (x):=E_m{}^a|_{\q=0}$
and 
$S(x):=\cS|_{\q=0}$.
Integrating out 
the auxiliary field $S$ 
turns the action into
\bea
S_{\rm SG}
=
\frac{1}{\k}
\int\rd^3x\, e\,
\Big{\{}
\frac{1}{2}\cR
-\L_{\rm AdS}
\Big{\}}
+{\rm fermions}
~, \qquad \L_{\rm AdS} =- 4\l^2~.
\eea

\subsection{Two-form supergravity}\label{section2.4}

In this section we introduce a variant formulation for $\cN=1$ AdS supergravity 
which is obtained by replacing the conformal compensator 
$\varphi^4$ with a two-form multiplet.\footnote{In the case of Minkowski superspace, 
the two-form multiplet was described in \cite{GGRS}.}

Let us first consider a massless two-form multiplet coupled to conformal 
supergravity. 
It is described by a
real scalar superfield
defined by 
\bea
L=\cD^\a\L_\a~,
\label{2.17}
\eea
where the prepotential $\L_\a$ is a  primary real spinor superfield of dimension 3/2,
\bea
\d_\s\L_\a=\frac{3}{2}\s\L_\a
~.
\label{2266}
\eea
This super-Weyl transformation implies that  $L$ is primary of dimension of 2,
\bea
\d_\s L=2\s L~.
\label{2277}
\eea
The superfield $L$ defined by \eqref{2.17} is a gauge-invariant field strength 
with respect to gauge transformations of the form 
\bea
\d_\z \L_\a = \frac{\ri}{2} \cD^\b \cD_\a \z_\b + 2 \cS \z_\a~,\qquad 
\cD^\a \d_\z \L_\a =0~,
\label{2.20}
\eea
where the gauge parameter $\z_\a$ is an arbitrary real spinor superfield.
The gauge invariance of $L$ follows from the identity 
\bea
\cD^\b \cD_\a \cD_\b = 4\ri \cS \cD_\a -\frac{8\ri}{3}(\cD^{\b}\cS)M_{\a\b}
-2\ri C_{\a\b\g}M^{\b \g}~.
\eea

The gauge parameter in \eqref{2.20} is defined modulo arbitrary shifts 
of the form 
\bea
\z_\a \to \z'_\a = \z_\a + \ri \cD_\a \x~, \qquad \bar \x = \x~,
\eea
in the sense that $\d_{\z'} \L_\a =\d_\z \L_\a $.
This property means that the two-form multiplet is 
a gauge theory with linearly dependent generators, in accordance with
the terminology of the Batalin-Vilkovisky quantisation \cite{BV}. 

We now assume $L$ to be nowhere vanishing, 
such that $L^{-1} $ exists.
Then $L$ can be used as a conformal compensator corresponding 
to a variant formulation of AdS supergravity.
Upon replacement $\varphi \to L^{1/4}$, 
the supergravity action \eqref{N=1-scalar} turns into 
\bea
S_{\rm SG}&=&
- \frac{4 } {\k}\ri \int  \rd^3x\rd^2\q
\,E \, \sqrt{L} \Big\{
\frac{\ri}{16} \cD^\a\ln L  \,\cD_\a \ln L 
-2\cS \Big\}~.
\label{2.23}
\eea
The supersymmetric cosmological  term in \eqref{N=1-scalar} 
does not contribute, since 
$\varphi^4 $ turns into  $L=\cD^\a\L_\a$, which is  a total derivative.
Hence, the $\cN=1$ two-form supergravity does not allow for a 
supersymmetric cosmological term.
This is analogous to the new minimal formulation for $\cN=1$ supergravity in four dimensions \cite{new,SohniusW2,SohniusW3}.
However, the difference from the new minimal supergravity 
is that a cosmological terms is now generated dynamically.

For the theory with action \eqref{2.23}, the equation of motion for the compensator is
\bea
\cD_\a {\mathbb S} =0~, \qquad 
{\mathbb S}:= L^{-\frac{3}{4} } \Big( \frac{\ri}{2} \cD^2 +\cS\Big) L^{\frac{1}{4} }~,
\label{2299}
\eea
and therefore 
\bea
{\mathbb S} =\l ={\rm const}~. 
\eea
If a solution with $\l \neq 0$ is chosen, it describes an AdS 
background. Unlike the supergravity formulation \eqref{N=1-scalar},
the action \eqref{2.23} does not contain a free parameter. 
The negative cosmological constant is generated dynamically. 
It should be pointed out that  the equation of motion for the gravitational superfield,
which corresponds to \eqref{2.23}, is obtained from \eqref{2.16b}
by replacing $\vf \to L^{\frac{1}{4}}$.


\subsection{Superform formulation for the two-form multiplet}

In this subsection we present a superform formulation for the three-form 
multiplet coupled to conformal supergravity, 
as an extension of the flat-superspace construction given in  \cite{GGRS}.
Let us consider a gauge super 2-form
\bea
B_2=\hf \rd z^N\wedge \rd z^M B_{MN}=\hf  E^B\wedge E^AB_{AB}
~,
\eea
which is defined modulo gauge transformations of the form 
\bea
B_2\to B_2+\rd A_1~,\qquad  A_1=\rd z^NA_N=E^BA_B~,
\label{2.34}
\eea
where the gauge parameter $A_1$
is an arbitrary super 1-form.
Associated with the potential $B_2$ is the gauge-invariant field strength
\bea
H_3:=\rd B_2
&=&\frac{1}{2}\rd z^{P}\wedge\rd z^{N}\wedge  \rd z^M
\pa_M B_{NP}
\non \\
&=&
\frac{1}{2} E^{C}\wedge E^{B}\wedge E^A\Big\{\cD_{A}B_{BC}
-T_{AB}{}^DB_{DC}\Big\}
~.
\eea
By construction, $H_3$ is an exact super 3-form, and
 hence it is closed, $\rd H_3=0$.

We are interested in a closed super 3-form $H_3$ such that (i) its components 
are descendants of a scalar primary superfield $L$;  and 
(ii) its lowest non-zero component is constrained to be $H_{a\b\g}=\ri(\g_a)_{\b\g}L$.
It turns out that the closure condition,  $\rd H_3=0$, completely determine the entire 
super 3-form to be
\bea
H_3[L] &=&
\frac{\ri}{2}E^\g \wedge E^\b\wedge E^a
(\g_a)_{\b\g}L
+\frac{1}{4}E^\g\wedge E^b\wedge E^a\ve_{abc}
(\g^c)_{\g}{}^\d\cD_{\d}L
\non\\
&&
-\frac{1}{24}E^c\wedge E^b\wedge E^a\ve_{abc}
\big(\ri\cD^2+8\cS\big)L~,
\label{H_3-1}
\eea
which is obtained from \eqref{2.13} by replacing $\cL \to L$.
In general, if $L$ is an arbitrary scalar superfield,  the superform
$H_3$ given by \eqref{H_3-1} is closed but not exact.
However, if we choose $L:=\cD^\a\L_\a$ in \eqref{H_3-1} 
then $H_3$ turns out to be exact.
In fact, the following super 2-form 
\bea
B_2[\L_\a]&=&
-\ri E^\b\wedge E^a (\g_a)_{\b}{}^{\g}\L_\g
-\frac{1}{4} E^b\wedge E^a \ve_{abc}(\g^c)^{\r\t}\cD_{\r}\L_\t
~,
\label{2.37}
\eea
 is such that 
 \bea
 \rd B_2 [\L_\a]=H_3[\cD^\a\L_\a]~.
 \eea
This proves that, if we consider the two- and three-forms
\bsubeq
\bea
B_{a b}
&=&
-\hf\ve_{abc}(\g^c)^{\r\t}\cD_{\r}\L_\t
~,
\label{2.11a}
\\
H_{abc}&=&
-\frac{1}{4}\ve_{abc}\big(\ri\cD^2+8\cS\big)
\cD^\d \L_\d
~,
\label{2.11b}
\eea
\esubeq
the latter is the field strength of the former,
\bea
H_{abc}
&=&
3\cD_{[a}B_{bc]}
+2\ve_{abc}(\cD^{\a}\cS)\L_\a
~.
\label{2.12}
\eea

Using the super-Weyl transformation laws \eqref{2.14ab}
and \eqref{2266}, one can show that the superform \eqref{2.37}
is super-Weyl invariant, 
\bea
\d_\s B_2[\L_\a] =0 \quad \Longrightarrow \quad 
\d_\s H_3[\cD^\a \L_\a] =0~.
\label{2422}
\eea
This result will be important for our analysis in section \ref{GS1}.

Choosing $B_2$ in the form \eqref{2.37} corresponds to a partial fixing of the gauge 
freedom \eqref{2.34}. The residual gauge freedom is given by 
\bea
\d_\z B_2[\L_\a] = B_2[\d_\z \L_\a] \quad \Longrightarrow \quad
\rd \,\d_\z B_2[\L_\a] =0~,
\quad 
\eea
where $\d_\z \L_\a$ is defined by \eqref{2.20}.


\section{Two-form multiplets in $\cN=2$ supergravity}
\setcounter{equation}{0}

It is well-known that the 3D AdS  group is reducible, 
$$\rm SO(2,2) \cong \Big( SL(2, {\mathbb R}) \times SL( 2, {\mathbb R}) \Big)/{\mathbb Z}_2~,$$ 
and so are its supersymmetric extensions,  
${\rm OSp} (p|2; {\mathbb R} ) \times  {\rm OSp} (q|2; {\mathbb R} )$.
This implies that  $\cN$-extended AdS supergravity exists in several versions  \cite{AT}. 
These are known as the  $(p,q)$ AdS supergravity theories
where the  non-negative integers $p \geq q$ are such that 
$\cN=p+q$.\footnote{For any values of $p$ and $q$ allowed, 
the pure  $(p,q)$ AdS supergravity was constructed in \cite {AT}
as a Chern-Simons theory with the gauge group
 ${\rm OSp} (p|2; {\mathbb R} ) \times  {\rm OSp} (q|2; {\mathbb R} )$.}   
In this section we choose $\cN=2$ and describe four off-shell formulations 
for (1,1) AdS supergravity and one 
for (2,0)  AdS supergravity.  Only one of these five off-shell supergravity theories 
is new, the so-called complex two-form supergravity, the others 
were presented in \cite{KT-M11}.

\subsection{Conformal  supergravity}
\label{geometry}

We consider a curved $\cN=2$ superspace, 
$\cM^{3|4}$,  parametrised by
local bosonic ($x^m$) and fermionic ($\q^\m, \bar \q_\m$)
coordinates  $z^{{M}}=(x^{m},\q^{\mu},{\bar \q}_{{\mu}})$,
where $m=0,1,2$ and  $\mu=1,2$.
The Grassmann variables $\q^{\mu} $ and $\teb_{{\mu}}$
are related to each other by complex conjugation:
$\overline{\q^{\mu}}=\teb^{{\mu}}$.
The supervielbein $E^A =(E^a, E^\a, \bar E_\a)$ and its inverse 
$E_A= (E_a, E_\a, \bar E^\a )$ are defined similarly to \eqref{beins}.

Within the superspace formulation for $\cN=2$ conformal supergravity 
proposed in \cite{HIPT} and fully developed in \cite{KLT-M11},
the  structure group is  ${\rm SL}(2,{\mathbb{R}})\times {\rm U(1)}$.
The covariant derivatives
have the form
\bea
\cD_{{A}}=(\cD_{{a}}, \cD_{{\a}},\cDB^\a)
=E_{{A}}+{\bm\O}_{{A}}~,~~~~~~
{\bm\O}_{{A}}:=
\O_{{A}}+\ri \,\F_{{A}}\cJ~.
\label{CovDev}
\eea
We recall that 
 the Lorentz connection $\O_A$ can be written in several equivalent forms
 \eqref{2.444}.
The ${\rm U(1)}$ generator
acts on the covariant derivatives as follows:
\bea
&{[}\cJ,\cD_{\a}{]}
=\cD_{\a}~,
\qquad
{[}\cJ,\cDB^{\a}{]}
=-\cDB^\a~.
\eea

In general, the covariant derivatives have graded commutation relations of the form
\bea
{[}\cD_{{A}},\cD_{{B}}\}&=&
T_{ {A}{B} }{}^{{C}}\cD_{{C}}
+{\bm R}_{{A} {B}}
~,~~~~~~
{\bm R}_{{A} {B}}:=
\hf R_{{A} {B}}{}^{{cd}}M_{{cd}}
+\ri \,R_{ {A} {B}}\cJ
~.
\label{algebra}
\eea
In order to describe the multiplet of conformal supergravity, 
certain constraints should be imposed on the torsion tensor \cite{HIPT}.
Solving these constraints leads to the following algebra of covariant derivatives:
\bsubeq \label{algebra-final}
\bea
\{\cD_\a,\cD_\b\}
&=&
-4\bar{R}M_{\a\b}
~,
\label{N=2-alg-1}
\\
\{\cD_\a,\cDB_\b\}
&=&
-2\ri (\g^c)_{\a\b} \cD_c
-2\cC_{\a\b}\cJ
-4\ri\ve_{\a\b}\cS\cJ
+4\ri\cS M_{\a\b}
-2\ve_{\a\b}\cC^{\g\d}M_{\g\d}
~,
\label{2.7b}  \\
{[}\cD_{a},\cD_\b{]}
&=&
\ri\ve_{abc}(\g^b)_\b{}^{\g}\cC^c\cD_{\g}
+(\g_a)_\b{}^{\g}\cS\cD_{\g}
-\ri(\g_a)_{\b\g}\bar{R}\cDB^{\g}
-(\g_a)_\b{}^{\g}{\bm C}_{\g\d\r}M^{\d\r}
\non\\
&&
-\frac{1}{3}\Big(
2\cD_{\b}\cS
+\ri\cDB_{\b}\bar{R}
\Big)M_a
-\frac{2}{3}\ve_{abc}(\g^b)_{\b}{}^{\a}\Big(
2\cD_{\a}\cS
+\ri\cDB_{\a}\bar{R}
\Big)M^c
\non\\
&&
-\hf\Big(
(\g_a)^{\a\g}{\bm C}_{\a\b\g}
+\frac{1}{3}(\g_a)_\b{}^{\g}\big(
8\cD_{\g}\cS
+\ri\cDB_{\g}\bar{R}
\big)
\Big)\cJ ~,
\label{2.7c}
\\
{[}\cD_a,\cD_b]{}
&=&
\hf\ve_{abc}(\g^c)^{\a\b}\ve^{\g\d}\Big(
-\ri\bar{{\bm C}}_{\a\b\d}
+\frac{4\ri}{3}\ve_{\d(\a}\cDB_{\b)}\cS
+\frac{2}{3}\ve_{\d(\a}\cD_{\b)} R
\Big)\cD_\g
\non\\
&&
+\hf\ve_{abc}(\g^c)^{\a\b}\ve^{\g\d}
\Big(
-\ri {\bm C}_{\a\b\d}
+\frac{4\ri}{3}\ve_{\d(\a}\cD_{\b)}\cS 
-\frac{2}{3}\ve_{\d(\a}\cDB_{\b)}\bar{R}
\Big)\cDB_\g
\non\\
&&
-  \ve_{abc}\Big(\,\frac{1}{4}  (\g^c)^{\a\b}(\g_d)^{\t\d}
\big(\ri\cD_{(\t}\bar{{\bm C}}_{\d\a\b)}
+\ri\cDB_{(\t} {\bm C}_{\d\a\b)}\big)
+\frac{1}{6}
(\cD^2 R+\cDB^2 \bar{R})
\non\\
&&~~~~~~~~~
+\frac{2}{3} \ri 
\cD^\a\cDB_{\a}\cS
-4\cC^{c}\cC_d
-4\cS^2
-4\bar{R}R
\Big)M^d
\non\\
&&
+\ri  \ve_{abc} \Big(
\frac{1}{2}  (\g^c)^{\a\b}[\cD_\a,\cDB_{\b}]\cS
-\ve^{cef}\cD_{e}\cC_{f}
-4\cS\cC^c
\Big)\cJ
~.
\label{2.7e}
\eea
\end{subequations}
The algebra involves four dimension-1 torsion superfields:
a real scalar $\cS$, a
complex scalar $R$ and its conjugate $\bar{R}$,  and a real vector $\cC_a$.
The ${\rm U(1)}$ charge of $R$ is $-2$. 
These torsion superfields obey differential constraints implied by the Bianchi identities,
which are: 
\begin{subequations}
\bea
\cDB_\a R&=&0~,  \label{2.12a}
\\
(\cDB^2-4R)\cS
&=&0~, 
 \label{2.12b} \\
\cD_{\a}\cC_{\b\g}
&=&
\ri {\bm C}_{\a\b\g}
-\frac{1 }{ 3}\ve_{\a(\b}\big(
 \cDB_{\g)}\bar{R}
+4\ri \cD_{\g)}{\cS}
\big)~.  \label{2.12c}
\eea
\end{subequations}
In this paper  we make use of the definitions 
\bea
\cD^2 := \cD^\a \cD_\a~, \qquad \bar \cD^2 := \bar \cD_\a \bar \cD^\a~.
\label{388}
\eea
As follows from \eqref{2.12c}, the complex dimension-3/2 symmetric spinor  
$ {\bm C}_{\a\b\g}$, which appears in \eqref{algebra-final}, 
is a descendant of the torsion three-vector $\cC_a$,   
$ {\bm C}_{\a\b\g} = -\ri \cD_{(\a}\cC_{\b\g )} $.

The definition of the torsion and curvature tensors, eq.  \eqref{algebra},
can be recast in the  superform notation, 
which will be used in section \ref{section4}. 
Associated with the connection ${\bm\O}_A$, eq. \eqref{CovDev},
is the connection one-form ${\bm\O} = E^C {\bm\O}_C$.
Its action on a real super-vector 
\bea
V_A =(V_a , \J_\a , \bar \J^\a)~, \qquad  \cJ \J_\a = \J_\a
\eea
is given by 
\bea
{\bm\O} V_A ={\bm\O}_A{}^B V_B= \O_A{}^B V_B + \ri \, \F_A{}^B V_B
~,
\label{3.8connection}
\eea
with $\O_A{}^B $ and $\F_A{}^B$ being 
 the  Lorentz 
 and  U(1) connections, respectively.
 Using the definitions given, the torsion and curvature  two-forms are 
\bsubeq
\bea
T^C&:=&\hf E^B\wedge E^AT_{AB}{}^C
=-\rd E^C+E^B\wedge{\bm\O}_B{}^C
~,
\label{torsion-def}
\\
{\bm R}_C{}^D&:=&
\hf E^B\wedge E^AR_{AB}{}_C{}^D
=\rd{\bm\O}_C{}^D-{\bm\O}_C{}^E\wedge{\bm\O}_E{}^D
~.
\eea
\esubeq

The important property of the 
algebra
\eqref{algebra-final}
is that its  form is preserved under super-Weyl transformations
of the covariant derivatives \cite{KLT-M11,KT-M11}
\bsubeq  \label{2.3}
\bea
\d_\s\cD_\a&=&\hf\s\cD_\a+\cD^{\g}\s M_{\g\a}- \cD_{\a } \s \cJ~,
\\
\d_\s \cDB{}_{\a}&=& {\hf\s} \cDB_{\a}+\cDB^{\g}\s {M}_{\g\a}
+\cDB_{\a}\s \cJ ~,
\\
\d_\s\cD_{a}
&=&
\s\cD_{a}
-\frac{\ri}{2}(\g_a)^{\g\d} \cD_{(\g}\s \cDB_{\d)}
-\frac{\ri}{2}(\g_a)^{\g\d} \cDB_{(\g}\s \cD_{\d)}
 \non \\
&&
+\ve_{abc} \cD^b\s M^c
-\frac{\ri}{8} (\g_a)^{\g\d} {[}\cD_{\g},\cDB_{\d}{]}\s \cJ
\eea
and the torsion tensors
\bea
\d_\s\cS&=&
\s\cS
+\frac{\ri}{4}\cD^\a\cDB_\a\s
~,
 \label{2.11d} \\
\d_\s\cC_{a}&=&
\s\cC_{a}
+\frac{1}{8}(\g_a)^{\g\d}  [\cD_{\g},\cDB_{\d}]\s
~,
\\
\d_\s R &=&
\s R
+\frac{1}{4} \cDB^2\s ~. 
\label{2.11f}
\eea
\esubeq
Here the super-Weyl parameter $\s$ is an unconstrained real scalar superfield.
It follows from \eqref{2.3} that the super-Weyl transformation law 
of the supervielbein is 
\begin{subequations}\label{N=2vielbein-super-Weyl}
\bea
\d_\s E^a &=&-\s E^a ~, \\
\d_\s E^\a &=& -\hf\s E^\a +\frac{\ri}{2} E^b (\g_b)^{\a\g} \bar \cD_\g \s~, 
\non \\
\d_\s \bar E_\a &=& -\hf\s\bar E_\a +\frac{\ri}{2} E^b (\g_b)_{\a\g}  \cD^\g \s~.
\eea
\end{subequations}
The group of super-Weyl transformations must be a subgroup of  
the supergravity gauge group
in order for the superspace geometry under consideration to describe 
the multiplet of $\cN=2$ conformal supergravity.

A tensor superfield $T $ of a ${\rm U(1)}$ charge $q$, $\cJ T = q T$, 
is said to be super-Weyl primary if its super-Weyl transformation law is 
\bea
\d_\s T = w \s T~,
\eea
for some constant parameter $w$ which will be referred to as 
the super-Weyl weight of $T$.

The action for $\cN=2$ conformal supergravity 
was constructed for the first time by Ro\v{c}ek and van Nieuwenhuizen 
\cite{RvanN86} using the $\cN=2$ superconformal tensor calculus. More recently, 
it was re-formulated within the superform approach  \cite{BKNT-M2}.
The interested reader is referred to these publications for the technical details.


\subsection{Supersymmetric actions}

As in the 4D $\cN=1$ case, 
there are two (closely related) locally supersymmetric and super-Weyl invariant actions
in 3D $\cN=2$ supergravity  \cite{KLT-M11}.

Given a real scalar Lagrangian $\cL =\bar \cL$ with the 
 super-Weyl transformation law
\bea
\d_\s\cL=\s\cL~,
\eea
the action
\bea
S&=&\int \rd^3x\rd^2\q\rd^2\qb  \,E \,\cL
~,~~~
\qquad E= {\rm Ber}(E_M{}^A)
~,
\label{N=2Ac}
\eea
is invariant under the supergravity gauge group. 
It is also  super-Weyl invariant due to the transformation law
\bea
\d_\s E=-\s E~.
\label{SW-Ber}
\eea

Given a covariantly chiral scalar Lagrangian $\cL_{\rm c}$ of super-Weyl weight two, 
\bea
\cDB_\a\cL_{\rm c}=0~,\qquad \cJ \cL_{\rm c} = -2 \cL_{\rm c}~, \qquad
\d_\s\cL_{\rm c}=2\s\cL_{\rm c}~,
\eea
the following {\rm chiral} action
\bea
S_{\rm c}=\int\rd^3x\rd^2\q\, \cE \,\cL_{\rm c} 
\label{3.14}
\eea
is locally supersymmetric and super-Weyl invariant. 
Action
(\ref{3.14}) involves integration over the chiral subspace
of the full superspace, with $\cE$ the chiral density possessing the properties
\bea
\cJ  \cE = 2 \cE~, \qquad 
\d_\s\cE=-2\s\cE~.
\eea
The explicit expression for $\cE$ in terms of the supergravity prepotentials is given in 
\cite{Kuzenko12}.  
Alternatively, the chiral density can be read off using the general formalism 
of integrating out fermionic dimensions, which was developed in 
\cite{KT-M-2008-2}.

The two actions, \eqref{N=2Ac} and \eqref{3.14}, 
are related to each other as follows
\bea
\int\rd^3x\rd^2\q\rd^2\qb\, E \,\cL
= \int\rd^3x\rd^2\q\, \cE \,\cL_{\rm c} ~, \qquad \cL_{\rm c} :=
-\frac{1}{4}(\cDB^2-4R)\cL~.
\label{3.16}
\eea
This relation shows that the chiral action, or its conjugate antichiral action,  
is more fundamental than  \eqref{N=2Ac}.

The chiral projection operator in \eqref{3.16} defined by 
\bea
\bar{\D}:=-\frac{1}{4}(\cDB^2-4R)
\eea
 plays a fundamental role in $\cN=2$ supergravity.
Among its most important properties is the following:
given a primary complex scalar $\j$ satisfying
\bea
\cJ \j=(2-w) \j~, \qquad \d_\s \j=(w-1)\s\j~,
\label{3.17}
\eea
for some constant super-Weyl weight $w$,
its descendant 
\bea
\f=\bar{\D}\j
\eea
 is a primary chiral superfield of super-Weyl  weight $w$,
\bea
{\bar \cD}_\a \f=0 ~, \qquad
\cJ\f=-w\f~,\qquad 
\d_\s\f=w \s\f~.
\label{3.199}
\eea
For every primary chiral scalar superfield, its super-Weyl weight $w$ and 
${\rm U(1)}$ charge $q$ are related to each other as $w+q=0$, 
in accordance with  \cite{KLT-M11}. 
Any superfield $\f$ with the properties \eqref{3.199}
will be referred to as a weight-$w$ chiral scalar.

The chiral action, eq.  (\ref{3.14}), can be represented as an integral over the full superspace, 
\bea
S_{\rm c} = \int\rd^3x\rd^2\q\rd^2\qb\, E \, {\frak C} \cL_{\rm c}~,
\eea
if we make use of 
an {\it improved complex linear} superfield ${\frak C} $
 defined by the two properties: (i) ${\frak C} $ obeys the constraint 
\begin{subequations} \label{322}
\bea
\bar \D {\frak C} =1~;
\eea
(ii) the transformation properties of ${\frak C} $ are 
\bea
\d_\s {\frak C} = - \s {\frak C} ~, \qquad 
\cJ {\frak C} = 2 {\frak C} ~.
\eea
\end{subequations}

A possible choice for ${\frak C} $ is 
\bea
{\frak C}  = \frac{\bar \eta} {\bar \D \bar \eta} ~, \qquad \bar \cD_\a \eta =0~, 
\qquad \d_\s \eta = \hf \s \eta~,
\eea 
for some covariantly chiral superfield $\eta$ 
such that $\D \eta $ is nowhere 
vanishing. In case ${\frak C} $ is not required to be super-Weyl primary, 
it can be identified with $R^{-1}$, 
\bea
S_{\rm c}=\int\rd^3x\rd^2\q\rd^2\qb\, E \,\frac{\cL_{\rm c}}{R}~,
\eea
provided $R$ is nowhere vanishing.
This representation is analogous to that discovered by Siegel \cite{Siegel}
and Zumino \cite{Zumino78} in 4D $\cN=1$ supergravity.

The chiral action can also be described using the super 3-form
constructed in \cite{KLRST-M13}
\bea
{\X}_3 [\cL_{\rm c}]
&=&
-2 \bar{E}_\g \wedge \bar{E}_\b \wedge E^a\,  (\g_a)^{\b\g} \cL_{\rm c}
 -\frac{\ri}{2} \bar{E}_\g \wedge E^b \wedge E^a\,\ve_{abd} (\g^d)_{\g\d}  \cD^\d \cL_{\rm c}
 \non\\
 &&
+\frac{1}{24}E^c \wedge E^b \wedge E^a \, \ve_{abc}(\cD^2 -16 \bar{R})  \cL_{\rm c} ~.
\label{Sigma_3}
\eea
This superform is closed  and 
 super-Weyl invariant,
 \bea
 \rd \, \X_3 [\cL_{\rm c}] =0~, \qquad  \d_\s \X_3 [\cL_{\rm c}]=0~. 
\eea
 The chiral action is equivalently represented as  
 \bea 
S_{\rm c} = \int_{\cM^3} \X_3[\cL_{\rm c}]~,
\label{3300}
\eea
where the integration is carried out over a spacetime  $\cM^3$ being homotopic  to the bosonic body 
of the curved superspace $\cM^{3|4}$ obtained by switching off the Grassmann variables.


\subsection{AdS supergravity}

There are two off-shell formulations for (1,1) AdS supergravity
developed in  \cite{KT-M11}, minimal and non-minimal ones, which do not have  
gauge two-forms in the sector of auxiliary fields.

\subsubsection{(1,1) AdS supergravity} \label{section3.3.1}

In the minimal case, the conformal compensators are
a  weight-1/2 chiral scalar $\F$, $\bar \cD_\a \F =0$, and its conjugate $\bar \F$.
 Of course, $\F$ has to be nowhere vanishing, such that $\F^{-1}$ exists, 
 in order to serve as a conformal compensator.
 The supergravity action is
\bea
S_{\text{(1,1)\,SG}}^{\text{minimal}} = -\frac{4}{\k} \int {\rm d}^3x \rd^2\q\rd^2\qb
\,E\, \bar \F  \F
+ \Big\{ \frac{\m}{\k} \int {\rm d}^3x {\rm d}^2 \q \,\cE\,   \F^4  + {\rm c.c.} \Big\} ~,
\label{3.19}
 \eea
where $\m$ is a complex parameter.
The second terms in the action is the
supersymmetric cosmological term. 
 Using the component results of \cite{KLRST-M13}, for the cosmological constant 
 one obtains
 \bea
 \L_{\rm AdS} = -4 |\m|^2~.
 \label{N=2cc}
 \eea

The above minimal formulation for (1,1) AdS supergravity 
(which was called  type I  minimal supergravity in  \cite{KT-M11}) 
 is the 3D analogue of the old minimal formulation for 4D $\cN=1$ supergravity \cite{WZ,old1,old2}.

For the supergravity theory with action \eqref{3.19}, 
the equation of motion for the chiral compensator is
\begin{subequations} \label{N=2equations}
\bea
{\mathbb R}=\m~, \qquad 
{\mathbb R}:= \F^{-3} \bar \D \bar \F~.
\label{3.28a}
\eea
We also reproduce the equation of motion for the $\cN=2$ gravitational 
superfield\footnote{The $\cN=2$ gravitational 
superfield was introduced in  \cite{ZupnikPak,Kuzenko12}.}
\bea
{\mathbb C}_{\a\b}=0~, \qquad 
{\mathbb C}_{\a\b} :=  - \frac{1}{4} 
\Big([\cD_{(\a} , \bar \cD_{\b)}]-4\cC_{\a\b} \Big) (\Phi \bar \Phi)^{-1} \ ,
\label{328b}
\eea
\end{subequations}
see \cite{KNT-M15} for the technical details. 
The specific feature of $\mathbb R$ and ${\mathbb C}_{\a\b}$ is that they 
are super-Weyl invariant. The super-Weyl and local 
${\rm U(1)}$ transformations can be used to choose the gauge 
$\F =1$, which implies that $\cS =0$ and 
 $\mathbb R$ and ${\mathbb C}_{\a\b}$ coincide with  
the torsion superfields $R$ and $C_{\a\b}$, respectively.  
In this gauge, every solution to the equations \eqref{N=2equations}  is locally 
diffeomorphic to  
the (1,1) AdS superspace \cite{KLT-M12}.

Within the non-minimal formulation for (1,1) AdS supergravity \cite{KT-M11}, 
the conformal compensators are an improved complex linear scalar $\G$ and 
its conjugate $\bar \G$. The former has the transformation properties
\begin{subequations}
\bea
\d_\s \G= -\s\G~,\qquad
\cJ\G=2\G
\eea
and  obeys  the improved linear constraint 
\bea
\bar \D 
\Gamma = \m ={\rm const}~,
\eea
\end{subequations}
compare with \eqref{322}.
The supergravity action is
\bea
S_{\text{(1,1)\,SG}}^{\text{non-minimal}} = -\frac{2}{\k} \int \rd^3x\rd^2\q\rd^2\qb 
\,E\,
{ 
{ (\bar \G \, \G)} 
}^{-1/2}~.
\eea
As demonstrated in \cite{KT-M11}, 
this theory is dual to the minimal AdS supergravity, eq. (\ref{3.19}).
The theory under consideration 
is the 3D analogue of the 
non-minimal $\cN=1$ AdS supergravity in four dimensions
\cite{BKdual}.
Both the formulations lead to the (1,1) AdS superspace  \cite{KT-M11,KLT-M12}
as the maximally supersymmetric solution.


\subsubsection{(2,0) AdS supergravity}

The conformal compensator for (2,0) AdS supergravity 
is a {\it linear multiplet}  \cite{HIPT,KLT-M11,KT-M11} describing the field strength 
of an Abelian vector multiplet. 
It is realised in terms  of a real scalar superfield $L=\bar L$ subject to the constraint
\bea
\bar \D
L=0 \quad \Longleftrightarrow \quad \D L=0~,
\label{N=2realLinear}
\eea
which is consistent with the super-Weyl transformation law
\bea
\d_\s L=\s L~.
\label{3.20}
\eea
The constraint (\ref{N=2realLinear}) is solved in terms of 
a real unconstrained prepotential $V$, 
\bea
L= \ri \cD^\a {\bar \cD}_\a V~, \qquad \bar V =V~,
\label{G-prep}
\eea
which is defined modulo gauge transformations of the form 
\bea
\d_\l V = \l + \bar \l~,\qquad \bar \cD_\a \l =0~.
\label{G-prep-gauge}
\eea
To reproduce the super-Weyl transformation \eqref{3.20}, it suffices to choose
\bea
\d_\s V =0~.
\eea

In order to be used as a conformal compensator, $L$ has to be nowhere vanishing,
such that $L^{-1}$ exists.
The action for (2,0) AdS supergravity was constructed in \cite{KT-M11}. 
It is 
\bea
S_{\text{(2,0)\,SG}} =\frac{4}{\k} \int {\rm d}^3x \rd^2\q\rd^2\qb
\,E\, 
\Big\{ L \ln L - 4V \cS
+4\x VL \Big\}
~,
\label{Type-II-AdS}
\eea
where the parameter $\x$ determines the cosmological constant.
The equations of motion for this theory can be written in the form \cite{KNT-M15}
\begin{subequations} \label{SCcompositeN=2} 
\bea
{\mathbb S} &=& \x ~, \qquad 
{\mathbb S} := - \frac{\ri}{4 } L^{-1} \Big( \cD^\g \bar \cD_\g \ln L + 4 \ri \cS \Big) \ ,  \label{ScompositeN=2} \\
{\mathbb C}_{\a\b}&=&0~, \qquad
{\mathbb C}_{\a\b} := 
- \frac{1}{4} \Big([\cD_{(\a} , \bar \cD_{\b)}] - 4 \cC_{\a\b} \Big) {L}^{-1} \label{CcompositeN=2} \ ,
\eea
\end{subequations}
with ${\mathbb S}$ and ${\mathbb C}_{\a\b} $ being super-Weyl invariant.\footnote{The
vector superfield  ${\mathbb C}_{\a\b} $ should not be confused with \eqref{328b}.}
The super-Weyl gauge freedom can be used to set
$L=1$, which implies $R=0$,
and then  ${\mathbb S}$ and ${\mathbb C}_{\a\b} $ turns into 
the torsion superfields $\cS$ and $C_{\a\b}$, respectively. 
Under the gauge condition chosen, 
every solution to the equations \eqref{SCcompositeN=2} is locally 
diffeomorphic to the (2,0) AdS superspace \cite{KT-M11,KLT-M12}.

The above supergravity theory (called type II minimal supergravity in  \cite{KT-M11}) 
is the 3D analogue of the new minimal for  
$\cN=1$ supergravity in four dimensions \cite{new,SohniusW2,SohniusW3}.
The latter theory is known to allow no supersymmetric cosmological term.
Such a supersymmetric cosmological term does exist in the 3D case, 
and it is given by the Chern-Simons $\x$-term in \eqref{Type-II-AdS}.
For $\x\neq 0$ the theory possesses 
a maximally supersymmetric solution, 
which is the (2,0) AdS superspace \cite{KT-M11,KLT-M12}
 corresponding to the (2,0) AdS supersymmetry
\cite{AT}.


\subsection{Two-form supergravity}

There is one more variant off-shell formulation for (1,1) AdS supergravity proposed in 
\cite{KT-M11}. Its conformal compensator is the so-called two-form multiplet, 
which is the 3D cousin of the well-known three-form multiplet 
in 4D $\cN=1$ supersymmetry, which was  proposed by Gates \cite{Gates}
and reviewed in \cite{GS,GGRS}.

In curved superspace, the two-form multiplet is described by a real unconstrained
scalar prepotential $P =\bar P$ which enters any action functional,
$S=S[\P, \bar \P]$,
only via the covariantly chiral descendant 
\bea
\P=
\bar \D
P
\label{F4-P}
\eea
and its conjugate $\bar \P$.
In order for $\P$ to be a primary superfield,
the prepotential $P$ 
should possess
the super-Weyl transformation law
\bea
\d_\s P = \s  P~,
\label{s-weyl-P}
\eea
which implies 
\bea
\d_\s \Pi=2\s\P~,  \qquad\cJ\Pi=-2\Pi~.
\label{345}
\eea
The chiral scalar \eqref{F4-P} is a gauge-invariant field strength
for gauge transformations of the form 
\bea
\d_L P = L~, 
 \qquad 
\bar \D
L
=0~, \qquad \bar L =L~.
\label{gauge-inv-P}
\eea
Here the linear gauge parameter can be expressed via an unconstrained 
superfield $V$ as in  \eqref{G-prep}.
Since $V$ is defined modulo gauge transformations \eqref{G-prep-gauge},
we conclude that any system with action $S=S[\P, \bar \P]$, which describes the 
dynamics of the two-form multiplet, is 
a gauge theory with linearly dependent generators.

Lagrangian quantisation of the two-form multiplet can be carried out 
similarly to that of the 4D $\cN=1$ three-form multiplet 
coupled to supergravity \cite{BK88} (see \cite{Ideas} for a review).

Upon replacement $\F^4 \to \P$ in \eqref{3.19}
the supergravity action turns into
\bea
S_{\text{(1,1)\,SG}}^{\text{two-form}} &=&-\frac{4}{\k} \int \rd^3x\rd^2\q\rd^2\qb
\,E\,\Big\{
\big(\bar \P \P\big)^{\frac{1}{4}}
- \hf mP
\Big\} \non \\
&=& -\frac{4}{\k} \int \rd^3x\rd^2\q\rd^2\qb
\,E\,
\big(\bar \P \P\big)^{\frac{1}{4}}
 + \Big\{ \frac{m}{\k} \int {\rm d}^3x {\rm d}^2 \q \,\cE\,   \P  + {\rm c.c.} \Big\}
~,
\label{3-form_sugra}
\eea
where $m$ is a real parameter.
In the second form, the action 
is manifestly invariant under gauge transformations \eqref{gauge-inv-P}.
The equation of motion for the compensator is
\bea
{\mathbb R} +\bar {\mathbb R}=2m~, \qquad 
{\mathbb R}:= \P^{-{3}/{4}} \bar \D \bar \P^{{1}/{4}}~,
\eea
and therefore 
\bea
{\mathbb R} =\m =\text{const}~.
\label{3.455}
\eea

The action for type I minimal supergravity \eqref{3.19} involves two real parameters, 
${\rm Re} \,\m$ and ${\rm Im}\, \m$, which appear in the supersymmetric cosmological term. The action for two-form supergravity \eqref{3-form_sugra} contains only one 
real parameter, $m$, which determines the corresponding supersymmetric cosmological term. As is seen from \eqref{3.455}, the second parameter 
${\rm Im} \,\m$ is generated dynamically. At the component level, 
the cosmological constant in the theory \eqref{3.455} is given by \eqref{N=2cc}.

The two-form supergravity theory described above is the 3D analogue 
of the variant formulation for 4D $\cN=1$ supergravity known as three-form 
supergravity. The latter was proposed for the first time by Gates and Siegel 
 \cite{GS} and fully developed at the component level in
 \cite{Binetruy:1996xw,OvrutWaldram}. The super-Weyl invariant formulation 
 for the three-form supergravity was given in \cite{KMcC}. 
 Our formulation of the 3D two-form supergravity
is similar to \cite{KMcC}. 


\subsection{Superform formulation for the two-form multiplet}

We now present a geometric formulation for the two-form multiplet
used in the previous section.
Let us introduce a super 2-form $B_2$  defined by 
\bea
B_2[P]&=&
-\bar{E}_\a\wedge E^\a P
+\frac{\ri}{2} E^\b\wedge E^a  (\g_a)_{\b\g}\cD^{\g}P
+\frac{\ri}{2} \bar{E}_\b\wedge E^a (\g_a)^{\b\g}\cDB_{\g}P
\non\\
&&
-\frac{1}{16}\ve_{abc}E^b\wedge E^a \big(
(\g^c)^{\r\t}{[}\cD_{\r},\cDB_{\t}{]}
-8\cC^c\big)P
\label{B2-real}
\eea
and consider its exterior derivative $H_3:= \rd B_2$.
It is not difficult to check that $H_3$ is  given by  the following expression:
\bea
H_3[\P]
&=&
-\ri\bar{E}_\g \wedge \bar{E}_\b \wedge E^a\, (\g_a)^{\b\g}\P
-\ri E^\g \wedge E^\b \wedge E^a\, (\g_a)_{\b\g}\bar{\P}
\non\\
&&
+\frac{1}{4}\bar{E}_\g \wedge E^b \wedge E^a\,  \ve_{abd} (\g^d)^{\g\d} \cD_\d \P
-\frac{1}{4} E^\g \wedge E^b \wedge E^a\,\ve_{abd} (\g^d)_{\g\d} \bar \cD^\d \bar{\P}
\non\\
&&
+\frac{\ri}{48}E^c \wedge E^b \wedge E^a \,\ve_{abc} \Big(
(\cD^2 -16 \bar{R})\P
-(\bar \cD^2 -16 R)\bar{\P}
\Big)
~,
\label{H_3F4}
\eea
and, hence, it is 
constructed solely in terms of
the compensator $\P$ and its conjugate $\bar \P$, 
with $\P$ being   related to $P$ as in~\eqref{F4-P}.

The relation $H_3[\P]=\rd B_2[P]$
implies that 
the top components of $B_2[P]$ and $H_3[\P]$,  
\bsubeq
\label{BBHH_ab}
\bea
B_{a b}
&=&
-\frac{1}{8}\ve_{abc}\Big(
(\g^c)^{\r\t}{[}\cD_{\r},\cDB_{\t}{]}
-8\cC^c\Big)P
~,
\label{B_2_ab}
\\
H_{abc}&=&-\frac{\ri}{8} \ve_{abc} \Big(
(\bar \cD^2 -16 R)\P
-(\cD^2 -16 \bar{R})\bar{\P}
\Big)
~,
\label{H_3_abc}
\eea
\esubeq
are connected to each other as
\bea
H_{abc}
&=&
3\cD_{[a}B_{bc]}
+\ve_{abc}\big(\ri\cD^{\a} R-2\cDB^{\a}\cS\big)\cD_{\a}P
+\ve_{abc}\big(\ri\cDB^{\a}\bar{R}+2\cD^{\a}\cS \big)\cDB_{\a}P
~.
\label{H=DB}
\eea
Equations \eqref{BBHH_ab} and \eqref{H=DB} tell us  that the imaginary part of the top component field of the chiral superfield $\P$,
defined by $F=-\frac 14 \cD^2\P|$,
is the field strength of a gauge two-form.

The gauge transformation \eqref{gauge-inv-P}  of the prepotential $P$ is equivalent to 
the following transformation of the super 2-form \eqref{B2-real}:
\bea
\d_L B_2[P]=B_2[L] \quad \Longrightarrow \quad \d_L H_3[\P] =0~.
\label{3.54}
\eea
This allows us to interpret $B_2[P]$ as a gauge two-form and 
$H_3[\P]$ as its gauge-invariant field strength. 
The closed super 2-form $B_2[L]$ in \eqref{3.54}
is actually exact, $B_2[L] = \rd A_1$, 
where $A_1$ is the gauge potential of a vector multiplet.

Using the super-Weyl transformation laws 
 \eqref{N=2vielbein-super-Weyl} and \eqref{s-weyl-P}, 
one can check that the superform \eqref{B2-real}
 is  invariant under
arbitrary  super-Weyl transformations, 
\bea
\d_\s B_2 [P] =0\quad \Longrightarrow \quad
\d_\s H_3[\P]=0~.
\label{355}
\eea
This property will be important for our analysis in section \ref{GS2}.

Let us recall the closed super 3-form $\X_3 [\cL_{\rm c}]$, 
defined by eq.  \eqref{Sigma_3},
which generates the supersymmetric invariant \eqref{3300}.
If we choose  $\cL_{\rm c}={\P}$, 
with $\P$ given 
by \eqref{F4-P}, then the exact super 3-form $H_{3}[\P]$ proves to be the imaginary part ${\X}_3 [\P]$,
\bea
H_{3}[\P]=\frac{\ri}{2}\X_3 [\P]-\frac{\ri}{2}\bar{\X}_3[\bar{\P}]
~.
\eea
The real part of ${\X}_3 [\P]$, on the other hand, is not exact and 
generates a non-trivial supersymmetric invariant, which 
may be realised as the 
full superspace integral \eqref{N=2Ac},
with $P$ playing the role of the Lagrangian $\cL$.

The local U(1) and super-Weyl transformations may be used to 
choose the gauge $ \P=1$.
This condition implies that
$\cS=0$ and the algebra of covariant derivatives reduces to that of type I minimal supergravity~\cite{KLT-M11,KT-M11}
with one extra constraint: 
the imaginary part of $R$ is now 
 the divergence of a vector (related, by Poincar\'e duality,  to a two-form potential).
To see this it suffices to write the super 3-form $H_3[\P]$ in the gauge $\P=1$
\bea
H_3
&=&
-\ri E^\g \wedge E^\b \wedge E^a\, (\g_a)_{\b\g}
-\ri\bar{E}_\g \wedge \bar{E}_\b \wedge E^a\,(\g_a)^{\b\g} 
\non\\
&&
+\frac{1}{3} E^c \wedge E^b \wedge E^a \, \ve_{abc}\,\ri\big(R-\bar{R}\big)
~,
\label{Hgauge1}
\eea
keeping in mind that $H_3=\rd B_2[P]$.
Note that a similar constraint appears
in the case of the 4D $\cN=1$ three-form supergravity where $\ri(R-\bar{R})$ is also the divergence of a vector~\cite{Binetruy:1996xw,OvrutWaldram,Binetruy:2000zx}.


\subsection{Complex two-form supergravity}

In the framework of 4D $\cN=1$ Poincar\'e supersymmetry, 
the complex three-form multiplet was introduced by 
Gates and Siegel~\cite{GS}
as
a conformal compensator 
for the Stelle-West formulation for 4D $\cN=1$ supergravity \cite{old1}, in which
 the complex auxiliary field $F$ was realised as  
the field strength of a complex gauge three-form.  
The name ``complex three-form multiplet'' was coined in \cite{GGRS}.
This multiplet was recently used in \cite{FLMS} 
(under the name of ``double three-form multiplet'')
to construct a super-Weyl invariant formulation for the complex three-form
supergravity of \cite{old1}, in the spirit of the super-Weyl invariant formulation \cite{KMcC} 
for  three-form supergravity \cite{GS,OvrutWaldram}.
Here we propose 
a 3D $\cN=2$ cousin of the complex 
three-form multiplet.

A complex two-form multiplet coupled to conformal supergravity
 is described in terms of a covariantly chiral scalar $\U$ and 
its conjugate $\bar \U$, with $\U$ being defined by 
\bea
\U =
\bar \D
 \bar \S~,
\label{3.42}
\eea
where $\S $ is a {\it complex linear} superfield constrained by 
\bea 
\bar \D
\S=0~.
\eea
In general, if $\S$ is chosen to transform homogeneously under the 
super-Weyl transformations, 
 its  ${\rm U(1)}$ charge is determined by the super-Weyl weight \cite{KLT-M11} 
\bea
\d_\s \S=w_\S \s\S \quad \Longrightarrow \quad \cJ\S=(1-w_\S) \S
~.
\label{3.44}
\eea
We wish the chiral scalar $\U$ to be super-Weyl primary, 
which means
\bea
\d_\s \U=w_\U \s\U~, \qquad
\cJ \U=-w_\U \U~ ,
\label{3.45}
\eea
in accordance with \eqref{3.199}. The transformation properties \eqref{3.44} and 
\eqref{3.45} are consistent with \eqref{3.42} only if $w_\S =1$, 
and therefore 
\bea
\d_\s \U=2 \s\U~, \qquad
\cJ \U=-2 \U~ .
\label{3.46}
\eea

The chiral scalar $\U$ defined by \eqref{3.42} is 
a gauge-invariant field strength under gauge transformations  
 of the form
\bea
\d_L \bar \S = L_1 + \ri L_2~, \qquad \bar \D
L_i
=0~, \qquad \bar L_i =L_i
\label{3.63}
\eea
For many purposes such as Lagrangian quantisation, 
it is advantageous to work with unconstrained superfields.  
The anti-linear superfield $\bar \S$ can always be represented as
\bea
\bar \S = \cD^\a \J_\a~,
\label{3.64}
\eea
for some  unconstrained complex spinor prepotential $\J_\a$.
The chiral scalar $\U$ defined by \eqref{3.42} is a gauge-invariant field strength
under gauge transformations of the form
\bea
\d \J_\a = \bar \cD_\a Z + \cD^\b \L_{\a\b}~, \qquad \L_{\a\b}=\L_{\b\a}~,
\eea
with unconstrained complex gauge parameters $Z$ and $\L_{(\a\b)}$.
Here the gauge transformation generated by $\L_{\a\b}$ leaves
the superfield \eqref{3.64} invariant. 
The gauge transformation generated by $Z$ is equivalent to \eqref{3.63}
when acting on $\bar \S$.
Any dynamical system with action $S[\U, \bar \U]$, which is 
realised in terms of the  unconstrained
prepotentials $\J_\a$ and $\bar \J_\a$, is a gauge theory with linearly dependent 
generators of an infinite  stage of reducibility, following 
the terminology of the Batalin-Vilkovisky quantisation \cite{BV}.

Upon replacement $\F^4 \to \U$ in \eqref{3.19}
the supergravity action turns into
\bea
S_{\text{(1,1)\,SG}}^{\text{complex\,two-form}} =-\frac{4}{\k} \int \rd^3x\rd^2\q\rd^2\qb
\,E\,
\big(\bar \U \U\big)^{\frac{1}{4}}~.
\eea
This complex two-form supergravity allows no supersymmetric cosmological term, 
and the action involves no free parameter, unlike the actions for 
type I supergravity \eqref{3.19} and two-form supergravity \eqref{3-form_sugra}. 
However, the equation of motion for $\J_\a$ is
\bea
\cD_\a {\mathbb R} =0~, \qquad 
{\mathbb R}:= \U^{-{3}/{4}} \bar \D \bar \U^{{1}/{4}}~,
\eea
and it implies that ${\mathbb R} =\m =\text{const}$. 
Thus the complex cosmological parameter $\m$ is generated dynamically.


\subsection{Superform formulation for the complex two-form multiplet}

Similarly to the real two-form multiplet, the complex two-form multiplet
has a geometric superform origin. 
Let us consider the following
complex super 2-form:
\bea
C_2[\bar\S]&=&
2\ri\bar{E}_\a\wedge E^\a\, \bar\S
+E^\b\wedge E^a  (\g_a)_{\b\g}\cD^{\g}\bar\S
+\bar{E}_\b\wedge E^a (\g_a)^{\b\g}\cDB_{\g}\bar\S
\non\\
&&
+\frac{\ri}{8}\ve_{abc}E^b\wedge E^a \Big(
(\g^c)^{\r\t}{[}\cD_{\r},\cDB_{\t}{]}
-8\cC^c\Big)\bar\S
~.
\label{B2-complex}
\eea
All coefficients $C_{AB}$ of $C_2[\bar\S] =\hf  E^B\wedge E^A C_{AB}$ are 
 descendants of $\bar \S$.
 For the exterior derivative of $ C_2[\bar \S]$
we get
\bea
\rd C_2[\bar\S]
&=&
-2\bar{E}_\g \wedge \bar{E}_\b \wedge E^a(\g_a)^{\b\g} \U
-\frac{\ri}{2}\bar{E}_\g \wedge E^b \wedge E^a\ve_{abd} (\g^d)^{\g\d} \cD_\d \U
\non\\
&&
+\frac{1}{24} E^c \wedge E^b \wedge E^a\ve_{abc}(\cD^2-16\bar R)\U
\equiv \X_3[\U]
~.
\label{H_3F4-2}
\eea
Thus, all coefficients of $\rd C_2[\bar\S]$ are descendants of $\U$.

The expression for $ \X_3[\U]$ 
is obtained from 
\eqref{Sigma_3} by replacement  $\cL_{\rm c} \to \U$. 
Since both $\cL_{\rm c} $ and $ \U$ are chiral primary superfields of the same weight, 
we conclude that  $ \X_3[\U]$ 
is super-Weyl  invariant, $\d_\s \X_3[\U]=0$. 
A stronger result is that the superform \eqref{B2-complex}
is also super-Weyl invariant 
\bea
\d_\s C_2[\bar\S] =0~.
\eea

Our result
$\Xi_3[\U]=\rd C_2[\bar\S]$
implies that 
the top components of the superforms $C_2[\bar\S]$ and $\Xi_3[\U]$, 
\bsubeq
\label{CCC}
\bea
C_{a b}
&=&
\frac{\ri}{4}\ve_{abc}\Big(
(\g^c)^{\r\t}{[}\cD_{\r},\cDB_{\t}{]}
-8\cC^c\Big)\bar\S
~,
\label{B_2_ab-2}
\\
\Xi_{abc}&=&\frac{1}{4} \ve_{abc}(\cD^2 -16 \bar{R})\U
~,
\label{H_3_abc-2}
\eea
\esubeq
are related to each other as
\bea
\Xi_{abc}
&=&
3\cD_{[a}C_{bc]}
+2\ve_{abc}\big(\cD^{\a}R+2\ri\cDB^{\a}\cS \big)\cD_{\a}\S
+2\ve_{abc}\big(\cDB^{\a} \bar R-2\ri\cD^{\a}\cS\big)\cDB_{\a}\S
~.
\label{H=DB-2}
\eea
This confirms that the $F$-component of $\U$ is the field strength of a complex two-form.

The gauge transformation of $\bar \S$, eq. \eqref{3.63}, 
can be viewed as the following superform transformation 
\bea
\d_L C_2[\bar\S] = C_2[ L_1 + \ri L_2] \quad \Longrightarrow \quad 
\d_L  \X_3[\U]=0~.
\eea
This allows us to interpret $C_2[\bar \S]$ as a gauge complex two-form and 
$\X_3[\U]$ as the corresponding gauge-invariant field strength. 

The local U(1) and super-Weyl transformations may be used to choose  
the gauge  $\U=1$. 
In this gauge, $\cS=0$ and the algebra of covariant derivatives reduces to that of type I minimal supergravity~\cite{KT-M11,KLT-M11}
with one extra constraint: the torsion $R$ 
 is the divergence of a vector 
 (related, by Poincar\'e duality, to a complex two-form potential).
This follows from the fact that 
$\Xi_3[\U]$ in the gauge $\U=1$ is given by 
\bea
\Xi_3
&=&
\rd C_2=
-2\bar{E}_\g \wedge \bar{E}_\b \wedge E^a(\g_a)^{\b\g}
-\frac{2}{3} E^c \wedge E^b \wedge E^a\ve_{abc}\bar R
~.
\label{Hgauge1-2}
\eea


\section{Green-Schwarz superstrings coupled to two-form supergravity}
\label{section4}

In this section we will show that  the $\cN=1$ and $\cN=2$ two-form 
supergravity theories
provide consistent backgrounds for the Green-Schwarz superstring.

\subsection{3D $\cN=1$ Green-Schwarz superstring in  curved superspace}
\label{GS1}

In the case of 3D $\cN=1$ Green-Schwarz superstring, 
we draw on the results obtained by  Bergshoeff {\it et al.} 
\cite{Bergshoeff:1985su}.
To describe the dynamics of a superstring propagating in a two-form supergravity
background, we propose the following action
\bea
S=T_2\int\rd^2\x\,\Big\{
\hf\sqrt{-\g}\g^{ij}LE_i{}^aE_j{}^b\eta_{ab}
-\e^{ij}E_i{}^BE_j{}^AB_{AB}\Big\}
~.
\label{GS-string-1}
\eea
Here $\x^i=(\t,\s)$ are the world-sheet coordinates, $\g_{ij}$ is 
 the world-sheet metric,
$\g=\det \g_{ij}=\hf\e^{ij}\e^{kl}\g_{ik}\g_{jl}$ with $\e^{12}=\e_{21}=1$. 
Both the kinetic and Wess-Zumino terms in \eqref{GS-string-1}
involve certain target space fields 
associated with two-form supergravity,
which are the supervielbein $E_{M}{}^A$ entering the action via 
the pull-back supervielbein
\bea
E_i{}^A=\pa_i z^M E_{M}{}^A
~,
\label{4.2}
\eea
the super 2-form  $B_{AB}$ and the compensator $L= \cD^\a L_\a$
(the dilaton superfield).

The classical consistency of the Green-Schwarz superstring action 
 requires that it be invariant
under gauge fermionic transformations ($\k$-symmetry) 
of the form
\bea
\d E^a&=&0
~,~~~
\d E^\a=2(\g_a)^{\a\b}L^{\frac{1}{4}}E_i^a\k^i_\b
~,
\label{k-sym-1}
\eea
where we have defined $\d E^A:= \d z^M E_M{}^A$.
 The gauge parameter  $\k^i_\a$
is a real 3D spinor and also a 2D vector satisfying the self-duality condition
$(\g^{ij}-(-\g)^{-\hf}\e^{ij})\k_\a{}_j=0$.

It can be shown that the action \eqref{GS-string-1} 
is invariant under the gauge transformation \eqref{k-sym-1} provided 
the super 3-form $H_3=\rd B_2$  
is given by eq.~\eqref{H_3-1} and the world-sheet metric transforms as 
\bea
\d(\sqrt{-\g}\g^{ij})
&=&
-2\sqrt{-\g} L^{\frac{1}{4}} \Big(
4\ri  E_k{}^\a
-L^{-1}(\g_{c})^{\a\b}E_k{}^c\cD_\b L\Big)
\big(2\g^{k(i}\g^{j)l}- \g^{ij}\g^{kl}\big)\k_{l\a}
~.~~~
\label{k-sym-2}
\eea
Let us point out that one can absorb the factor of $L^{1/4}$  into $\k_\a^i$.
After this redefinition, the action~\eqref{GS-string-1}  
and the $\k$-transformations \eqref{k-sym-1} and  \eqref{k-sym-2} 
become similar to those  in  \cite{Bergshoeff:1985su}.

The action \eqref{GS-string-1} is invariant under arbitrary super-Weyl transformations 
of the target space, as a consequence of the relations 
\eqref{N=2vielbein-super-Weyl}, \eqref{2266} and \eqref{2422}.
The super-Weyl gauge freedom may be fixed by setting $L=1$. 


\subsection{3D $\cN=2$ Green-Schwarz superstring in  curved superspace}
\label{GS2}

Now we turn to constructing the covariant action for 
 the 3D $\cN=2$  superstring in a two-form supergravity background, 
and make use of 
the results by Grisaru {\it et al.} \cite{Grisaru:1985fv}
concerning the 10D $\cN=2$ superstring. 
We propose the following superstring action
\bea
S=T_2\int\rd^2\x \,\Big\{
\hf\sqrt{-\g}\g^{ij} (\F\bar{\F})^2E_i{}^aE_j{}^b\eta_{ab}
-\hf\e^{ij}E_i{}^BE_j{}^AB_{AB}\Big\}
~,
\label{GS-string-22}
\eea
where the pull-back supervielbein $E_i{}^A$ is defined similarly to \eqref{4.2}.
The dilaton $(\F\bar{\F})^2$ is constructed in terms of the 
conformal compensator described by  
a weight-1/2 chiral scalar superfield $\F$ and its conjugate $\bar \F$.
The concrete structure of $\F$ depends on the supergravity formulation chosen.
In the case of three-form supergravity, the conformal compensator 
is the three-from multiplet, and then $\F^4=\Pi=\bar\D P$.  
On the other hand, the choice $\F^4=\U=\bar\D \bar\S$  corresponds to 
complex three-form supergravity.

Both the real and complex two-form supergravities 
possess a real super 2-form $B_2$ 
which can be used as the Kalb-Ramond field $B_{AB}$ 
in the action \eqref{GS-string-22}.
For two-form supergravity, the  choice
of $B_2$ is unique, modulo an overall numerical factor, 
and is  given by $B_2[P]$,  eq.~\eqref{B2-real}.
In the case of complex two-form supergravity, there is a whole family 
of possible super 2-forms that can be put in a one-to-one correspondence with 
a circle U(1). However, all these choices are equivalent. 
For concreteness, we choose $B_2$ 
to be  the real or imaginary part of 
the super 2-form  $C_2[\bar\S]$ given by eq.~\eqref{B2-complex}.

Let us show that the action \eqref{GS-string-22}
 is $\k$-symmetric once we consider a  background of
 real or complex two-form supergravity.
We postulate the following $\k$-symmetry transformation
\bea
\d E^a=0
~,~~~
\d E^\a
=\F^{\frac{3}{2}}\bar{\F}^{-\frac{1}{2}}
E_i^a(\g_a)^{\a\b}\Big(\g^{ij}\k_{j \b}-(-\g)^{-\hf}\e^{ij}\bar{\k}_{j \b}\Big)
~,
\label{k-symm-2}
\eea
where $\d E^A:= \d z^M E_M{}^A$, 
$\d \bar{E}_\a$ is given by the complex conjugate of $\d E^\a$ and 
$
\overline{\k_{i}^{\a}}
\equiv \bar{\k}_{i}^{\a}
$.
We point out the relation
\bea
\d E_i{}^A
&=&
\pa_i \d E^A
-2\d E^CE_i{}^B{\bm\O}_{[BC)}{}^A
+\d E^CE_i{}^BT_{BC}{}^A
~,
\eea
where we have used the definitions \eqref{3.8connection} and \eqref{torsion-def}. 
Then
it is not difficult to show that the variation of the action 
 is given by the following
expression (compare with \cite{Grisaru:1985fv})
\bea
\d S&=&
T_2\int\rd^2\x \,\Big\{
\hf \d(\sqrt{-\g}\g^{ij})(\F\bar{\F})^2E_i{}^aE_j{}^b\eta_{ab}
-\sqrt{-\g}\g^{ij}(\F\bar{\F})^2E_i{}^B\d E^AT_{AB}{}^cE_j{}^d\eta_{cd}
\non\\
&&
+\sqrt{-\g}\g^{ij} E_i{}^aE_j{}^b\eta_{ab}
\big(\F\bar{\F}^2\d E^\a\cD_\a\F
+\F^2\bar{\F}\d\bar{E}_\ad\cDB^\ad\bar{\F}\big)
+\e^{ij}E_i{}^CE_j{}^B\d E^AH_{ABC}
\Big\}
~,~~~~~~~~~
\label{var-1}
\eea
where 
$H_3:=\frac{1}{6}E^C\wedge E^B\wedge E^AH_{ABC}=\rd B_2$.

Let us 
first consider the case of two-form supergravity, with
$\F^4={\Pi}$.
To show that the variation \eqref{var-1} vanishes, 
we have to make use of the geometrical data specific for the two-form supergravity.
The only non-vanishing torsion appearing in the variation~\eqref{var-1} is the dimension-zero torsion
which is
\bea
T_{\a}{}^{\b}{}^c=-2\ri(\g^c)_\a{}^{\b}
\label{New1}
~.
\eea
The non-trivial components of the super 3-form $H_3$ given by eq.~\eqref{H_3F4},
which enter the variation~\eqref{var-1}, are 
\bea
H_{\a\b c}=-2\ri(\g_c)_{\a\b}\bar{\F}^4
~,~~~
H_{ab\g}=-\hf\ve_{abd} (\g^d)_{\g\d} \bar \cD^\d \bar{\F}^4
\label{New2}
\eea
together with their complex conjugates.
Substituting the expressions \eqref{New1} and  \eqref{New2} into the variation~\eqref{var-1} and using the identities 
\bea
(\g_a)_{\a}{}^\g(\g_b)_{\g}{}^\b=\eta_{ab}\d_\a^\b+\ve_{abc}(\g^c)_{\a}{}^\b
~,~~~~~~
\g^{i[j}\g^{k]l}=\hf \e^{jk}\e^{il}\g^{-1}
~,
\eea 
one can show that the Green-Schwarz action is indeed invariant provided 
the $\k$-transformation law of the world-sheet metric is postulated to be 
\bea
&&\d(\sqrt{-\g}\g^{ij})=
2\sqrt{-\g}\Big(2\g^{k(i}\g^{j)l}-\g^{ij}\g^{kl}\Big)
\F^{\frac{3}{2}}\bar{\F}^{-\frac{1}{2}}\Big(
2\ri\bar{E}_k{}_\a
+\F^{-1}(\g_c)_{\a\b}E_k{}^c\cD^\b\F
\Big)\k_{l}^{ \a} 
\non\\
&&\qquad -2\Big(\e^{k(i}\g^{j)l}+\ve^{l(i} \g^{j)k}\Big)
\F^{-\frac{1}{2}}\bar{\F}^{\frac{3}{2}}\Big(
2\ri E_k{}^\a
+\bar{\F}^{-1}(\g_c)^{\a\b}E_k{}^c\cDB_\b\bar{\F}
\Big)\k_{l \a}
+{\rm c.c.}
\eea

The superstring action constructed is invariant under arbitrary super-Weyl transformations
of the background fields, as follows from the transformation laws
\eqref{N=2vielbein-super-Weyl}, \eqref{345} and \eqref{355}.

It is clear that the analysis in the case of the complex two-form supergravity 
is identical to the one presented above with the only difference that $\F^4$  is replaced with 
$\U$ instead of $\Pi$.
In fact, in proving the $\kappa$-invariance of the action
only the real closed super 3-form $H_3$ enters the computations rather than its 
potential $B_2$.
Therefore we have proven that both the real and complex two-form supergravities are consistent backgrounds for the 3D $\cN=2$ Green-Schwarz superstring.


\section{Goldstino superfields coupled to supergravity}\label{section5}

In this section we present various models for spontaneously broken local 
$\cN=1$ and $\cN=2$ supersymmetry that are obtained by coupling 
the off-shell supergravity theories, which have been described in the previous
sections, to nilpotent Goldstino superfields.
It should be pointed out that the first model 
for spontaneously broken local $\cN=1$ supersymmetry was constructed
in 1977 \cite{DD}
by coupling on-shell $\cN=1$ supergravity to the Volkov-Akulov action.

We often make use of the notion of 
reducible and irreducible Goldstino superfields introduced in \cite{BHKMS}.
By definition, an irreducible Goldstino superfield contains Goldstone 
spin-$\hf$ fermion(s) as the only independent component field(s). 
Every reducible Goldstino superfield also contains some auxiliary field(s)
along with the Goldstone spin-$\hf$ fermion(s).

\subsection{$\cN=1$ Goldstino superfields}\label{section5.1}

A reducible Goldstino multiplet is described by a real scalar superfield $X$ 
subject to the nilpotency constraint 
\bea
X^2=0~.
\label{5.1}
\eea
We also require $\cD^2 X$ to be nowhere vanishing so that 
$(\cD^2 X)^{-1}$ is well defined, and therefore \eqref{5.1} implies
\bea
X = -\frac{\cD^\a X \cD_\a X}{\cD^2 X}~.
\label{N=1composite}
\eea
As a result, $X$ has two independent component fields, 
a spinor $\j_\a (x) $ and an auxiliary scalar $F(x)$,  that may be defined 
as $\ri \j_\a = \cD_\a X|$ and $\ri F = -\frac{1}{4} \cD^2 X|$, 
where $F^{-1}$ is well defined.
The lowest component of the Goldstino superfield, $X|$, 
is a composite field as a consequence of \eqref{N=1composite}.

We postulate $X$ to be super-Weyl primary of weight $1/2$, 
which means the super-Weyl transformation law of $X$ is
\bea
\d_\s X=\hf \s X~.
\eea
The Goldstino superfield action is 
\bea
S_X&=&
\ri \int  \rd^3x\rd^2\q
\,E \,\Big\{ \frac{\ri}{2} \cD^\a X \cD_\a X +2 f'  \varphi^3 X\Big\}
~,
\label{5.2}
\eea
for some non-zero parameter $f'$ which characterises the scale of supersymmetry breaking. The second term in the action involves the compensator, $\vf$, 
of $\cN=1$ AdS supergravity, see section \ref{section2.3}.
The action is super-Weyl invariant. 

The nilpotency constraint \eqref{5.1} is invariant under local arbitrary re-scalings of $X$, 
\bea
X ~\to ~ \tilde X =\re^{\r} X~, 
\label{5.3}
\eea
for any real scalar $\r$. 
Such a re-scaling \eqref{5.3} acts on the component fields of $X$ as 
\begin{subequations}\label{N=1rescale}
\bea
\j_\a ~& \to &  ~ \tilde \j_\a= \re^{\r|} \Big( \j_\a + \frac{\j^2}{4F} (\cD_\a \r )|\Big)~,
\label{N=1rescale.b}\\
F ~& \to & ~ \tilde F =\re^{\r| } \Big( F -\hf \j^\a (\cD_\a \r)| - \frac{\j^2}{16F} (\cD^2 \r)| \Big)~. \label{N=1rescale.c}
\eea 
\end{subequations}
Each of these transformations is a local re-scaling accompanied by a nilpotent shift
of the field under consideration, and therefore
$\tilde F^{-1}$ is well defined.
Requiring the action \eqref{5.2} to be stationary under \eqref{5.3}
(following the 4D works \cite{KMcAT-M,BK17}) 
gives the constraint 
\bea
f' \vf^3 X = \frac{\ri}{2} X \cD^2 X= X \D X~, \qquad
\D:=  \frac{\ri}{2} \cD^2 +\cS
~.
\label{5.4}
\eea
Here $X\D X$ is  manifestly a super-Weyl primary.
As follows from \eqref{N=1rescale.b} and \eqref{N=1rescale.c},
the $F$-component of the nonlinear constraint \eqref{5.4} is equivalent 
to a sum of the equation of motion for $F$ and a linear combination of the 
equations of motion for $\j_\a$.

Consider an irreducible Goldstino superfield $\cX$ constrained by 
\bea
\cX^2=0~, \qquad f' \vf^3 \cX = \cX \D \cX~,
\label{5.5}
\eea
with $\D \cX $ being nowhere vanishing. 
This superfield is irreducible because the Goldstino $\j_\a =
-\ri \cD_\a X|$ is the only independent component field of $\cX$.
Indeed, the second constraint in \eqref{5.5} proves to express the auxiliary field 
$F$ in terms of the Goldstino, see Appendix 
\ref{AppendixA}. 
The dynamics of $\cX$ is described by the action 
\bea
S_\cX&=&
 \ri f' \int  \rd^3x\rd^2\q
\,E \,  \varphi^3 \cX~,
\label{5.6}
\eea
which is obtained from \eqref{5.2} by making use of the nonlinear constraint 
obeyed by $\cX$. The Goldstino theories \eqref{5.2} and \eqref{5.6}
prove to be equivalent, which may be shown by extending the 4D analyses given in  
\cite{KS,KT0,BFKVP2}.\footnote{Ref. \cite{KT0} is a considerably 
generalised and extended version of 
\cite{Kuzenko:2010ef}.}  
This issue is discussed in more detail in Appendix 
\ref{AppendixA}. The flat-superspace limit of our Goldstino theory 
defined by eqs. \eqref{5.5} and \eqref{5.6} is analogous to the 2D $\cN=1$ 
Goldstino model pioneered by Ro\v{c}ek \cite{Rocek}.

It is not difficult to check that the constraints \eqref{5.5} are 
satisfied if 
\bea
\cX = f' \vf ^3 \frac{X}{\D X}~,
\label{5.7}
\eea
where $X$ is only subject to the nilpotency constraint \eqref{5.1}.
The important property of $\cX$ defined by \eqref{5.7} is that it is invariant 
under arbitrary local re-scalings \eqref{5.3},
\bea
\d_\r X =\r X \quad \Longrightarrow \quad \d_\r \cX =0~,
\eea
for arbitrary real superfield $\r$, compare with the 4D analysis in \cite{BK17}.
This remarkable property actually can be explained by recalling 
at the component transformation law 
\eqref{N=1rescale.c} implied by  \eqref{5.3}.
The point is the superfield transformation  \eqref{5.3} implies 
an arbitrary local re-scaling of  the auxiliary field of $X$, $F \to \re^{\r|} F$.
Since $\cX$ does not contain an independent auxiliary field, it should remain 
invariant under  \eqref{5.3}.

Let us consider the model for spontaneously broken local supersymmetry 
which is obtained by coupling the supergravity theory \eqref{N=1-scalar} to the 
Goldstino superfield $X$. The dynamics of this system is described 
by the action 
\bea
S=S_{\rm SG} +S_X~.
\label{5111}
\eea
The component structure of this theory will be discussed elsewhere.
Here we only present the corresponding cosmological constant. 
It is obtained upon eliminating all the auxiliary fields, and is given by 
\bea
\L=\hf f'^2 \k +\L_{\rm AdS}
=\hf f'^2 \k - 4\l^2
~.
\eea

The supergravity-matter system \eqref{5111} may be reformulated as
a model for nilpotent supergravity. 
Varying  \eqref{5111} with respect to the compensator $\vf$ gives the equation
\bea
{\mathbb S} -\l = -\frac{3}{8} f' \k \frac{X}{\vf} ~,
\label{5133}
\eea
where $\mathbb S$ is defined by \eqref{N=1equations.a}.
Since $X$ is nilpotent, the equation of motion implies 
\bea
({\mathbb S} -\l )^2=0~.
\eea
Making use of \eqref{5133} in order to express $X$ in terms of the supergravity fields, 
the action \eqref{5111} can be recast as a higher-derivative supergravity theory 
\bea
S&=&
 \frac{8\ri}{3\k}   \int  \rd^3x\rd^2\q
\,E \,  \varphi^4 \Big\{ {\mathbb S}+\frac{\l}{2} \Big\}
-\frac{32}{ (3f' \k)^2 }  \int  \rd^3x\rd^2\q
\,E \,  \varphi^2 \cD^\a {\mathbb S} \cD_\a {\mathbb S}~.
\eea 
In four dimensions, various approaches to nilpotent $\cN=1$ supergravity were 
developed, e.g., in \cite{ADFS,DFKS,K15,BMST,KMcAT-M,AM,CDFP}. Our presentation here is similar to \cite{K15}.

To conclude this subsection,  we note that the nilpotent Goldstino superfield $X$
can also be coupled to the two-form supergravity constructed 
in section \ref{section2.4}.
For this we should simply replace the action 
\eqref{5.2}
with 
\bea
\widetilde{S}_X&=&
\ri \int  \rd^3x\rd^2\q
\,E \,\Big\{ \frac{\ri}{2} \cD^\a X \cD_\a X +2 f'  L^{3/4} X\Big\}
~.
\eea
Then, the equation of motion \eqref{5133} turns into 
 \bea
\cD_\a\Big( {\mathbb S} +\frac{3}{8} f' \k \frac{X}{L^{1/4}} \Big) =0~,
\eea
where $\mathbb S$ is now defined as in \eqref{2299}.


\subsection{Reducible $\cN=2$ Goldstino superfields}

The family of nilpotent  $\cN=2$ Goldstino superfields, 
both reducible and irreducible, is more populous than 
in the $\cN=1$ case.\footnote{One can also introduce spinor Goldstino superfields, 
by analogy with the 4D $\cN=1$ constructions given in \cite{IK78,IK82,SW}. 
However such superfields are not particularly useful in the supergravity framework.}  
However practically all 3D $\cN=2$ Goldstino superfields can be obtained 
from the known 4D $\cN=1$ Goldstino supermultiplets by dimensional reduction,
at least in the flat superspace case. This is why our discussion 
of nilpotent  $\cN=2$ Goldstino superfields will be reasonably concise. 
We will try to emphasise only conceptual constructions and those results 
that are truly new or have not received much discussion in the 4D case.

\subsubsection{Nilpotent chiral scalar superfield}\label{section5.2.1}
 
 To begin with, we consider a 3D $\cN=2$ locally supersymmetric  counterpart of  
the reducible Goldstino superfield introduced by  Casalbuoni {\it et al.} 
\cite{Casalbuoni} and independently by Komargodski and Seiberg \cite{KS}.
We choose it to be a covariantly chiral scalar $X$ of super-Weyl weight +1/2, 
\bea
\bar \cD_\a X =0~, \qquad 
\d_\s X=\hf  \s X  \quad \Longrightarrow \quad \cJ X=-\hf X~,
\label{chiral}
\eea
which is subject to the nilpotency constraint
\bea
 X^2 =0~,
\label{NC_chiral}
\eea
in conjunction with the requirement that  the descendant 
$\cD^2  X $ is nowhere vanishing.
The nilpotency condition implies that $X$ has two independent component 
fields, a complex Goldstino $\j_\a(x)$ and a complex auxiliary field $F(x)$, 
which we define as $\psi_{\alpha}= \frac{1}{\sqrt{2}} \cD_{\alpha} X|$ and 
$F=-\frac{1}{4} \cD^2 X|$, respectively.

The constraints on $X$ do not make use of any supergravity compensator, 
which means that $X$  is defined in any  conformal supergravity background. 
However, a compensator is required in order to define an action functional 
for the  Goldstino superfield. 
Here we choose the chiral compensator $\F$ corresponding to the 
minimal (1,1) AdS supergravity described in section \ref{section3.3.1}.
The dynamics of this supermultiplet is described by the action 
\bea
S_X = \int \rd^3 x \rd^2 \q  \rd^2 \bar{\q} \, E\,\bar X X 
-  \Big\{f  \int \rd^3 x \rd^2 \q \, \cE\,
	\F^3 X + {\rm c.c.} \Big\}~,
\label{GA_chiral}
\eea
in which the parameter supersymmetry breaking, $f$, may be chosen to be real. 

We now consider a model for spontaneously broken $\cN=2$ local supersymmetry 
which is obtained by coupling  the Goldstino superfield $X$ to the minimal (1,1) AdS supergravity reviewed in section \ref{section3.3.1}. The complete action is 
\bea
S = S_{\text{(1,1)\,SG}}^{\text{minimal}} +S_X~,
\label{5.22}
\eea
where the supergravity action $S_{\text{(1,1)\,SG}}^{\text{minimal}} $ is given by eq. \eqref{3.19}. This theory  proves to generate the following cosmological constant 
\bea
\L = f^2 \k + \L_{\rm AdS} = f^2 \k- 4 |\m|^2~.
\label{5.23}
\eea
Varying the action \eqref{5.22}  with respect to the chiral compensator gives the equation of motion 
\bea
{\mathbb R} -\m = -\frac{3}{4} f \k \frac{X}{\F}~,
\label{5.24}
\eea
where the super-Weyl neutral chiral scalar $\mathbb R$ is defined by \eqref{3.28a}.
Since $X$ is nilpotent, the above equation implies 
\bea
({\mathbb R} -\m )^2=0~,
\label{5.25}
\eea
and thus the torsion superfield $({\mathbb R} -\m)$ becomes nilpotent.
Eq. \eqref{5.24} can be used to eliminate $X$ and $\bar X$ from the action 
\eqref{5.22}, resulting with the following geometric higher-derivative 
supergravity action
\bea
S&=& 
-\frac{4}{3\k} \int {\rm d}^3x \rd^2\q\rd^2\qb
\,E\, \bar \F  \F
- \Big\{ \frac{\m}{3\k} \int {\rm d}^3x {\rm d}^2 \q \,\cE\,   \F^4  + {\rm c.c.} \Big\} \non \\
&&+ \Big( \frac{4}{3f \k}\Big)^2 \int \rd^3 x \rd^2 \q  \rd^2 \bar{\q} \, E\,
\bar \F \F\, |{\mathbb R}-\m|^2
\label{5.26}
\eea
Here the expression in the first line differs from  the supergravity action 
\eqref{3.19} only by new values for the parameters involved, 
$\k \to 3 \k $ and $\m \to -\m$. The functional form of the action \eqref{5.26} 
differs from its 4D $\cN=1$ counterpart derived in \cite{K15} (see also \cite{KMcAT-M})
in the sense that the supersymmetric Einstein-Hilbert term completely cancelled out 
in the latter case. 

The nilpotency condition \eqref{NC_chiral} is preserved if $X$ is locally rescaled, 
\bea
X ~\to ~\re^{\t} X~, \qquad \bar \cD_\a \t=0~,
\label{re-scaling_chiral}
\eea
where the parameter $\t$ is 
neutral under  ${\rm U(1)}$. 
Requiring the action \eqref{GA_chiral} to be stationary under such re-scalings  of $X$ 
(compare with \cite{KMcAT-M}) gives the nonlinear equation 
\bea
X \bar \D \bar X
= f \F^3 X~.
\label{Nonlinear_chiral}
\eea
This nonlinear constraint proves to express the  auxiliary field $F$
in terms of the Goldstini $\j_\a$ and $\bar \j_\a$ and their derivatives, 
see Appendix \ref{AppendixB}.

The constraints  \eqref{chiral}, \eqref{NC_chiral} and \eqref{Nonlinear_chiral} define 
an irreducible Goldstino superfield $\cX$, 
\bea
\bar \cD_\a \cX =0~, \qquad 
\d_\s \cX=\hf  \s \cX~,  \qquad
\cX^2=0~, \qquad \cX \bar \D \bar \cX
= f \F^3 \cX~.
\eea
It is the 3D $\cN=2$ analogue
of the 4D $\cN=1$ Goldstino superfield used by Lindstr\"om and Ro\v{c}ek \cite{LR}
in their off-shell model for  spontaneously broken $\cN=1$ local 
supersymmetry.\footnote{Ref. \cite{LR} is the first work on off-shell de Sitter supergravity in four dimensions. 
Terminology ``de Sitter supergravity'' was introduced  by 
Bergshoeff {\it et al.} \cite{BFKVP}. The only difference between the supergravity models put forward in \cite{BFKVP} and \cite{LR} is that they made use of different Goldstino superfields -- the 4D $\cN=1$ analogues of $X$ and $\cX$, respectively. 
The two supergravity models are equivalent on-shell \cite{BHKMS}.
However, the power of the approach advocated in \cite{BFKVP} is that the nilpotency 
condition $X^2=0$ is model independent, which implies that 
the Goldstino superfield can be readily coupled to matter multiplets.}  
The corresponding action can be given in two different but equivalent forms
\bea
S_\cX = -\int \rd^3 x \rd^2 \q  \rd^2 \bar{\q} \, E\,\bar \cX \cX 
= -  f  \int \rd^3 x \rd^2 \q \, \cE\,  
	\F^3 \cX ~.
\label{5300}
\eea

So far we have considered the coupling of the nilpotent Goldstino superfield
$X$ to the minimal (1,1) AdS supergravity. Its coupling to the two-form (or complex 
two-form) supergravity is obtained simply by replacing the chiral compensator 
$\F$ in \eqref{GA_chiral} with $\P^{1/4}$ ($\U^{1/4}$ in the case of 
complex three-form supergravity). However, there is a different universal 
approach to couple a nilpotent chiral supermultiplet to any off-shell supergravity. 
It consists in replacing $X$, defined by \eqref{chiral} and \eqref{NC_chiral},
with a super-Weyl primary scalar $\frak X$ with the properties
\bea
\bar \cD_\a {\frak X} =0~, \qquad \d_\s  {\frak X} = 2\s  {\frak X} ~,
\qquad  {\frak X}^2 =0~.
\eea
The action  \eqref{GA_chiral}  has to be replaced with 
\bea
S_{\frak X} = \int \rd^3 x \rd^2 \q  \rd^2 \bar{\q} \, E\, \frac{\bar {\frak X}{\frak X}}{ W}
-  \Big\{f  \int \rd^3 x \rd^2 \q \, \cE\, {\frak X } + {\rm c.c.} \Big\}~,
\eea
where ${W}$ is a real scalar primary superfield of weight $+3$
such that (i) it is nowhere vanishing;  and 
(ii) it is a composite of the supergravity compensators.
In particular, $W = (\bar \F \F)^3$ in the case of minimal (1,1) AdS supergravity,
$W = L^3$ for (2,0) AdS supergravity,
$W=(\bar \P \P)^{3/4}$ for the two-form supergravity,  and so on.


\subsubsection{Nilpotent real scalar  superfield}\label{section5.2.2}

We now introduce a 3D $\cN=2$ analogue of the reducible Goldstino 
superfield proposed in \cite{KMcAT-M}. It is a real scalar superfield 
subject to the nilpotency conditions:
\begin{subequations} \label{NC_real}
\bea
V^2&=&0~, \\
V \cD_A \cD_B V &=&0~,\\
V \cD_A \cD_B \cD_C V &=&0~. 
\eea
\end{subequations}
The super-Weyl transformation of $V$ is postulated to be 
\bea
\d_\s V = \s V~.
\eea
We also require that the descendant $\hf \{\D, \bar \D\} V$ is nowhere vanishing.
The nilpotency conditions \eqref{NC_real} imply that $V$ has three independent 
component fields (see Appendix \ref{AppendixB} for more details) that may be chose as follows: the complex Goldstino 
$\c_\a (x) = \frac{1}{\sqrt{2}} \cD_\a \bar \D V|$, its conjugate $\bar \c_\a (x)$
and a real auxiliary field $D(x) = \hf \{\D, \bar \D\} V|$, 
with $D^{-1}$ being well defined.

The constraints \eqref{NC_real} imposed on $V$
 do not make use of any supergravity compensator, 
which means that $V$  is defined in any  conformal supergravity background. 
However, a compensator is required in order to formulate an action functional 
for the  Goldstino superfield. As in the previous section,
here we again choose the chiral compensator $\F$ corresponding to the 
minimal (1,1) AdS supergravity (minimal type I supergravity) described in section \ref{section3.3.1}.
The dynamics of the nilpotent superfield $V$ is described by the action
\bea
S_V= \int \rd^3 x \rd^2 \q  \rd^2 \bar{\q} \, E\,\Big\{
\frac{ | \D V |^2}{(\bar \F \F)^3}
- 2f  
V \Big\}~,
\label{GA_real}
\eea
with $f$ the supersymmetry breaking parameter.\footnote{Had 
we chosen $V$ to be an {\it unconstrained} real scalar superfield, 
the action \eqref{GA_real} would have described the dynamics of a two-form multiplet 
(with a linear superpotential) coupled to the minimal type I supergravity.} 

The constraints \eqref{NC_real} are 
preserved if $V$ is locally rescaled, 
\bea
V ~\to ~\re^{\r} V~, 
\eea
for any real scalar $\r$. 
Requiring the action \eqref{GA_real} to be stationary under such re-scalings  of $V$ 
gives the nonlinear equation 
\begin{subequations}\label{Nonlinear_real}
\bea
\hf V \Big\{ \F^{-3} \bar \D , \bar \F^{-3} \D\Big\} V
= f V~.
\eea
Due  to the constraints \eqref{NC_real}, this may equivalently be rewritten as 
\bea
V  \F^{-3} \bar \D ( \bar \F^{-3} \D V) 
= V \bar \F^{-3} \D ( \F^{-3} \bar \D V) = fV ~.
\eea
\end{subequations}
This nonlinear constraint proves to express the auxiliary field $D$ in terms of 
the Goldstini.

The constraints  \eqref{NC_real} and \eqref{Nonlinear_real} define an irreducible 
Goldstino superfield $\cV$. It is a 3D $\cN=2$ counterpart of the Goldstino superfield 
 introduced in \cite{BHKMS}.  
 The corresponding action can be written in two equivalent forms
 \bea
S_\cV=- \int \rd^3 x \rd^2 \q  \rd^2 \bar{\q} \, E\,
\frac{ | \D \cV |^2}{(\bar \F \F)^3}
= - f  \int \rd^3 x \rd^2 \q  \rd^2 \bar{\q} \, E\,\cV ~.
\label{5388}
\eea
The Goldstino models \eqref{GA_real} and \eqref{5388} are equivalent on the mass shell.

The irreducible Goldstino superfields $\cX$ and $\cV$ are related to each other as
follows
\begin{subequations} 
\label{5399}
\bea
f \cV &=& \bar \cX \cX~, \label{5399a} \\
\cX &=& \F^{-3} \bar \D \cV~. \label{5399b}
\eea
\end{subequations}
These relations are analogous to those given in \cite{LR} 
in the 4D case.


\subsubsection{Relating $X$ and $V$}

Starting from the nilpotent chiral superfield $X$ described in section \ref{section5.2.1},
we  define 
\bea
f V := \bar X X~,
\label{XtoV}
\eea
as a generalisation of \eqref{5399a}.
The superfield $V$ introduced satisfies all the requirements imposed 
on the nilpotent Goldstino superfield $V$
in section \ref{section5.2.2}. One of the two auxiliary fields of $X$ does not contribute 
to the right-hand side of \eqref{XtoV}.

Implementing the field redefinition \eqref{XtoV} in the Goldstino superfield action \eqref{GA_real} leads to the following higher-derivative action
\bea
S_{\rm HD}[X, \bar X]= \int \rd^3 x \rd^2 \q  \rd^2 \bar{\q} \, E\,\Big\{
\frac{1}{ f^2} 
\frac{ |X \D X |^2}
{(\bar \F \F)^3}
- 2  \bar X X \Big\}~.
\label{5.41}
\eea
Its important property is 
\bea
S_{\rm HD}[\cX, \bar \cX]= S_\cX~,
\eea
with $S_\cX$ given by \eqref{5300}.
Unlike the Goldstino action \eqref{GA_chiral}, 
 \eqref{5.41} is invariant under the discrete transformation 
$X \to -X$. 
The model \eqref{5.41} will be studied in more detail in Appendix \ref{AppendixC}.


\subsubsection{Nilpotent two-form Goldstino superfield}

As a generalisation of the 4D $\cN=1$ models proposed in  \cite{FKRR,BK17},  
we introduce a nilpotent two-form Goldstino multiplet. It is described by  
a chiral scalar superfield
\begin{subequations} \label{5.43}
\bea
Y = -\frac{1}{4} (\bar \cD^2 -4R) U ~,\qquad \bar U =U~,
\label{5.43a}
\eea
which is constrained to be nilpotent,
\bea
Y^2=0~.
\label{5.43b}
\eea 
\end{subequations}
The prepotential $U$ in  \eqref{5.43a} 
is defined modulo gauge transformations of the form
\bea
\d_L U = L~, 
 \qquad 
\bar \D L
=0~, \qquad \bar L =L~,
\eea
and $Y$ and $\bar Y$ are gauge-invariant field strengths.

The super-Weyl transformation of the prepotential $U$ is
\bea
\d_\s U = \s U~,
\eea
which implies
\bea
\d_\s Y = 2 \s Y~.
\eea
To describe dynamics of the nilpotent two-form multiplet, we propose the action
\bea
S_{Y} = \int \rd^3 x \rd^2 \q  \rd^2 \bar{\q} \, E\, \frac{\bar Y Y}{ (\bar \F \F)^3}
-  \Big\{f  \int \rd^3 x \rd^2 \q \, \cE\, {Y } + {\rm c.c.} \Big\}~.
\label{5.4777}
\eea
The component structure of this model will be discussed in Appendix B.4. 
Here we would like just to point out that the Goldstino superfield $Y$ contains two independent
auxiliary fields, $F= H+\ri G$, 
of which $H$  is a scalar and $G$ is the divergence of a vector. 
In supergravity, both $H$ and $G$ produce positive contributions to the cosmological constant.  
While the contribution from $H$ is {\it universal} and
uniquely determined by the parameter of the supersymmetry breaking $f$ in \eqref{5.4777}, 
the contribution from $G$ is {\it dynamical}. 
We believe that the latter may be used to neutralise the negative contribution from the supersymmetric cosmological term.


\subsection{Irreducible $\cN=2$ Goldstino superfields}

Using the nilpotent chiral superfield $X$ described in section \ref{section5.2.1},
we introduce a composite superfield 
\bea
\S = 
f  \frac{\bar X}{\bar \D \bar X} ~.
\eea
It has the following transformation properties
\bea
\d_\s \S= -\s\S~,\qquad
\cJ\S=2\S~,
\eea
as well as it  identically satisfies the improved linear constraint 
\begin{subequations} \label{5.47}
\bea
\bar \D \S =f~,
\eea
compare with \eqref{322}. By construction, it is nilpotent, 
\bea
\S^2 =0~.
\eea
It also obeys the nonlinear constraint 
\bea
f \cD_\a \S =  
-\frac{1}{4}\S(\bar \cD^2 - 4 R)\cD_\a\S~,
\eea
\end{subequations}
which is equivalent to 
\bea
f\cD_\a\S = - \ri\S\cD_{\a\b}\cDB^\b {\S}
~.
\eea
Thus $\S$ is a 3D $\cN=2$ counterpart of the irreducible Goldstino superfield introduced in \cite{KTyler}. Unlike other irreducible Goldstino superfields, 
such as $\cX$ and $\cV$, the constraints obeyed by $\S$, eq. \eqref{5.47}, 
do not make use of any supergravity compensator. 
In other words, $\S$ couples to conformal supergravity, 
and this feature makes $\S$ pretty unique in the family of 
irreducible Goldstino superfields.

The remarkable feature of $\S$ and its conjugate is that these superfields are invariant
under local re-scalings of $X$ and its conjugate, eq. \eqref{re-scaling_chiral}, 
\bea
\d_\t X = \t X \quad \Longrightarrow \quad \d_\t \S =0~,  \qquad \bar \cD_\a \t=0~,
\eea
compare with \cite{BK17}.
In complete analogy with the 4D $\cN=1$ case \cite{KTyler},
 every irreducible Goldstino superfield is a descendant of $\S$ and $\bar \S$,
 for instance 
 \bea
 f \cV = (\bar \F \F)^3 \bar \S \S~.
\label{5.53}
 \eea
 Therefore  we conclude that 
all irreducible Goldstino superfields are invariant under local re-scalings  
\eqref{re-scaling_chiral}.

As pointed out above, the Goldstino superfields $\S$ and $\bar \S$ 
couple to conformal supergravity.
Relation \eqref{5.53} clearly shows that the conformal compensators have to be used
in order to define $\cV$ as a composite superfield constructed from 
$\S$ and $\bar \S$.


\section{Concluding comments} 

The results obtained in this work may lead to several interesting developments including the following: 

\begin{itemize}

\item The work by Ovrut and Waldram \cite{OvrutWaldram} provided 
membrane solutions in the 4D $\cN=1$ three-form supergravity. 
 In a similar way,  the two-form supergravity theories described in the present 
 paper should possess string solutions. It is of interest to derive such solutions 
 explicitly.  

\item 
In three dimensions,  consistent models for massive supergravity can be constructed
by adding certain higher-derivative terms to the standard supergravity action. 
These include $\cN=1$ and $\cN=2$ topologically massive 
\cite{DK,Deser,KLRST-M13}
and new massive 
\cite{Andringa:2009yc,BHRST10,BOS14,KNT-M15} supergravity theories. 
Coupling these theories to the Goldstino superfields described in section
\ref{section5} should give consistent models for spontaneously broken massive 
supergravity.

\item It is of interest to construct $\cN=3$ and $\cN=4$
Goldstino superfields, as an extension of the 4D results 
given in \cite{KMcAT-M,KT-M17}. The $\cN=3$ case is especially interesting 
since it has no 4D analogue.

\item Since we formulated the 3D Green-Schwarz superstring action, with 
$\cN=1$ and $\cN=2$  spacetime supersymmetry, in off-shell supergravity 
backgrounds, the quantum superstring analysis given in \cite{MT1,MT2} may be extended 
from the Minkowski superspace to other maximally supersymmetric backgrounds
including the AdS one. 

\end{itemize}

\noindent
{\bf Acknowledgements:}\\
E.I.B. would like to thank the Centre for Theoretical Physics at Tomsk State Pedagogical University
where some of this work was done for warm hospitality.
The work of E.I.B. was  supported  by the ARC Future Fellowship FT120100466.
The work of JH is supported by an Australian Government Research Training Program (RTP) Scholarship.
The work of SMK is supported in part by the Australian 
Research Council, project No. DP160103633.
The work of GT-M was supported by
the Interuniversity Attraction Poles Programme initiated by the Belgian Science Policy (P7/37) 
and the C16/16/005 grant of the KULeuven.


\appendix


\section{Component structure of $\cN=1$ Goldstino models}
\label{AppendixA}


In this appendix we will discuss the component actions 
for the $\cN=1$ Goldstino models introduced in section \ref{section5.1}.
For simplicity we will perform our analysis in flat superspace.

Here we specialise the superspace $\cM^{3|2}$ of section \ref{section2.1}
to be the standard $\cN=1$ Minkowski superspace ${\mathbb M}^{3|2}$
parameterised by  Cartesian real coordinates $z^A= (x^a, \theta^{\alpha})$. 
The covariant derivatives $\cD_{A}= (\cD_a, \cD_\a )$  on $\cM^{3|2}$, 
defined by eq. \eqref{23cd}, become the flat-superspace ones
\be 
D_A =(\pa_a , D_\a)~, \qquad 
D_{\alpha}= \partial_{\alpha}+ \ri (\gamma^a)_{\a \b} \theta^{\b} \partial_a =  \partial_{\alpha}+ \ri  \theta^{\b} \partial_{\a \b}~.
\label{B1}
\ee
Making use of the anti-commutation relation 
\bea
\{ D_{\a}, D_{\b} \}= 2 \ri \pa_{\a \b}
\eea
allows us to obtain a number of useful properties including the following:
\bea
 D_{\a } D_{\b} =\ri \pa_{\a \b} +\frac{1}{2} \ve_{\a \b} D^2~,\qquad
D^\a D_\b D_\a =0
~, \qquad D^2 D^2 =-4 \Box\,.
\label{B2}
\eea
We recall that  $D^2= D^{\a} D_{\a}$. Given a supersymmetric action 
\bea
S= \ri \int \rd^3 x \rd^2 \theta \, \cL ~, \qquad \bar \cL =\cL~,
\eea
with some superfield Lagrangian $\cL$, the component action is computed by the rule
\bea
S= -\frac{\ri}{4} \int \rd^3 x \, D^2 \cL|~.
\eea
As usual,  the bar-projection is defined by $U| := U|_{\q=0}$, 
for any superfield $U(x,\q)$.

Let us now consider a real scalar superfield $X$.  We define its {\it real} component fields $\f (x)$, $\j_\a (x)$ and $F(x)$ as
\be 
\phi= X|\,, \qquad  \ri \psi_{\alpha}= D_{\alpha} X|\,, \qquad \ri F = -\frac{1}{4} D^2 X|
~.
\label{B3}
\ee
Introducing a free supersymmetric model with action
\be 
S_X= \ri \int \rd^3 x \rd^2 \theta \,\Big\{\frac{\ri}{2} D^{\a} X D_{\a} X + 2 f' X\Big\}\,,
\label{B4}
\ee
at the component level we obtain 
\be 
S_X= -\int \rd^3 x \, \Big\{\hf \pa^a \f \pa_a \f + 
\frac{\ri}{2} \psi^{\a} \pa_{\a \b} \psi^{\b}
  -2 F^2 +2 f' F
\Big\}\,.
\label{freeaction}
\ee

Let us turn to the component analysis of the $\cN=1$ Goldstino model \eqref{5.2}
in flat superspace.
To describe a reducible Goldstino multiplet, we subject $X$ to the nilpotency condition 
\be
X^2=0 
\label{B4.1}
\ee
and assume that $F^{-1}$ is well defined. The nilpotency constraint 
allows us to solve for $\phi$ in terms of the Goldstino $\psi$ and the auxiliary field $F$:
\be 
\phi=\frac{\ri \psi^2}{4 F}\,. 
\label{B5}
\ee
With the constraint \eqref{B4.1} imposed, the supersymmetric action \eqref{B4} defines a nonlinear interacting theory. Making use of \eqref{freeaction} and \eqref{B5}
leads to the following action:
\be 
S_X= -\int \rd^3 x \, \Big\{\frac{\ri}{2} \psi^{\a} \pa_{\a \b} \psi^{\b}+
\frac{1}{32} \frac{\psi^2}{F} \Box \frac{\psi^2}{F}   -2 F^2 +2 f' F
\Big\}\,.
\label{B6}
\ee
The equation of motion for $F$ is 
\be 
\frac{\d S_X}{\d F}= 4 F+ \frac{1}{16} \frac{\psi^2}{F^2} \Box \frac{\psi^2}{F} - 2f'=0\,.
\label{B10}
\ee
This equation can be solved by repeated substitution which gives
\be 
F = \frac{f'}{2} -\frac{1}{8 f'^3} \psi^2 \Box \psi^2\,. 
\label{B11}
\ee
Substituting it back into~\eqref{B6} gives the following action for the Goldstino
\bea 
\widetilde{S}_X = -\int \rd^3 x \, \Big\{
\frac{f'^2}{2} +\frac{\ri}{2} \psi^{\a} \pa_{\a \b} \psi^{\b} +\frac{1}{8 f'^2} \psi^2 \Box \psi^2
\Big\}\,. 
\label{B12}
\eea
Since the auxiliary field possesses a non-vanishing expectation value, 
$\langle F \rangle = \hf f'$, the supersymmetry is spontaneously broken.
The constant term in the integrand \eqref{B12} generates a positive contribution 
to the cosmological constant in supergravity.

Our next goal is to study the component structure of the irreducible 
Goldstino model $S_\cX$, eq.  \eqref{5.6}, in Minkowski superspace. 
We recall that it is obtained from the reducible Goldstino theory defined by eqs. \eqref{B4} and 
\eqref{B4.1} by requiring 
the action~\eqref{B4} to be stationary under 
local re-scalings $X \to \re^{\rho} X$. 
This gives the constraint
\be 
\frac{\ri}{2} X D^2 X = f' X\, ,
\label{B7}
\ee
which allows one to solve for the auxiliary field $F$ in terms of $\psi$. 
Evaluating the top component of 
\eqref{B7} 
gives
\be 
F - \frac{f'}{2}-  \frac{\ri}{4 F} \psi^{\alpha} \pa_{\a \b} \psi^{\b} 
-\frac{1}{64} \frac{\psi^2}{F^2} \Box \frac{\psi^2}{F} =0\,. 
\label{B13.0}
\ee
This equation can be solved by repeated substitution to result with 
\be 
F = \frac{f'}{2} + \frac{\ri}{2 f'} \psi^{\alpha} \pa_{\a \b} \psi^{\b} -\frac{1}{4 f'^3} \psi^2 \pa_{\a \b} \psi^{\b} \pa^{\a \g} \psi_{\g}
+\frac{1}{8 f'^3} \psi^2 \Box \psi^2\,. 
\label{B13}
\ee
Plugging this into \eqref{B6} leads to the component action 
\bea
S_\cX = -\int \rd^3 x \,\Big\{
\frac{f'^2}{2} +\frac{\ri}{2} \psi^{\a} \pa_{\a \b} \psi^{\b}-\frac{1}{4 f'^2} \psi^2 \pa_{\a \b} \psi^{\b} \pa^{\a \gamma} \psi_{\gamma} +\frac{1}{8 f'^2} \psi^2 \Box \psi^2\Big\}\,.
\label{A188}
\eea

Comparing the two expressions for $F$, 
which are given by eqs. \eqref{B11} and \eqref{B13} and 
which correspond to the models $S_X$ and $S_\cX$, respectively, 
we see that they are different. The final Goldstino actions 
\eqref{B12} and \eqref{A188} also have different quartic terms.
Nevertheless, the two models are equivalent. 
Indeed, it was pointed out in section \ref{section5.1}
that the top component of \eqref{B7} 
is equivalent to a sum of the equation of motion for $F$ and a linear combination of the 
equations of motion for $\j_\a$, both equations of motion corresponding to 
the action \eqref{B6}. One can readily check that the left-hand side of 
\eqref{B13.0} can be represented as 
\bea
F - \frac{f'}{2}-  \frac{\ri}{4 F} \psi^{\alpha} \pa_{\a \b} \psi^{\b} 
-\frac{1}{64} \frac{\psi^2}{F^2} \Box \frac{\psi^2}{F} 
=
\frac 14 \Big( \frac{\d S_X}{\d F} + \frac{ \j^\a}{F} \frac{\d S_X}{\d \j^\a } \Big)~,
\label{AAA1}
\eea 
and therefore the two expressions for $F$ coincide on the mass shell. 
Moreover, it may be shown that every solution to the equation of motion 
for the Goldstino action \eqref{B12} is a solution to the equation of motion for
 \eqref{A188} and vice versa. This follows from the identity
\bea
S_\cX = \widetilde{S}_X 
+\frac{1}{4f'^2} \int \rd^3 x \, \j^2 
\ve^{\a\b} \frac{ \d \widetilde{S}_X }{\d \j^\a} 
\frac{\d \widetilde{S}_X }{\d \j^\b} ~.
\label{AAA2}
\eea



\section{Component structure of $\cN=2$  Goldstino models}
\label{AppendixB}


In this appendix we will discuss the component actions for  ${\cal
N}=2$ Goldstino models in flat superspace. 
We specialise the superspace $\cM^{3|4}$ of section \ref{geometry}
to be the standard $\cN=2$ Minkowski superspace ${\mathbb M}^{3|4}$
parameterised by  Cartesian  coordinates $z^A= (x^a, \theta^{\alpha}, \bar \q_\a)$,
with $\bar \q^\a$ being the complex conjugate of $\q^\a$. 
The covariant derivatives $\cD_{A}= (\cD_a, \cD_\a , \bar \cD^\a)$  on $\cM^{3|4}$, 
defined by eq. \eqref{CovDev}, become the flat-superspace ones
$D_{A}= (\pa_a, D_\a , \bar D^\a)$.
Here
the spinor covariant derivatives have the form 
\bea
D_{\alpha}= \partial_{\alpha} + \ri{\bar \theta}^{\beta} (\gamma^a)_{\alpha \beta}
\partial_a =\partial_{\alpha} + \ri{\bar \theta}^{\beta}
\partial_{\alpha \beta}\,, \quad
\bar D_{\alpha}= -\bar \partial_{\alpha} - \ri{ \theta}^{\beta}
\partial_{\alpha \beta}
\eea
and obey the anti-commutation relations 
\bea
\{D_{\alpha}, D_{\beta}\}=0\,, \qquad
\{\bar D_{\alpha}, \bar D_{\beta}\}=0\,, \qquad
\{ D_{\alpha}, \bar D_{\beta}\} =-2 \ri \partial_{\alpha \beta}\,.
\label{A2}
\eea

Given a supersymmetric action 
\bea
S=  \int \rd^3 x \rd^2 \q \rd^2 \bar \q  \, \cL 
+\Big\{ \int \rd^3 x \rd^2 \q\, \cL_{\rm c} +{\rm c.c.}\Big\}
~, \qquad \bar \cL =\cL~,\qquad \bar D_\a \cL_{\rm c} =0~,
\eea
with some real  $\cL$ and chiral $\cL_{\rm c}$ superfield Lagrangians, 
the component action is computed using the formula
\bea
S= \frac{1}{16} \int \rd^3 x \, D^2 \bar D^2  \cL|
-   \Big\{ \frac{1}{4}   \int \rd^3 x  \, D^2 \cL_{\rm c}| +{\rm c.c.}\Big\}~.
\eea
The contractions $D^2$ and $\bar D^2$ are defined as in \eqref{388}.


\subsection{The ${\cal N}=2$ chiral scalar Goldstino superfield}


Let us consider a model of a chiral scalar superfield $X$ satisfying
\begin{equation}
\bar D_{\alpha} X=0\,, \quad X^2=0\,.
\label{A3}
\end{equation}
This model defines a reducible Goldstino superfield model analogous to
the 4D ${\cal N}=1$ chiral model studied  in~\cite{Casalbuoni, KS}.
Hence, our analysis will be similar to those in~\cite{KS,KT0,BFKVP2}.
A general chiral superfield can be written as
\begin{equation}
X =\phi + \sqrt{2} \theta^{\alpha} \psi_{\alpha}+ \theta^2 F\,,
\label{A4.1}
\end{equation}
so that the components can be defined as
\begin{equation}
\phi= X|\,, \quad  \psi_{\alpha}= \frac{1}{\sqrt{2}} D_{\alpha} X|\,, \quad
F=-\frac{1}{4} D^2 X|\,.
\label{A4.2}
\end{equation}
The nilpotency condition $X^2=0$ gives
\begin{equation}
\phi= \frac{\psi^2}{2 F}\,, \quad \bar \phi= \frac{\bar \psi^2}{2 \bar F}\,.
\label{A5}
\end{equation}
The action for $X$ follows from
\eqref{GA_chiral} 
\begin{equation}
S_X= \int \rd^3 x \rd^2 \theta \rd^2 \bar \theta \, \bar X X - 
\Big\{ f \int \rd^3 x \rd^2 \theta \, X +{\rm c.c.}  \Big\}
\,.
\label{A6}
\end{equation}
The integral over $\theta$ and $\bar \theta$ can be performed using~\eqref{A4.2}, \eqref{A5}
to give the following component action
\begin{equation}
S= \int \rd^3 x \, \Big[ -\frac{1}{2} (\langle u \rangle + \langle \bar u \rangle)
+\frac{\bar \psi^2}{2 \bar F} \Box \frac{\psi^2}{2 F} + F \bar F
- f (F +\bar F)\Big]\,,
\label{A7}
\end{equation}
where we have defined
\begin{equation}
\langle u \rangle = \ri \psi^{\alpha} \partial_{\alpha \beta}
\bar \psi^{\beta}\,, \quad
\langle \bar u \rangle =- \ri  \partial_{\alpha \beta}
\psi^{\beta}\bar \psi^{\alpha}\,.
\label{A8}
\end{equation}
The superfield $X$ defined by~\eqref{A3} describes a reducible multiplet  containing
 the Goldstino $\psi_{\alpha}$ and an auxiliary field $F$. 

As was explained in the previous Appendix there are two approaches to define an irreducible Goldstino multiplet. 
We can eliminate $F$ and $ \bar F$ from the action~\eqref{A7}
using the equations of motion
\begin{equation}
F= f+ \frac{\bar \psi^2}{2 \bar F^2}\Box \frac{ \psi^2}{2  F}\,,
\quad
\bar F= f+ \frac{ \psi^2}{2 F^2}\Box \frac{ \bar \psi^2}{2 \bar  F}\,.
\label{A10}
\end{equation}
Solving equations~\eqref{A10} by repeated substitution yields
\begin{equation}
F = f  + \frac{1}{4} f^{-3} \bar \psi^2 \Box \psi^2
-\frac{3}{16} f^{-7} \psi^2 \bar \psi^2 (\Box \psi^2) (\Box \bar \psi^2)\,.
\label{A11}
\end{equation}
Then the Goldstino action becomes
\begin{equation}
S= -\int \rd^3 x \,\Big[f^2 +  \frac{1}{2} (\langle u \rangle + \langle \bar u \rangle)
+\frac{1}{4 f^2}(\partial^a \bar \psi^2)  (\partial_a  \psi^2)
+\frac{1}{16 f^6} \psi^2 \bar \psi^2 (\Box \psi^2)(\Box \bar \psi^2)
 \Big]\,.
\label{A11.1}
\end{equation}

Alternatively, 
we can require that the action be stationary under re-scalings $X \to e^{\tau}X$, $ {\bar D}_{\a} \tau =0$
which gives the constraint
\begin{equation}
-\frac{1}{4} X {\bar D}^2 \bar X = f X\,.
\label{A9}
\end{equation}
Eqs.~\eqref{A3}, \eqref{A9} define a Goldstino model~\eqref{Nonlinear_chiral} as was discussed in Section 5. 
From~\eqref{A9} we find the following equation for the auxiliary field
\be
F= f + \ri {\bar F}^{-1} {\bar \psi}^{\a} \pa_{\a \b} \psi^{\b} -\frac{1}{4} {\bar F}^{-2} {\bar \psi}^2 \Box ( F^{-1} \psi^2)\,. 
\label{A10.00A}
\ee
Solving it by repeated substitution we obtain
\bea
F & = & f + f^{-1} \langle \bar u \rangle - f^{-3} (  \langle  u \rangle \langle \bar u \rangle +\frac{1}{4} \bar \psi^2 \Box \psi^2) 
+ f^{-5} ( \langle  u \rangle^2 \langle \bar u \rangle + \langle \bar u \rangle^2  \langle  u \rangle) 
\nonumber \\
& + & \frac{1}{4} f^{-5} (  \langle \bar u \rangle \psi^2 \Box \bar \psi^2 +2  \langle  u \rangle \bar \psi^2 \Box \psi^2  + \bar \psi^2 \Box ( \psi^2  \langle \bar u \rangle))
\nonumber \\
& - & 3 f^{-7} (\langle  u \rangle^2  \langle \bar u \rangle^2  +\frac{1}{4} \psi^2 \bar \psi^2  
\Box (  \langle  u \rangle^2 - \langle  u \rangle \langle \bar u \rangle +  \langle \bar u \rangle^2)
+\frac{1}{16} \psi^2 \bar \psi^2  \Box \psi^2 \Box \bar \psi^2
)\,,
\label{A10.0}
\eea
where $\langle u \rangle$ is given in eq.~\eqref{A8}.
Comparing eqs.~\eqref{A11} and~\eqref{A10.0} we see that the solution for $F$ is different in our two approaches 
but the difference is related to the equation of motion for the Goldstino as was explained at the end of the previous Appendix.


\subsection{The ${\cal N}=2$ real scalar Goldstino superfield}


The  real scalar Goldstino superfield is defined 
to obey the constraints
\begin{equation}
V^2=0\,, \quad V D_{A} D_{B} V=0\,, \quad V D_{A} D_{B}D_C V=0\,.
\label{A12}
\end{equation}
We will start with a general ${\cal N}=2$ real scalar superfield
\begin{equation}
V= v + \sqrt{2} \theta^{\alpha} \lambda_{\alpha} +
\sqrt{2} \bar \theta_{\alpha} \bar \lambda^{\alpha} + \theta^2 F +\bar \theta^2 \bar F
+ \theta^{\alpha} \bar \theta^{\beta} A_{\alpha \beta}
+ \sqrt{2} \bar \theta^2 \theta^{\alpha} \varrho_{\alpha} 
+ \sqrt{2} \theta^2 \bar \theta_{\alpha} \bar \varrho^{\alpha} +\theta^2 \bar \theta^2
 {\mathbb D}\,.
 \label{A13}
\end{equation}
Here $A_{\alpha \beta}$ describes both a vector $\tilde{A}^a$ and a scalar $\varphi$:
\begin{equation}
A_{\alpha \beta}= (\gamma_a)_{\alpha \beta} \tilde{A}^a + \ri \epsilon_{\alpha \beta}
\varphi\,.
\label{A13.1}
\end{equation}
Imposing conditions~\eqref{A12} we find that $v, \lambda_{\alpha},
\bar \lambda_{\alpha}, A_{\alpha \beta}, F, \bar F$ can be solved
in terms of $\varrho_{\alpha}, \bar \varrho_{\alpha}, {\mathbb D}$ as follows
\begin{eqnarray}
&&
v= \frac{\varrho^2 \bar \varrho^2}{4 {\mathbb D}^3}\,, \quad
\lambda_{\alpha}= \frac{\varrho_{\alpha} \bar \varrho^2}{2 {\mathbb D}^2}\,,
\quad
\bar \lambda_{\alpha}= \frac{\bar \varrho_{\alpha} \varrho^2}{2 {\mathbb D}^2}\,,
\nonumber \\
&&
F= \frac{\bar \varrho^2}{2 {\mathbb D}}\,, \quad
\bar F= \frac{ \varrho^2}{2 {\mathbb D}}\,, \quad
A_{\alpha \beta} =\frac{2 \varrho_{\alpha} \bar \varrho_{\beta}}{{\mathbb D}}\,.
\label{A15}
\end{eqnarray}
Hence, we have explicitly shown that the model~\eqref{A12} describes a reducible Goldstino
multiplet $(\varrho_{\alpha}, {\mathbb D})$ consisting of the Goldstino $\varrho_{\alpha}$
and an auxiliary field ${\mathbb D}$.

Alternatively, we can define the Goldstino as follows. Let
\begin{equation}
W_{\alpha}=-\frac{1}{4} {\bar D}^2 D_{\alpha} V\,.
\label{A16}
\end{equation}
Let us define
\begin{equation}
\psi_{\alpha}= \frac{1}{\sqrt{2}} W_{\alpha}|\,, \quad
D=-\frac{1}{4} D^{\alpha} W_{\alpha}|\,.
\label{A17}
\end{equation}
Since  $W_{\alpha}$ satisfies
\begin{equation}
D^{\alpha} W_{\alpha} =\bar D_{\alpha} \bar W^{\alpha}
\label{A18}
\end{equation}
we see that $D$ is real.
Using eqs.~\eqref{A13}, \eqref{A15}, \eqref{A16}, \eqref{A17}
we obtain
\begin{eqnarray}
&&
\psi_{\alpha}= \varrho_{\alpha} -\frac{\ri}{2} \partial_{\alpha \beta} \bar \lambda^{\beta}=
\varrho_{\alpha} - \frac{\ri}{4}\partial_{\alpha \beta}\Big(
\frac{\bar \varrho^{\beta}\varrho^2}{{\mathbb D}^2}\Big)\,,
\nonumber \\
&&
D = {\mathbb D}- \frac{1}{4} \Box v = {\mathbb D}- \frac{1}{16}\Box
\Big(
\frac{\varrho^2 \bar \varrho^{2}}{{\mathbb D}^3}\Big)\,.
\label{A19}
\end{eqnarray}
From here we can derive the following useful relations
\begin{equation}
\varrho^2 \bar \varrho^{\alpha}= \psi^2 \bar \psi^{\alpha}\,, \quad
\bar \varrho^2  \varrho^{\alpha}= \bar \psi^2  \psi^{\alpha}\,, \quad
\varrho^2 \bar \varrho^2 =\psi^2 \bar \psi^2\,,
\label{A20}
\end{equation}
which, in turn, allow us to invert~\eqref{A19} to get
\begin{equation}
\varrho_{\alpha}=
\psi_{\alpha} + \frac{\ri}{4}\partial_{\alpha \beta}\Big(
\frac{\bar \psi^{\beta}\psi^2}{D^2}\Big)\,, \qquad
{\mathbb D}= D+ \frac{1}{16}\Box
\Big(
\frac{\psi^2 \bar \psi^{2}}{D^3}\Big)\,.
\label{A21}
\end{equation}
Substituting~\eqref{A21} into~\eqref{A15} we obtain the components of $V$ in terms
of $(\psi_{\alpha}, D)$
\begin{eqnarray}
&&
v = \frac{\psi^2 \bar \psi^2}{4 D^3}\,, \quad
\lambda_{\alpha}= \frac{\psi_{\alpha} \bar \psi^2}{2 D^2}\,,
\quad
\bar \lambda_{\alpha}= \frac{\bar \psi_{\alpha} \psi^2}{2 D^2}\,,
\nonumber \\
&&
F = \frac{\bar \psi^2}{2 D} + \frac{\bar \psi^2}{4 D^3} \langle u \rangle    \,, \quad
\bar F= \frac{ \psi^2}{2 D} +\frac{ \psi^2}{4 D^3} \langle \bar u \rangle  \,,
\nonumber \\
&&
A_{\alpha \beta}= \frac{2\psi_{\alpha} \bar \psi_{\beta}}{D}
-\frac{\ri}{2 D^3} \psi^2 \bar \psi^{\gamma}( \partial_{\alpha \gamma} \bar \psi_{\beta})
+\frac{\ri}{2 D^3} \bar \psi^2  (\partial_{\beta \gamma}  \psi_{\alpha})
 \psi^{\gamma}
\nonumber \\
&&
 \ \ \ \ \ \   -  \frac{1}{8 D^5} \psi^2 \bar \psi^2 \partial_{\alpha \gamma}
\partial_{\beta \delta} (\psi^{\delta} \bar \psi^{\gamma}) -
\frac{1}{4 D^5} \psi^2 \bar \psi^2 \partial_a \psi_{\alpha}
\partial^a \bar \psi_{\beta}
\,.
\label{A22}
\end{eqnarray}
Either $(\varrho_{\alpha}, {\mathbb D})$ or $(\psi_{\alpha}, D)$ can be used
to describe a reducible Goldstino multiplet in this model. Relations~\eqref{A19}
and~\eqref{A21} allow one to quickly transform from one description to another.
Since the components of $V$ are simpler when written in terms
of $(\varrho_{\alpha}, {\mathbb D})$ below we will use this pair of fields.

The action for the Goldstino superfield can be taken as the flat superspace limit
of~\eqref{GA_real}
\begin{equation}
S= \frac{1}{16} \int \rd^3 x \rd^2 \theta \rd^2 \bar \theta \, D^2 V \bar D^2 V -
2f \int \rd^3 x \rd^2 \theta \rd^2 \bar \theta \,V\,.
\label{A23}
\end{equation}
Using the nilpotency conditions~\eqref{A12} the first term of the action~\eqref{A23}
can also be written as
\begin{equation}
-\frac{1}{4 }\int \rd^3 x \rd^2 \theta \rd^2 \bar \theta \,V D^{\alpha} W_{\alpha}=
\frac{1}{4} \int \rd^3 x \rd^2 \theta \,W^{\alpha} W_{\alpha}\,.
\label{A24}
\end{equation}
However, we find that eq.~\eqref{A23} is more convenient to use.
In terms of $(\varrho_{\alpha}, {\mathbb D})$ the action~\eqref{A23} reads
\begin{eqnarray}
S  =  \int \rd^3 x \, \Big[& {\mathbb D}^2 & -2 f {\mathbb D} -\frac{\ri}{2} \varrho^{\alpha}
(\partial_{\alpha \beta} \bar \varrho^{\beta}) +
\frac{\ri}{2}
(\partial_{\alpha \beta} \varrho^{\beta})\bar \varrho^{\alpha}
 -  \frac{\varrho^{\alpha}\bar \varrho^{ \beta}}{4 {\mathbb D}} \partial_{\alpha \beta}
\partial_{\gamma \delta} \Big(\frac{\varrho^{\gamma} \bar \varrho^{\delta}}{{\mathbb D}}\Big)
\nonumber \\
& + & \frac{1}{8} (\Box {\mathbb D} )\frac{\varrho^2 \bar \varrho^2}{{\mathbb D}^3}
-\frac{1}{4} \varrho^{\alpha} \Box \Big( \frac{\varrho_{\alpha} \bar \varrho^2}{{\mathbb D}^2}\Big)
-\frac{1}{4} \bar \varrho_{\alpha} \Box \Big( \frac{\bar \varrho^{\alpha} \varrho^2}{{\mathbb D}^2} \Big)
+\frac{1}{4 {\mathbb D}} \varrho^2 \Box \Big( \frac{\bar \varrho^2}{{\mathbb D}}\Big)
\nonumber \\
& + & \frac{\ri}{16} \partial_{\alpha \beta} \Big( \frac{\bar \varrho^{\beta} \varrho^2}{{\mathbb D}^2}
\Big) \Box ( \frac{ \varrho^{\alpha} \bar \varrho^2}{{\mathbb D}^2}\Big)
+ \frac{1}{256 {\mathbb D}^6} \varrho^2 \bar \varrho^2 \Box^2 (\varrho^2 \bar \varrho^2)\Big]~.
\label{A25}
\end{eqnarray}
Again, there are two approaches to define an irreducible Goldstino multiplet. We can
eliminate ${\mathbb D}$ using its equation of motion:
\begin{eqnarray}
{\mathbb D} & = & f -\frac{1}{4} \frac{\varrho^{\alpha} \bar \varrho^{\beta}}{{\mathbb D}^2}
\partial_{\alpha \beta} \partial_{\gamma \delta}
\Big(\frac{\varrho^{\gamma} \bar \varrho^{\delta}}{{\mathbb D}}\Big) -
\frac{1}{16} \Box \Big( \frac{\varrho^2 \bar \varrho^2}{{\mathbb D}^3}\Big)
+\frac{3}{16} (\Box {\mathbb D}) \frac{\varrho^2 \bar \varrho^2}{{\mathbb D}^4}
\nonumber \\
&-& \frac{1}{4}\frac{\varrho^{\alpha} \bar \varrho^2}{{\mathbb D}^3} (\Box \varrho_{\alpha})
-\frac{1}{4}\frac{\bar \varrho_{\alpha}  \varrho^2}{{\mathbb D}^3} (\Box \bar \varrho^{\alpha})
+\frac{1}{8 {\mathbb D}^2} \varrho^2 \Box \Big( \frac{\bar \varrho^2}{{\mathbb D}}\Big) +
\frac{1}{8 {\mathbb D}^2} \bar \varrho^2 \Box \Big( \frac{ \varrho^2}{{\mathbb D}}\Big)
\nonumber \\
& - &
\frac{\ri}{16} \frac{\varrho^{\alpha} \bar \varrho^2}{{\mathbb D}^3}
\partial_{\alpha \beta}\Box \Big( \frac{\bar \varrho^{\beta} \varrho^2}{{\mathbb D}^2}\Big)
-
\frac{\ri}{16} \frac{\bar \varrho^{\alpha}  \varrho^2}{{\mathbb D}^3}
\partial_{\alpha \beta}\Box \Big( \frac{ \varrho^{\beta} \bar \varrho^2}{{\mathbb D}^2}\Big)
+
\frac{3}{256 {\mathbb D}^7} \varrho^2 \bar \varrho^2 \Box^2 (\varrho^2 \bar \varrho^2)
~,
\label{A26}
\end{eqnarray}
which can be solved by repeated substitutions. 
The second approach is to require that the action~\eqref{A23} be stationary under local re-scalings $V \to e^{\rho} V$
which yields the constraint
\be 
\frac{1}{32} V \{ D^2, \bar D^2\} V = f V\,,
\label{A26.0}
\ee
as was discussed in Section 5. 
Here for simplicity we will follow the first approach and eliminate ${\mathbb D}$ using the equation of motion~\eqref{A26}. 
From eq.~\eqref{A26} we see that,
the solution for ${\mathbb D}$
has to be of the following form
\begin{equation}
{\mathbb D}= f + \varrho^{\alpha} \bar \varrho^{\beta} {\cal A}_{\alpha \beta} +
\varrho^2 {\cal B} +\bar \varrho^2 {\bar {\cal B}}
+ \bar \varrho^2 \varrho^{\alpha} {\cal C}_{\alpha} +
 \varrho^2 \bar \varrho_{\alpha} {\bar {\cal C}}^{\alpha}+ \varrho^2 \bar \varrho^2 {\cal F}\,,
\label{A27}
\end{equation}
where ${\cal A}, \ {\cal B},\  {\cal C}, \ {\cal F}$ depend on $\varrho$ only through derivatives.
Note that in eq.~\eqref{A27} there are no  terms linear in Goldstino.
Examining eqs.~\eqref{A25}, \eqref{A27}  one can show that the last three terms in~\eqref{A27}
do not contribute to the action
and, hence, can be ignored. Keeping this in mind, 
we obtain
\begin{equation}
{\mathbb D}= f-\frac{1}{4 f^3} \varrho^{\alpha} \bar \varrho^{\beta}
\partial_{\alpha \beta} \partial_{\gamma \delta} (\varrho^{\gamma} \bar \varrho^{\delta})
-
\frac{1}{16 f^3}  \Box(\varrho^2 \bar \varrho^2 )
+\frac{1}{8 f^3} \varrho^2 \Box (\bar \varrho^2 ) +
\frac{1}{8 f^3} \bar \varrho^2 \Box ( \varrho^2) + \dots\,,
\label{A28}
\end{equation}
where the ellipsis stands for the terms which do not contribute to the action.
Substituting eq.~\eqref{A28} into~\eqref{A25} we find the following action for the Goldstino
\begin{eqnarray}
S  &=&  -\int \rd^3 x  \,\Big[  f^2  +\frac{1}{2}( \langle w \rangle +\langle \bar w \rangle)
+\frac{1}{4 f^2} \varrho^{\alpha} \Box (\varrho_{\alpha} \bar \varrho^2)
+\frac{1}{4 f^2} \bar \varrho_{\alpha} \Box (\bar \varrho^{\alpha}  \varrho^2)
-\frac{1}{4 f^2} \varrho^2 \Box (\bar \varrho^2)
\nonumber \\
&   & + 
\frac{1}{4 f^2} (\langle w \rangle - \langle {\bar w} \rangle)^2
+\frac{\ri}{16 f^4} \varrho^{\alpha} \bar \varrho^2 \partial_{\alpha \beta} \Box (\bar \varrho^{\beta}\varrho^2)
\nonumber \\
&&+\frac{1}{64 f^6} \varrho^2 \bar \varrho^2 \Box (\langle w \rangle - \langle {\bar w} \rangle)^2
+\frac{1}{32 f^6} \varrho^{\alpha} \bar \varrho^{\beta} \partial_{\alpha \beta}
\partial_{\gamma \delta} (\varrho^{\gamma} \bar \varrho^{\delta}) \Box (\varrho^2 \bar \varrho^2)\Big]\,,~~~
\label{A29}
\end{eqnarray}
where $\langle w \rangle = \ri \varrho^{\alpha} \partial_{\alpha \beta} \bar \varrho^{\beta}$.


\subsection{From $V$ to equivalent two-form multiplet}


There is another possibility to study the model from the previous subsection.
For this we will introduce
\begin{equation}
\Psi= -\frac{1}{4} \bar D^2 V\,, \quad \bar \Psi= -\frac{1}{4} D^2 V\,.
\label{A3.1}
\end{equation}
The action in eq.~\eqref{A23} can be equivalently written as
\begin{equation}
S= \int \rd^3 x \rd^2 \theta \rd^2 \bar \theta \,\bar \Psi \Psi -
f \int \rd^3 x \rd^2 \theta \, \Psi -f \int \rd^3 x \rd^2 \bar \theta  \,\bar \Psi \,.
\label{A3.2}
\end{equation}
Since $\Psi$ is chiral we can define its components as
\begin{equation}
\phi= \Psi|\,, \quad  \chi_{\alpha}= \frac{1}{\sqrt{2}} D_{\alpha} \Psi|\,, \quad
F_1+ \ri F_2=-\frac{1}{4} D^2 \Psi|\,.
\label{A3.3}
\end{equation}
From eq.~\eqref{A12} it follows that $\Psi^2=0$, hence,
\begin{equation}
\phi= \frac{\chi^2}{2 (F_1+ \ri F_2)}\,.
\label{A3.4}
\end{equation}
Note that in addition to the Goldstino $\chi_{\alpha}$, $\Psi$ contains two auxiliary
fields $F_1$ and $F_2$. However, as we will see below $F_2$ is a function of the Goldstino and $F_1$.
Therefore, it is the pair $(\chi_{\alpha}, F_1)$ which describes a reducible Goldstino multiplet
which, of course, is equivalent to the ones studied in the previous subsection up to a non-linear transformation which we will derive below.
To express $F_2$ in terms of $\chi_{\alpha}$ and $ F_1$ we note that
\begin{equation}
D^2 \Psi - \bar D^2 \bar \Psi= \ri \partial^{\alpha \beta} [ D_{\alpha}, \bar D_{\beta}]V\,.
\label{A3.5}
\end{equation}
Using the fact that
\begin{equation}
A_{\alpha \beta}=\frac{1}{2}[ D_{\alpha}, \bar D_{\beta}]V|
\label{A3.6}
\end{equation}
which follows from~\eqref{A13} we find that
\begin{equation}
F_2= -\frac{1}{4} \partial^{\alpha \beta} A_{\alpha \beta} =
-\frac{1}{2}\partial^{\alpha \beta}\Big ( \frac{\varrho_{\alpha}
\bar\varrho_{\beta}}{{\mathbb D}}\Big)\,.
\label{A3.7}
\end{equation}
Hence, we see that $F_2$ is expressed in terms of the Goldstino and the remaining
auxiliary field. The relation between $(\varrho_{\alpha}, {\mathbb D})$
and $(\chi_{\alpha}, F_1)$ can be obtained using the defining equation~\eqref{A3.1}
as well as the definition of the component~\eqref{A13}, \eqref{A3.3}. We get
\begin{equation}
 \chi_{\alpha}= \varrho_{\alpha}+
\frac{\ri}{4}
\partial_{\alpha \beta}\Big(
\frac{\bar \varrho^{\beta}\varrho^2}{{\mathbb D}^2}\Big)\,,
\qquad
F_1= {\mathbb D}+  \frac{1}{16}\Box
\Big(
\frac{\varrho^2 \bar \varrho^{2}}{{\mathbb D}^3}\Big)\,.
\label{A3.8}
\end{equation}
Using the identities
\begin{equation}
\varrho^2 \bar \varrho^{\alpha}= \chi^2 \bar \chi^{\alpha}\,, \quad
\bar \varrho^2  \varrho^{\alpha}=  \bar \chi^2  \chi^{\alpha}\,, \quad
\varrho^2 \bar \varrho^2 = \chi^2 \bar \chi^2\,,
\label{A3.9}
\end{equation}
we can invert~\eqref{A3.8}:
\begin{equation}
\varrho_{\alpha}=
 \chi_{\alpha} - \frac{\ri }{4}\partial_{\alpha \beta}\Big(
\frac{\bar \chi^{\beta}\chi^2}{F_1^2}\Big)\,, \qquad
{\mathbb D}= F_1- \frac{1}{16}\Box
\Big(
\frac{\chi^2 \bar \chi^{2}}{F_1^3}\Big)\,.
\label{A3.10}
\end{equation}
Using the relations \eqref{A3.7} and  \eqref{A3.10}, 
we can express $F_2$ in terms of the fields
$\chi_{\alpha}$ and $ F_1$. The result is 
\begin{eqnarray}
F_2 & = & -\frac{1}{8}\partial^{\alpha \beta}\Big [  \frac{1}{F_1} 
\Big\{ 4\chi_{\alpha} \bar \chi_{\beta}
+ {\ri}  \chi_{\alpha} \partial_{\beta \gamma} \Big(\frac{\chi^{\gamma} \bar \chi^2}{F_1^2}\Big)
+ {\ri} \bar \chi_{\alpha} \partial_{\beta \gamma} \Big(\frac{\bar \chi^{\gamma} \chi^2}{F_1^2}\Big) \Big\}
\nonumber \\
&& \qquad \quad + \frac{1}{ F_1^5} \chi^2 \bar \chi^2 \Big(\partial_{\alpha \gamma} \partial_{\beta \delta} (\bar \chi^{\gamma} \chi^{\delta})
- \frac{1}{2} \partial^a \chi_{\alpha} \partial_a \bar \chi_{\beta} \Big) \Big].
\label{B.4777}
\end{eqnarray}

Since the action~\eqref{A3.2} is the same as the action for a chiral superfield $X$ in~\eqref{A6}
it has the identical component structure:
\begin{equation}
S= \int \rd^3 x\, \Big[ -\frac{1}{2} (\langle v \rangle + \langle \bar v \rangle)
+\frac{\bar \chi^2}{2 ( F_1-\ri F_2)} \Box \frac{\chi^2}{2 (F_1+\ri F_2)} + F_1^2
+ F_2^2
- 2 f F_1 \Big]\,,
\label{A3.11}
\end{equation}
where $\langle v \rangle= \ri \chi^{\alpha} \partial_{\alpha \beta} \bar \chi^{\beta}$
and $F_2$ is given by~\eqref{B.4777}.
We will not present the final action in terms of $\chi_{\alpha} $ since it is
substantially more complicated than the one in eq.~\eqref{A29}. 

Out of the
three possible Goldstino
fields $\varrho, \  \psi$ and $\chi$ it is $\varrho$ which has the simplest action.


\subsection{Nilpotent two-form Goldstino superfield}


Here we will discuss the component structure of the model introduced in Subsection 5.2.4. As before, we will take the flat space limit. 
The two-form Goldstino multiplet is described by a chiral scalar superfield $Y$ satisfying the following conditions
\be 
Y=-\frac{1}{4} \bar D^2  U\,, \qquad Y^2=0\,, 
\label{BB1}
\ee
where $U$ is an {\it unconstrained} real superfield. Since $Y$ is chiral we can define its components in the usual way
\be 
\phi =Y|\,, \qquad \xi_{\a}= \frac{1}{\sqrt{2}} D_{\a} Y|\,, \qquad F=- \frac{1}{4} D^2 Y|\,. 
\label{BB2}
\ee
From~\eqref{BB1} it follows that 
\be 
D^2 Y - \bar D^2 \bar Y= {\rm i} \pa^{\a \b} [D_{\a}, \bar D_{\b}] U\,. 
\label{BB3}
\ee
This means that the imaginary part of the auxiliary field $F$ is the divergence of a vector. Let us denote 
$F= H+ \ri G$. Then we have $G= \pa_{a} C^a$, where $C^a$ is an auxiliary vector field. The action for the superfield $Y$ is 
given by the flat space limit of eq.~\eqref{5.4777}:
%
\bea
S_{Y} = \int \rd^3 x \rd^2 \q  \rd^2 \bar{\q} \, \bar Y Y 
-  \Big\{f  \int \rd^3 x \rd^2 \q \, {Y } + {\rm c.c.} \Big\}~.
\label{BB4}
\eea
Just like in the theory of three-form multiplet in four dimensions this action has to be supplemented with the boundary term~\cite{Duff, Duncan:1989ug, Groh:2012tf}
\be 
B_Y= \frac{1}{4}\int \rd^3 x \rd^2 \theta \rd^2
\bar \theta  D^{\a} (Y D_{\a} U - U D_{\a} Y) + {\rm c.c} =
  -2 \int \rd^3 x \ \pa_a (C^a G) +\dots\,,
\label{BB6}
\ee
where the ellipsis stands for the boundary terms which do not play a role and can be set to zero. 

Since the action~\eqref{BB4} is the same as the action for a chiral superfield it is given by 
\be 
S_Y=\int \rd^3 x \ \Big[ \frac{\xi^2}{2 F} \Box \frac{\bar \xi^2}{2 \bar F}+ \frac{\ri}{2}  (\pa_{\a \b} \xi^{\a}) \bar \xi^{\b} 
- \frac{\ri}{2}  \xi^{\a} (\pa_{\a \b} \bar \xi^{\b})
+ H^2 +G^2 -2 f H\Big]\,, 
\label{BB5}
\ee
where we have used the fact that $Y^2=0$ and, hence, $\phi= \xi^2/(2F)$. Now we will eliminate the auxiliary fields using their equations of motion. 
Varying the action~\eqref{BB5} with respect to $H$ and $C_a$ gives the following equations:
\bea
&&
H- f - \frac{\bar \xi^2}{4 \bar F^2} \Box \frac{\xi^2}{2 F} -\frac{ \xi^2}{4  F^2} \Box \frac{\bar \xi^2}{2 \bar F} =0\,, 
\nonumber \\
&& 
\pa_a \Big[ G +\ri \frac{\bar \xi^2}{4 \bar F^2} \Box \frac{\xi^2}{2 F} - \ri \frac{\xi^2}{4  F^2} \Box \frac{\bar \xi^2}{2 \bar F} \Big]=0\,. 
\eea
The second equation implies that 
\be 
G +\ri \Big[ \frac{\bar \xi^2}{4 \bar F^2} \Box \frac{\xi^2}{2 F} -\frac{ \xi^2}{4  F^2} \Box \frac{\bar \xi^2}{2 \bar F} \Big]=g \,, 
\label{BB7}
\ee
where $g$ is an arbitrary constant. Hence, we find that 
\be 
F= h + \frac{\bar \xi^2}{2 \bar F^2} \Box \frac{\xi^2}{2 F} \,, \qquad h = f +\ri g\,. 
\label{BB8}
\ee
Solving this equation by repeated substitution yields
\be 
F = h \Big( 1 + \frac{1}{4} |h|^{-4} \bar \xi^2 \Box \xi^2 -\frac{3}{16} |h|^{-8} \xi^2 \bar \xi^2 \Box \xi^2 \Box \bar \xi^2 \Big)\,, \qquad |h|^2 = f^2 + g^2\,. 
\label{BB9}
\ee
The boundary term on the solution $G = g + \dots$ gives $ -2 \int \rd^3 x \ (g^2  + {\rm total} \ {\rm derivative})$.
Substituting eq.~\eqref{BB9} into the bulk action~\eqref{BB5} and combining the result with the contribution from the boundary term 
yields the following Goldstino action
\bea 
S_Y + B_{Y} =  - \int \rd^3 x \ &\Big[& |h|^2 - \frac{\ri}{2}  (\pa_{\a \b} \xi^{\a}) \bar \xi^{\b} 
+  \frac{\ri}{2}  \xi^{\a} (\pa_{\a \b} \bar \xi^{\b}) 
\nonumber \\
&+ &\frac{1}{4} \frac{f^2 + 3 g^2}{|h|^4} \pa^a \xi^2 \pa_a \bar \xi^2 +
\frac{1}{16} \frac{f^2 + 7 g^2}{|h|^8} \xi^2 \bar \xi^2 \Box \xi^2 \Box \bar \xi^2 \Big]\,.
\eea
%



\section{Goldstino multiplet from a higher-derivative theory}\label{AppendixC}


In this appendix we will analyse the higher-derivative model \eqref{5.41}
in Minkowski  superspace. We first consider the case when the dynamical variable 
$X$ is an unconstrained chiral superfield, $\bar D_\a X=0$,  which obeys no 
nilpotency condition.
Then 
 the model with action 
\be 
S= \int \rd^3 x \rd^2 \theta  \rd^2 \bar \theta \,
\Big\{ \frac{1}{16 f^2} X D^2 X \bar X \bar D^2 \bar X - 2 \bar X X \Big\}
\label{C1}
\ee
has two phases, one with unbroken supersymmetry, and the other with spontaneously 
broken one. In the unbroken phase, the equations of motion have free
massless solutions
\bea
D^2 X=0~.
\eea
However, the kinetic term in \eqref{C1} has a wrong sign and thus the theory is 
ill-defined at the quantum level. We therefore turn to the phase with spontaneously 
broken supersymmetry in which $F$ develops a non-zero expectation value,
$\langle F\rangle =f$. 

Defining the components of $X$ as in eq.~\eqref{A4.2}
we obtain the component action:
\bea
S &= & 2\int \rd^3 x \,[  \pa^a \phi \pa_a \bar \phi +  \ri \psi^{\a} \pa_{\a \b} \bar \psi^{\b} -F \bar F ]
\nonumber \\
& + & \frac{1}{f^2} \int \rd^3 x \,\Big\{ F \bar F (\bar \phi \Box \phi +  \phi \Box \bar \phi ) + \phi \bar \phi (\Box \phi) (\Box \bar \phi)+ (F \bar F)^2 
- \pa_{a} (\phi \bar F) \pa^a (\bar \phi F)
\nonumber \\
&\  & \ \ \  \ \ \ \ \ \ \ \ \ \ \   \  -\frac{3 \ri}{2} F \bar F \psi^{\a} \pa_{\a \b} \bar \psi^{\b} -  \frac{3 \ri}{2} F \bar F \bar \psi^{\a} \pa_{\a \b} \psi^{\b} 
-\frac{\ri}{2} \bar F \pa_{\a \b} F \psi^{\a} \bar \psi^{\b} + \frac{\ri}{2}  F \pa_{\a \b} \bar F \psi^{\a} \bar \psi^{\b}
\nonumber \\
&\  & \ \ \  \ \ \ \ \ \ \ \ \ \ \   \ - \phi  F \bar \psi_{\a} \Box \bar \psi^{\a} - \bar  \phi \bar  F  \psi^{\a} \Box \psi_{\a}
+ F \pa^{\a \g}\phi \bar \psi_{\g} \pa_{\a \b} \bar \psi^{\b} - \bar F \pa^{\a \g}\bar \phi  \psi_{\g} \pa_{\a \b}  \psi^{\b}
\nonumber \\
&\  & \ \ \  \ \ \ \ \ \ \ \ \ \ \   \ + \frac{\ri}{2} (\bar \phi \pa^{\a \g} \phi -\phi \pa^{\a \g} \bar \phi) \pa_{\a \b} \bar \psi^{\b} \pa_{\g \d} \psi^{\d}
+ \frac{\ri}{2} \phi \bar \phi (\pa_{\a \b} \bar \psi^{\b} \Box \psi^{\a} + \pa_{\a \b}  \psi^{\b} \Box \bar\psi^{\a})
\nonumber \\
&\  & \ \ \  \ \ \ \ \ \ \ \ \ \ \   \ -\ri (\phi \Box \bar \phi) \bar \psi^{\a} \pa_{\a \b}  \psi^{\b}
-\ri (\bar \phi \Box  \phi)  \psi^{\a} \pa_{\a \b} \bar \psi^{\b}- (\psi^{\a} \pa_{\a \b} \bar \psi^{\b}) (\bar \psi^{\g} \pa_{\g \d}  \psi^{\d}) 
\Big\}\,.
\label{C2}
\eea
The equation of motion for $\bar F$ is 
\bea
&&-2 F f^2 + 2 F^2 \bar F  +F  \bar \phi \Box \phi + F  \phi \Box \bar \phi  - 2 \ri  F \psi^{\a} \pa_{\a \b} \bar \psi^{\b} - \ri F \bar \psi^{\a} \pa_{\a \b} \psi^{\b}
\nonumber \\
&& + 
\phi  \Box (\bar \phi F) - \ri \pa_{\a \b} F \psi^{\a} \bar \psi^{\b} -  \bar \phi \psi^{\a} \Box \psi_{\a} 
- \pa^{\a \g} \bar \phi \psi_{\g} \pa_{\a \b} \psi^{\b}=0\,.
\label{C3}
\eea
It shows that $F$ and $\bar F$ are no longer auxiliary fields since they 
cannot be expressed in terms of the off-shell physical fields 
$\phi$ and $\psi_\a$ and their conjugates. 
One could try to look for $F$ as a series in powers of the fields 
$\phi$,  $\psi_\a$ and their derivatives, 
\be 
F= f + a_1 \phi \Box \bar \phi +\ a_2  \bar \phi \Box  \phi + a_3 \psi^{\a} \pa_{\a \b} \bar \psi^{\b} + a_4 \bar \psi^{\a} \pa_{\a \b} \psi^{\b} + \dots \,,
\label{C4}
\ee
where $a_1,  a_2, \dots $ are some constants which have to be found by  
substituting \eqref{C4} into~\eqref{C3} and working 
order by order in perturbation theory.
Such a solution would correspond to a supersymmetry breaking phase 
(note that  $F=f$ for  $\phi=0, \ \psi_{\a} =0$). 
However, it is not difficult to show that no solution 
for $F$ exists:  substitution of~\eqref{C4} into~\eqref{C3} 
yields inconsistent equations 
\be 
2 a_2 + 2 \bar a_1 + f^{-1}=0\,, \qquad   2 a_1 + 2 \bar a_2 + 2 f^{-1}=0\,,
\label{C4.1}
\ee
and similarly for $a_3, a_4$. This means that it is impossible to solve the equation of motion for 
the 
field  $F$ and substitute the solution into eq.~\eqref{C2} 
to find the action for the off-shell physical fields. 
The procedure of eliminating 
the auxiliary field
$F$ can be fulfilled only when the physical 
fields are also on-shell. 
In other words, the equation \eqref{C3} and its conjugate have
to be solved in conjunction with the equations of motion for the physical fields, 
and then the above inconsistencies do not occur.
In doing so, we will obtain correctly normalised kinetic terms for the physical fields.
Indeed, since in the supersymmetry 
breaking phase $F = f +\dots$, for the relevant terms in \eqref{C2}
we get 
\bea
 & & 2\int \rd^3 x \,[  \pa^a \phi \pa_a \bar \phi +  \ri \psi^{\a} \pa_{\a \b} \bar \psi^{\b}]
\nonumber \\
&  & +\frac{1}{f^2} \int \rd^3 x \,\Big\{ 
F \bar F \Big(\bar \phi \Box \phi +  \phi \Box \bar \phi 
-\frac{3 \ri}{2}  (\psi^{\a} \pa_{\a \b} \bar \psi^{\b}  
+
\bar \psi^{\a} \pa_{\a \b} \psi^{\b} )
\Big) 
+ \phi \bar F \Box (\bar \phi F)\Big\}
\nonumber \\
&=&  -\int \rd^3 x \,[  \pa^a \phi \pa_a \bar \phi +  \ri \psi^{\a} \pa_{\a \b} \bar \psi^{\b}]
+\dots
\label{C.7}
\eea
where the ellipsis stands for cubic and higher order terms in the fields 
$\f$, $\j_\a$ and their conjugates.

We now restrict our study to the case of model \eqref{C1} 
with $X$ chosen to be nilpotent, 
\bea
X^2=0~.
\label{C.8}
\eea
Then $\phi$ can be expressed as in~\eqref{A5} and we have a reducible Goldstino model. The component action of this model is given by~\eqref{C2}
with $\phi$ replaced according to eq.~\eqref{A5}. The equation for the auxiliary field now reads
\bea
&&
-2 f^2 F +\frac{f^2}{2}\frac{\bar \psi^2}{\bar F^2} \Box \frac{\psi^2}{F} +\frac{1}{4} \psi^2 \Box \frac{\bar \psi^2}{\bar F} -
\frac{1}{4} \frac{\bar \psi^2}{\bar F^2} \Box (\bar F \psi^2)
\nonumber \\
&&
-\frac{1}{8}\frac{\psi^2 \bar \psi^2}{F^2 \bar F^3} \Box \psi^2 \Box \bar \psi^2 +2 F^2 \bar F 
+\frac{1}{4} \frac{\psi^2}{F} \Box \frac{F \bar \psi^2}{\bar F} -\frac{1}{4} \frac{F \bar \psi^2}{\bar F^2} \Box \frac{\bar F \psi^2}{F}
\nonumber \\
&&
-\frac{3 \ri}{2} F \psi^{\a}\pa_{\a \b} \bar \psi^{\b} -\frac{3 \ri}{2} F \bar \psi^{\a}\pa_{\a \b}  \psi^{\b} 
-\frac{\ri}{2} \psi^{\a} \bar \psi^{\b} (\pa_{\a \b} F) -\frac{\ri}{2} \pa_{\a \b} (F \psi^{\a} \bar \psi^{\b} )
\nonumber \\
&&
-\frac{1}{2} \pa^{\a \g} \Big( \frac{\bar \psi^2}{\bar F}\Big) \psi_{\g} \pa_{\a \b} \psi^{\b}
-\frac{1}{2} \frac{\bar \psi^2}{\bar F^2}   \pa^{\a \g} (\bar F \psi_{\g} \pa_{\a \b} \psi^{\b} )
-\frac{\ri}{8} \frac{\bar \psi^2}{\bar F^2} \pa^{\a \g} \Big(\frac{\psi^2}{F}\Big) \pa_{\a \b} \bar  \psi^{\b}  \pa_{\g \d}  \psi^{\d} 
\nonumber \\
&&
-\frac{\ri}{8} \frac{\bar \psi^2}{\bar F^2} \pa^{\a \g}\Big(\frac{\psi^2}{F} \pa_{\a \b} \bar  \psi^{\b}  \pa_{\g \d}  \psi^{\d}   \Big)
-\frac{\ri}{8 F \bar F^2} \psi^2 \bar \psi^2 [\pa_{\a \b} \bar \psi^{\b} \Box\psi^{\a} + \pa_{\a \b} \psi^{\b} \Box \bar \psi^{\a}]
\nonumber \\
&&
+\frac{\ri}{4} \bar{\psi^2}{\bar F}^{-2} \Box \Big[ \frac{\psi^2}{F} \bar \psi^{\a} \pa_{\a \b} \psi^{\b} \Big] 
+\frac{\ri}{4}  \bar{\psi^2}{\bar F}^{-2}  \psi^{\a} \pa_{\a \b} \bar \psi^{\b} \Box  \frac{\psi^2}{F} =0\,. 
\label{C8.01}
\eea
One can show that just like in the case of eq.~\eqref{C3} it is not possible to solve this equation for $F$ in terms of the physical fields $\psi$ and $\bar \psi$. 
The procedure of eliminating the field $F$ can be performed only if the Goldstino is on-shell. 
Therefore, we will follow the other approach: 
instead of considering the equations of motion for $F$, 
we will
require that the action~\eqref{C1} be stationary under re-scaling $ X \to \re^{\tau} X$, which yields
\be 
\bar D^2 \Big( X \bar X D^2 X \bar D^2 \bar X - 16 f^2 X \bar X \Big)=0\,. 
\label{C6}
\ee
Since $\bar D^2 \bar X$ is nowhere vanishing, this condition is equivalent to 
\bea 
X \bar D^2 \big(  \bar X D^2 X \big) = 16 f^2 X\,. 
\label{C66}
\eea
The problem of solving eqs.~\eqref{C6}, \eqref{C66} can be reformulated as follows. Let us define the superfield $Y$ by the rule
\be 
-\frac{1}{4} X \bar D^2 \bar X= f Y\,. 
\label{CCC1}
\ee
It then follows from eq.~\eqref{C6} that $Y$ has the properties
\be 
\bar D_{\a}Y=0\,, \quad Y^2=0\,, \quad -\frac{1}{4} Y \bar D^2 \bar Y = f Y\,. 
\label{CCC2}
\ee
That is $Y$ defines an irreducible Goldstino multiplet whose auxiliary field is uniquely solved in terms of the Goldstini with the solution given in eq.~\eqref{A10.0}. 
Therefore, the problem can be stated as to find $X$ using eq.~\eqref{CCC1} given $Y$. Comparing eqs.~\eqref{CCC1} and~\eqref{CCC2} 
we see that there is an obvious solution $X=Y$.\footnote{Note that if $X$ is a solution to~\eqref{CCC1} then so is $-X$. Hence, we have two supersymmetry breaking phases. 
For concreteness we select the phase in which $\langle F  \rangle =f$.} 
However, this solution is not unique. To show it we will examine eq.~\eqref{C66} in components. 
Let us consider the equation for $F$ followed from~\eqref{C66}. We obtain
\bea
&&
-2 f^2 F + 2 F^2 \bar F -2  \ri  F \psi^{\a} \pa_{\a \b} \bar \psi^{\b} -2  \ri  F \bar \psi^{\a} \pa_{\a \b} \psi^{\b} -2 \ri 
(\pa_{\a \b} F) \psi^{\a} \bar \psi^{\b} 
\nonumber \\
&&
+\frac{1}{2} \frac{F \bar \psi^2}{\bar F} \Box \frac{\psi^2}{F} +\frac{1}{2} \frac{  \psi^2}{F} \Box \frac{F \bar \psi^2}{\bar F} 
+\psi^{\a} \pa_{\a \b} \Big( \frac{\bar \psi^2}{\bar F} \pa^{\b \g} \psi_{\g}\Big)=0\,. 
\label{C66.01}
\eea
Note that we cannot solve this equation by repeated substitution. 
However, we can solve it by expanding $F$ in powers in the Goldstino and its derivatives
\be 
F = f + a_1 (\ri  \psi^{\a} \pa_{\a \b} \bar \psi^{\b} )+ a_2 (\ri  \bar \psi^{\a} \pa_{\a \b}  \psi^{\b}) +\dots\,. 
\label{C66.02}
\ee
Since $\psi$ is nilpotent this expansion is finite. Substituting it into~\eqref{C66.01} we can fix the coefficients. From the analysis presented above 
we know that there is a solution for $F$ given by~\eqref{A10.0}. Therefore, we will look for a solution in the form of~\eqref{A10.0}:
\bea
F &= & f + a_1 \langle u \rangle  + a_2 \langle \bar u \rangle + a_3 \langle u  \rangle  \langle \bar u \rangle + 
a_4 \psi^2 \Box \bar \psi^2 + a_5 \bar \psi^2 \Box  \psi^2
\nonumber \\
& + & a_6 (\langle u \rangle^2  \langle \bar u \rangle  + \langle \bar u \rangle^2 \langle u \rangle ) + a_7 
\langle \bar u \rangle  \psi^2 \Box \bar \psi^2 + a_8  \langle u  \rangle \bar \psi^2 \Box  \psi^2
+ a_9 \bar \psi^2 \Box (\psi^2 \langle \bar u \rangle )
\nonumber \\
& + & a_{10} \langle u \rangle^2 \langle \bar u\rangle^2 + a_{11} \psi^2 \bar \psi^2 \Box (\langle u \rangle^2 - \langle u\rangle  \langle \bar u \rangle  + 
\langle \bar u\rangle^2) + a_{12} \psi^2 \bar \psi^2 \Box \psi^2 \Box \bar \psi^2\,. 
\label{CCC3}
\eea
Substituting this ansatz into~\eqref{C66.01} we find that the coefficients $a_1$, $ a_2$, $ a_3$,  $ a_6$,  $ a_8$,  $a_9$,  $a_{10}$,  $ a_{11}$
are fixed as in~\eqref{A10.0}, whereas the remaining coefficients satisfy 
\be 
a_4 = -\frac{1}{4} f^{-3}- a_5 \,, \quad a_7 = -\frac{1}{4} f^{-5} -2 f^{-2} a_5\,, \quad a_{12} =\frac{1}{2} f^{-4} a_5- a_5^2
\label{CCC4}
\ee
and cannot be fixed uniquely. The solution~\eqref{A10.0} corresponds to $a_4=0$, $ a_5= -\frac{1}{4} f^{-3}$, $  a_7 =\frac{1}{4} f^{-5}$, $ a_{12}= -\frac{3}{16} f^{-7}$.
The ambiguity that we can have more than one solution to~\eqref{C66.01} 
is expected  be related to the fact that we can add to $F$ and to the action terms proportional to the equations
of motion as in~\eqref{AAA1} and  \eqref{AAA2}, 
but we will not discuss this issue in detail in this paper.

Let us now clarify why eq.~\eqref{C66}, or equivalently eq. \eqref{C66.01}, 
has a solution for $F$ despite the fact that eq.~\eqref{C8.01} does not.
For this we will consider the equation of motion for the superfield $X$. Since $X$ is nilpotent to find it we have to add the term
\be 
\int \rd^3 x \rd^2 \theta\, \lambda X^2 + {\rm c.} {\rm c.}
\label{C6.0}
\ee
to the action~\eqref{C1},   where $\lambda$ is a Lagrange multiplier. 
Thus, we obtain the following equation of motion for $X$:
\be 
\bar D^2 [\bar X   D^2 X  \bar D^2 \bar X] + \bar D^2  D^2 [\bar X   X  \bar D^2 \bar X] 
- 32 f^2 \bar D^2 \bar X - 128 f^2 \l X =0\,. 
\label{C7}
\ee
Multiplying it by $X$ we get the constraint~\eqref{C6}. 
However, the equation of motion for $X$ contains not just the equation of motion for $F$~\eqref{C8.01} but also the equation for the Goldstino. 
Hence, in obtaining eq.~\eqref{C66.01} equations of motion for both $F$ and $\psi$ are taken into account and 
that is why it has a solution. 

With the nilpotency condition \eqref{C.8} imposed,  
the action \eqref{C1} can be rewritten as
\bea 
S= \int \rd^3 x \rd^2 \theta  \rd^2 \bar \theta \,
\Big\{ \frac{1}{16 f^2} D^\a X D_\a X  
\bar D_{\b} \bar X \bar D^{\b} \bar X - 2 \bar X X \Big\}~.
\label{C.11}
\eea
Similar supersymmetric higher derivative models have been considered 
in the literature in the case when $X$ is an unconstrained chiral superfield. 
In particular, an action of the type \eqref{C.11}
 was studied in~\cite{Farakos:2012qu}. In their case they could solve for the auxiliary field in terms of the off-shell physical scalar field provided the fermions were ignored. However, if we take into account the  fermions as well we can show 
that it is also impossible to solve for the auxiliary field unless the fermions are on-shell.
Unlike in our case, eq. \eqref{C.7},
in the model studied in~\cite{Farakos:2012qu}  the kinetic term for scalars completely canceled in the supersymmetry breaking phase.
Ref. \cite{FFKP} studied a model with canonically normalised kinetic term.
It is obtained from \eqref{C.11} by replacement $- 2 \bar X X \to  \bar X X$.
It was shown in \cite{FFKP} that the resulting model cannot 
break supersymmetry.


\begin{footnotesize}

\end{footnotesize}


\begin{thebibliography}{66}

\bibitem{GreenSchwarz} 
  M.~B.~Green and J.~H.~Schwarz,
  ``Covariant description of superstrings,''
  Phys.\ Lett.\  {\bf 136B}, 367 (1984).

\bibitem{GSW} 
  M.~B.~Green, J.~H.~Schwarz and E.~Witten,
  {\it Superstring Theory}, vols 1 \& 2,
  Cambridge University Press (1987) 
  (Cambridge Monographs On Mathematical Physics).

\bibitem{MT1} 
  L.~Mezincescu and P.~K.~Townsend,
  ``Anyons from strings,''
  Phys.\ Rev.\ Lett.\  {\bf 105}, 191601 (2010)
  [arXiv:1008.2334 [hep-th]].

\bibitem{MT2} 
  L.~Mezincescu and P.~K.~Townsend,
  ``Quantum 3D superstrings,''
  Phys.\ Rev.\ D {\bf 84}, 106006 (2011)
  [arXiv:1106.1374 [hep-th]].

\bibitem{deAL1} 
  J.~A.~de Azcarraga and J.~Lukierski,
    ``Supersymmetric particles with internal symmetries and central charges,''
 Phys.\ Lett.\  {\bf 113B}, 170 (1982).
 
\bibitem{deAL2}  
  J.~A.~de Azcarraga and J.~Lukierski,
  ``Supersymmetric particles in $N=2$ superspace: Phase space variables and Hamiltonian dynamics,''
  Phys.\ Rev.\ D {\bf 28}, 1337 (1983).

\bibitem{Siegel:1983hh}
  W.~Siegel, ``Hidden local supersymmetry in the supersymmetric particle action,''
  Phys.\ Lett.\  {\bf 128B} (1983) 397.


\bibitem{Sezgin:1993xg}
  E.~Sezgin,  ``Aspects of kappa symmetry,'' in {\it Salamfestschrift}, 
  A. Ali, J. Ellis and S. Randjbar-Daemi (Eds.), World Scientific, 1994, pp. 478--498,
  hep-th/9310126.
  
\bibitem{Sorokin} 
  D.~P.~Sorokin,
  ``Superbranes and superembeddings,''
  Phys.\ Rept.\  {\bf 329}, 1 (2000)
  [hep-th/9906142].
  

\bibitem{Henneaux:1984mh} 
  M.~Henneaux and L.~Mezincescu,
  ``A $\sigma$-model interpretation of Green-Schwarz covariant 
  superstring action,''
  Phys.\ Lett.\  {\bf 152B}, 340 (1985).

\bibitem{Witten:1985nt} 
  E.~Witten, ``Twistor-like transform in ten dimensions,''
  Nucl.\ Phys.\ B {\bf 266}, 245 (1986).

\bibitem{Grisaru:1985fv}
  M.~T.~Grisaru, P.~S.~Howe, L.~Mezincescu, B.~Nilsson and P.~K.~Townsend,
``N=2 superstrings in a supergravity background,''
  Phys.\ Lett.\  {\bf 162B} (1985) 116.

\bibitem{Bergshoeff:1985su} 
  E.~Bergshoeff, E.~Sezgin and P.~K.~Townsend,
``Superstring actions in $D=3$, 4, 6, 10 curved superspace,''
  Phys.\ Lett.\  {\bf 169B}, 191 (1986).

\bibitem{Wulff:2016tju} 
A.~A.~Tseytlin and L.~Wulff,
  ``Kappa-symmetry of superstring sigma model and generalized 10d supergravity equations,''
  JHEP {\bf 1606}, 174 (2016)
  [arXiv:1605.04884 [hep-th]].
  
\bibitem{KT-M11}
S.~M.~Kuzenko and G.~Tartaglino-Mazzucchelli,
  ``Three-dimensional N=2 (AdS) supergravity and associated supercurrents,''
JHEP {\bf 1112}, 052 (2011)
[arXiv:1109.0496 [hep-th]].
  
 \bibitem{ADFS}
I.~Antoniadis, E.~Dudas, S.~Ferrara and A.~Sagnotti,
  ``The Volkov-Akulov-Starobinsky supergravity,''
Phys.\ Lett.\ B {\bf 733}, 32 (2014) 
[arXiv:1403.3269 [hep-th]].
     
\bibitem{DFKS} 
 E.~Dudas, S.~Ferrara, A.~Kehagias and A.~Sagnotti,
 ``Properties of nilpotent supergravity,''
JHEP {\bf 1509}, 217 (2015) [arXiv:1507.07842 [hep-th]].

 \bibitem{BFKVP} 
E.~A.~Bergshoeff, D.~Z.~Freedman, R.~Kallosh and A.~Van Proeyen,
``Pure de Sitter supergravity,''
  Phys.\ Rev.\ D {\bf 92}, no. 8, 085040 (2015)
  Erratum: [Phys.\ Rev.\ D {\bf 93}, no. 6, 069901 (2016)]
  [arXiv:1507.08264 [hep-th]]. 
  
\bibitem{HY} 
  F.~Hasegawa and Y.~Yamada,
  ``Component action of nilpotent multiplet coupled to matter in 4 dimensional $ \mathcal{N}=1 $ supergravity,''
  JHEP {\bf 1510}, 106 (2015)
  [arXiv:1507.08619 [hep-th]].  
  
\bibitem{K15} 
S.~M.~Kuzenko,
``Complex linear Goldstino superfield and supergravity,''
JHEP {\bf 1510}, 006 (2015)
[arXiv:1508.03190 [hep-th]].

\bibitem{KW} 
  R.~Kallosh and T.~Wrase,
  ``de Sitter supergravity model building,''
  Phys.\ Rev.\ D {\bf 92}, no. 10, 105010 (2015)
  [arXiv:1509.02137 [hep-th]].
  
\bibitem{SvdWW} 
  M.~Schillo, E.~van der Woerd and T.~Wrase,
  ``The general de Sitter supergravity component action,''
  Fortsch.\ Phys.\  {\bf 64}, 292 (2016)
  [arXiv:1511.01542 [hep-th]].  

\bibitem{BMST} 
I.~Bandos, L.~Martucci, D.~Sorokin and M.~Tonin,
``Brane induced supersymmetry breaking and de Sitter supergravity,''
JHEP {\bf 1602}, 080 (2016)
[arXiv:1511.03024 [hep-th]].

\bibitem{FKRR} 
  F.~Farakos, A.~Kehagias, D.~Racco and A.~Riotto,
  ``Scanning of the supersymmetry breaking scale and the gravitino mass in supergravity,''
  JHEP {\bf 1606}, 120 (2016)
  [arXiv:1605.07631 [hep-th]].

\bibitem{BHKMS} 
I.~Bandos, M.~Heller, S.~M.~Kuzenko, L.~Martucci and D.~Sorokin,
``The Goldstino brane, the constrained superfields and matter in $ \mathcal{N}=1 $ supergravity,''    JHEP {\bf 1611}, 109 (2016)
  [arXiv:1608.05908 [hep-th]].

\bibitem{KMcAT-M} 
  S.~M.~Kuzenko, I.~N.~McArthur and G.~Tartaglino-Mazzucchelli,
  ``Goldstino superfields in N=2 supergravity,''   JHEP {\bf 1705}, 061 (2017)
[arXiv:1702.02423 [hep-th]].

\bibitem{BK17} 
  E.~I.~Buchbinder and S.~M.~Kuzenko,
  ``Three-form multiplet and supersymmetry breaking,''
JHEP {\bf 1709}, 089 (2017)
  [arXiv:1705.07700 [hep-th]].

\bibitem{LR}
U.~Lindstr\"om and M.~Ro\v{c}ek,
``Constrained local superfields,''
Phys.\ Rev.\  D {\bf 19}, 2300 (1979).


\bibitem{SW}
 S.~Samuel and J.~Wess,
``A superfield formulation of the non-linear
realization of supersymmetry and its coupling to supergravity,''  
    Nucl.\ Phys.\  B {\bf 221}, 153  (1983).

\bibitem{KTyler} 
  S.~M.~Kuzenko and S.~J.~Tyler,
  ``Complex linear superfield as a model for Goldstino,''
  JHEP {\bf 1104}, 057 (2011) [arXiv:1102.3042 [hep-th]].




\bibitem{VA}
D.~V.~Volkov and V.~P.~Akulov,
``Possible universal neutrino interaction,''
  {JETP Lett.\  {\bf 16}, 438 (1972)}   
  [Pisma Zh.\ Eksp.\ Teor.\ Fiz.\   {\bf 16},  621 (1972)]; 
  ``Is the neutrino a Goldstone particle?,''
  Phys.\ Lett.\  B {\bf 46}, 109 (1973).

\bibitem{AV}
V.~P. Akulov and D.~V. Volkov, ``Goldstone fields with spin 1/2,''
   Theor. Math. Phys. {\bf 18}, 28 (1974)  28 [Teor. Mat. Fiz. {\bf 18}, 39 (1974)].

\bibitem{VS} 
  D.~V.~Volkov and V.~A.~Soroka,
  ``Higgs effect for Goldstone particles with spin 1/2,''
  JETP Lett.\  {\bf 18}, 312 (1973)
  [Pisma Zh.\ Eksp.\ Teor.\ Fiz.\  {\bf 18}, 529 (1973)].

\bibitem{VS2}
D.~V. Volkov and V.~A. Soroka, ``Gauge fields for symmetry group with
  spinor parameters,''   Theor. Math. Phys. {\bf 20} (1974)  829
[Teor. Mat. Fiz. {\bf 20}, 291(1974)].

\bibitem{DZ} 
S.~Deser and B.~Zumino,
``Broken supersymmetry and supergravity,''
Phys.\ Rev.\ Lett.\  {\bf 38}, 1433 (1977).



\bibitem{GGRS}
S.~J.~Gates Jr., M.~T.~Grisaru, M.~Ro\v{c}ek and W.~Siegel,
{\it Superspace, or One Thousand and One Lessons in Supersymmetry},
Benjamin/Cummings (Reading, MA),  1983, hep-th/0108200.    

 \bibitem{KLT-M11}
S.~M.~Kuzenko, U.~Lindstr\"om and G.~Tartaglino-Mazzucchelli,
  ``Off-shell supergravity-matter couplings in three dimensions,''
  JHEP {\bf 1103}, 120 (2011)
  [arXiv:1101.4013 [hep-th]].
  
\bibitem{vanN85}
P.~van Nieuwenhuizen,
``D = 3 conformal supergravity and Chern-Simons terms,''
Phys.\ Rev.\  D {\bf 32}, 872 (1985).
  
\bibitem{Kuzenko:2012ew} 
  S.~M.~Kuzenko and G.~Tartaglino-Mazzucchelli,
``Conformal supergravities as Chern-Simons theories revisited,''
  JHEP {\bf 1303}, 113 (2013)
  [arXiv:1212.6852 [hep-th]].


\bibitem{BKNT-M2}
D.~Butter, S.~M.~Kuzenko, J.~Novak and G.~Tartaglino-Mazzucchelli,
 ``Conformal supergravity in three dimensions: Off-shell actions,''
  JHEP {\bf 1310}, 073 (2013)
  [arXiv:1306.1205 [hep-th]].
  

\bibitem{Castellani} 
  L.~Castellani, R.~D'Auria and P.~Fre,
{\it Supergravity and superstrings: A Geometric perspective. Vol. 2: Supergravity},
World Scientific,  Singapore, 1991, pp. 680--684. 



\bibitem{Hasler}
M.~F.~Hasler,
  ``The three-form multiplet in N=2 superspace,''
  Eur.\ Phys.\ J.\ C {\bf 1}, 729 (1998)
  [hep-th/9606076].
  
\bibitem{Ectoplasm} 
S.~J.~Gates Jr., ``Ectoplasm has no topology: The prelude,''
in {\it Supersymmetries and Quantum Symmetries},
 J. Wess and E. A. Ivanov (Eds.), Springer, Berlin, 1999, pp. 46--57, 
 arXiv:hep-th/9709104;
``Ectoplasm has no topology,''
 Nucl.\ Phys.\  B {\bf 541}, 615 (1999)
 [arXiv:hep-th/9809056].
 
\bibitem{GGKS}
S.~J.~Gates Jr., M.~T.~Grisaru, M.~E.~Knutt-Wehlau and W.~Siegel,
``Component actions from curved superspace: Normal coordinates and
ectoplasm,'' Phys.\ Lett.\  B {\bf 421}, 203 (1998)
[hep-th/9711151].

  \bibitem{Becker:2003wb}
 M.~Becker, D.~Constantin, S.~J.~Gates Jr., W.~D.~Linch III, W.~Merrell and J.~Phillips,
 ``M-theory on Spin(7) manifolds, fluxes and 3D, N = 1 supergravity,''
  Nucl.\ Phys.\  B {\bf 683}, 67 (2004)
  [arXiv:hep-th/0312040].



\bibitem{KNT-M15} 
  S.~M.~Kuzenko, J.~Novak and G.~Tartaglino-Mazzucchelli,
  ``Higher derivative couplings and massive supergravity in three dimensions,''
  JHEP {\bf 1509}, 081 (2015)
  [arXiv:1506.09063 [hep-th]].



\bibitem{KLT-M12} 
  S.~M.~Kuzenko, U.~Lindstr\"om and G.~Tartaglino-Mazzucchelli,
``Three-dimensional (p,q) AdS superspaces and matter couplings,''
  JHEP {\bf 1208}, 024 (2012)
  [arXiv:1205.4622 [hep-th]].


\bibitem{BV}
  I.~A.~Batalin and G.~A.~Vilkovisky,
  ``Quantization of gauge theories with linearly dependent generators,''
  Phys.\ Rev.\  {\bf D28},   2567 (1983).
  
\bibitem{new}
M.~F.~Sohnius and P.~C.~West,
``An alternative minimal off-shell version of N=1 supergravity,''
Phys.\ Lett.\  B {\bf 105}, 353 (1981).
  
\bibitem{SohniusW2} 
M.~F.~Sohnius and P.~C.~West,
 ``The new minimal formulation of N=1 supergravity and its tensor calculus,''    
 in {\it Quantum Structure of Space and Time}, M. J. Duff and C. J. Isham (Eds.), 
Cambridge University Press, Cambridge, 1982, pp. 187--222. 
    
\bibitem{SohniusW3} 
 M.~Sohnius and P.~C.~West,
 ``The tensor calculus and matter coupling of the alternative minimal auxiliary field formulation of $N=1$ supergravity,''  Nucl.\ Phys.\ B {\bf 198}, 493 (1982).    


\bibitem{AT}
  A.~Ach\'ucarro and P.~K.~Townsend,
  ``A Chern-Simons action for three-dimensional AdS supergravity
 theories,''
  Phys.\ Lett.\  B {\bf 180}, 89 (1986).
  
\bibitem{HIPT}
P.~S.~Howe, J.~M.~Izquierdo, G.~Papadopoulos and P.~K.~Townsend,
``New supergravities with central charges and Killing spinors in 2+1 dimensions,''
Nucl.\ Phys.\  B {\bf 467}, 183 (1996)
  [arXiv:hep-th/9505032].
  

\bibitem{RvanN86} 
  M.~Ro\v{c}ek and P.~van Nieuwenhuizen,
  ``N $\geq$ 2 supersymmetric Chern-Simons terms as d = 3 extended conformal supergravity,''
  Class.\ Quant.\ Grav.\  {\bf 3}, 43 (1986).
   

\bibitem{Kuzenko12} 
S.~M.~Kuzenko,
``Prepotentials for N=2 conformal supergravity in three dimensions,''
JHEP {\bf 1212}, 021 (2012)  [arXiv:1209.3894 [hep-th]].  


\bibitem{KT-M-2008-2}
 S.~M.~Kuzenko and G.~Tartaglino-Mazzucchelli,
``Different representations for the action principle in 4D N = 2 supergravity,''
  JHEP {\bf 0904}, 007 (2009)
  [arXiv:0812.3464 [hep-th]].
  


\bibitem{Siegel}
W.~Siegel,
``Solution to constraints in Wess-Zumino supergravity formalism,''
Nucl.\ Phys.\  B {\bf 142}, 301 (1978). 

\bibitem{Zumino78} B. Zumino, ``Supergravity and superspace,'' in 
{\it Recent Developments in Gravitation}, Carg\`ese 1978, 
M. L\'evy and S. Deser (Eds.), Plenum Press, New York, 1979, pp. 405--459.

\bibitem{KLRST-M13}
S.~M.~Kuzenko, U.~Lindstr\"om, M.~Ro\v{c}ek, I.~Sachs and G.~Tartaglino-Mazzucchelli,
``Three-dimensional $\mathcal{N} =$ 2 supergravity theories: 
From superspace to components,''
  Phys.\ Rev.\ D {\bf 89}, no. 8, 085028 (2014)
  [arXiv:1312.4267 [hep-th]].

  
 
 
\bibitem{WZ}
J.~Wess and B.~Zumino,
 ``Superfield Lagrangian for supergravity,''
 Phys.\ Lett.\  B {\bf 74}, 51 (1978).


\bibitem{old1}
K.~S.~Stelle and P.~C.~West,
``Minimal auxiliary fields for supergravity,''
Phys.\ Lett.\  B {\bf 74},  330 (1978).

\bibitem{old2}
S.~Ferrara and P.~van Nieuwenhuizen,
``The auxiliary fields of supergravity,''
Phys.\ Lett.\  B {\bf 74}, 333 (1978).
 
\bibitem{ZupnikPak} 
  B.~M.~Zupnik and D.~G.~Pak,
  ``Superfield formulation of the simplest three-dimensional gauge theories 
  and conformal supergravities,''
  Theor.\ Math.\ Phys.\  {\bf 77}, 1070 (1988)
  [Teor.\ Mat.\ Fiz.\  {\bf 77}, 97 (1988)].



\bibitem{BKdual}
 D.~Butter and S.~M.~Kuzenko,
  ``A dual formulation of supergravity-matter theories,''
  Nucl.\ Phys.\ B {\bf 854}, 1 (2012)
  [arXiv:1106.3038 [hep-th]].
 
 
\bibitem{Gates} 
  S.~J.~Gates Jr.,
  ``Super $p$-form gauge superfields,''
  Nucl.\ Phys.\ B {\bf 184}, 381 (1981).

\bibitem{GS}
S.~J.~Gates Jr. and W.~Siegel,
``Variant superfield representations,''
Nucl.\ Phys.\ B {\bf 187}, 389 (1981).

 

\bibitem{BK88}
I.~L.~Buchbinder and S.~M.~Kuzenko,
 ``Quantization of the classically equivalent theories in the superspace of simple supergravity and quantum equivalence,''
Nucl.\ Phys.\ B {\bf 308}, 162 (1988). 


 
 \bibitem{Ideas} 
I.~L. Buchbinder and S.~M. Kuzenko, {\it Ideas and Methods of Supersymmetry and
Supergravity, Or a Walk Through Superspace},
 IOP, Bristol, 1995 (Revised Edition 1998).

\bibitem{Binetruy:1996xw} 
  P.~Binetruy, F.~Pillon, G.~Girardi and R.~Grimm,
``The 3-form multiplet in supergravity,''
  Nucl.\ Phys.\ B {\bf 477}, 175 (1996)
  [hep-th/9603181].


\bibitem{OvrutWaldram}  
B.~A.~Ovrut and D.~Waldram,
``Membranes and three-form supergravity,''
Nucl.\ Phys.\ B {\bf 506}, 236 (1997)
[hep-th/9704045].


\bibitem{KMcC} 
  S.~M.~Kuzenko and S.~A.~McCarthy,
  ``On the component structure of N=1 supersymmetric nonlinear electrodynamics,''
  JHEP {\bf 0505}, 012 (2005)
  [hep-th/0501172].



\bibitem{Binetruy:2000zx} 
P.~Binetruy, G.~Girardi and R.~Grimm,
``Supergravity couplings: A Geometric formulation,''
  Phys.\ Rept.\  {\bf 343}, 255 (2001)
  [hep-th/0005225].

\bibitem{FLMS} 
F.~Farakos, S.~Lanza, L.~Martucci and D.~Sorokin,
``Three-forms in supergravity and flux compactifications,''
  arXiv:1706.09422 [hep-th].

\bibitem{DD} 
T.~Dereli and S.~Deser,
``Fermionic Goldstone-Higgs effect in (2+1)-dimensional supergravity,''
  J.\ Phys.\ A {\bf 11}, L27 (1978).


\bibitem{KS}
  Z.~Komargodski and N.~Seiberg,
  ``From linear SUSY to constrained superfields,''
  JHEP {\bf 0909}, 066 (2009)
  [arXiv:0907.2441].



\bibitem{KT0}
S.~M.~Kuzenko and S.~J.~Tyler,
``On the Goldstino actions and their symmetries,''
  JHEP {\bf 1105} (2011) 055
  [arXiv:1102.3043 [hep-th]].


\bibitem{BFKVP2} 
E.~Bergshoeff, D.~Freedman, R.~Kallosh and A.~Van Proeyen,
  ``Construction of the de Sitter supergravity,''
  arXiv:1602.01678 [hep-th].


\bibitem{Kuzenko:2010ef} 
  S.~M.~Kuzenko and S.~J.~Tyler,
  ``Relating the Komargodski-Seiberg and Akulov-Volkov actions: Exact nonlinear field redefinition,''
  Phys.\ Lett.\ B {\bf 698}, 319 (2011)
  [arXiv:1009.3298 [hep-th]].

\bibitem{Rocek} 
M.~Ro\v{c}ek, ``Linearizing the Volkov-Akulov model,''
  Phys.\ Rev.\ Lett.\  {\bf 41}, 451 (1978).

\bibitem{AM} 
  I.~Antoniadis and C.~Markou,
  ``The coupling of non-linear supersymmetry to supergravity,''
  Eur.\ Phys.\ J.\ C {\bf 75}, no. 12, 582 (2015)
  [arXiv:1508.06767 [hep-th]].  

  
\bibitem{CDFP} 
  N.~Cribiori, G.~Dall'Agata, F.~Farakos and M.~Porrati,
  ``Minimal constrained supergravity,''
  Phys.\ Lett.\ B {\bf 764}, 228 (2017)
  [arXiv:1611.01490 [hep-th]].  

\bibitem{IK78}
E.~Ivanov and A.~Kapustnikov, ``General relationship between linear and
 nonlinear realisations of supersymmetry,'' J.\ Phys.\  A {\bfseries 11}   (1978) 2375.


\bibitem{IK82}
E.~Ivanov and A.~Kapustnikov, ``The nonlinear realisation structure of models
  with spontaneously broken supersymmetry,''
   J.\ Phys.\ G  {\bf 8} (1982) 167. 


\bibitem{Casalbuoni}
R.~Casalbuoni, S.~De Curtis, D.~Dominici, F.~Feruglio, and R.~Gatto,
 ``{Non-linear realization of supersymmetry algebra from supersymmetric constraint},'' 
   Phys.\ Lett.\  B 
  {\bfseries 220},  569 (1989).
  

\bibitem{DK} 
S.~Deser and J.~H.~Kay,
``Topologically massive supergravity,''
Phys.\ Lett.\ B {\bf 120}, 97 (1983).


\bibitem{Deser}
  S.~Deser,
  ``Cosmological topological supergravity,''
 in {\it Quantum Theory of Gravity}, S. M. Christensen (Ed.), 
 Adam Hilger, Bristol, 1984, pp. 374-381. 
 
 

\bibitem{Andringa:2009yc} 
  R.~Andringa, E.~A.~Bergshoeff, M.~de Roo, O.~Hohm, E.~Sezgin and P.~K.~Townsend,
  ``Massive 3D supergravity,''
  Class.\ Quant.\ Grav.\  {\bf 27}, 025010 (2010)
  [arXiv:0907.4658 [hep-th]].

\bibitem{BHRST10} 
  E.~A.~Bergshoeff, O.~Hohm, J.~Rosseel, E.~Sezgin and P.~K.~Townsend,
  ``More on massive 3D supergravity,''
  Class.\ Quant.\ Grav.\  {\bf 28}, 015002 (2011)
  [arXiv:1005.3952 [hep-th]].
  

  \bibitem{BOS14} 
  G.~Alkac, L.~Basanisi, E.~A.~Bergshoeff, M.~Ozkan and E.~Sezgin,
 ``Massive $ \mathcal{N} $ = 2 supergravity in three dimensions,''
  JHEP {\bf 1502}, 125 (2015)
  [arXiv:1412.3118 [hep-th]].
  

\bibitem{KT-M17} 
  S.~M.~Kuzenko and G.~Tartaglino-Mazzucchelli,
  ``New nilpotent ${\cal N}= 2$ superfields,''
  arXiv:1707.07390 [hep-th].
 
  
   
\bibitem{Duff} 
  M.~J.~Duff,
  ``The cosmological constant is possibly zero, but the proof is probably wrong,''
  Phys.\ Lett.\ B {\bf 226}, 36 (1989)
  [Conf.\ Proc.\ C {\bf 8903131}, 403 (1989)].  


\bibitem{Duncan:1989ug}
  M.~J.~Duncan and L.~G.~Jensen,
  ``Four-forms and the vanishing of the cosmological constant,''
  Nucl.\ Phys.\ B {\bf 336} (1990) 100.
  
\bibitem{Groh:2012tf}
  K.~Groh, J.~Louis and J.~Sommerfeld,
  ``Duality and couplings of 3-form-multiplets in N=1 supersymmetry,''
  JHEP {\bf 1305} (2013) 001
  [arXiv:1212.4639 [hep-th]].
 
\bibitem{Farakos:2012qu}
F.~Farakos and A.~Kehagias,
``Emerging potentials in higher-derivative gauged chiral models coupled to N=1 supergravity,'' JHEP {\bf 1211} (2012) 077
  [arXiv:1207.4767 [hep-th]].



\bibitem{FFKP} 
  F.~Farakos, S.~Ferrara, A.~Kehagias and M.~Porrati,
  ``Supersymmetry breaking by higher dimension operators,''
  Nucl.\ Phys.\ B {\bf 879}, 348 (2014)
  [arXiv:1309.1476 [hep-th]].  
  

\end{thebibliography}
\end{document}